\documentclass[aps,prc,onecolumn,superscriptaddress,nofootinbib,showpacs,showkeys,preprintnumbers]{revtex4-1}
\usepackage[normalem]{ulem}  
\usepackage[utf8]{inputenc}
\usepackage[LY1]{fontenc}
\usepackage{graphicx}
\usepackage{yfonts}
\usepackage{amsmath,amssymb}
\usepackage{mathtools}
\usepackage{bm}
\usepackage{url}
\usepackage{hyperref}
\usepackage{color}
\usepackage[usenames,dvipsnames]{xcolor}
\usepackage{here}
\usepackage{caption}
\usepackage{subcaption}
\usepackage[percent]{overpic}
\usepackage{transparent}
\usepackage{tikz}  
\usetikzlibrary{shapes, arrows, positioning, shapes.multipart, fit, arrows.meta, calc, decorations.markings, math}
\usepackage{lineno}  
\usepackage{draftcopy}  

\newcommand{\vkt}{\ensuremath{\vec k_T}}
\newcommand{\vlt}{\ensuremath{\vec\ell_T}}
\newcommand{\operp}{{\perp}}
\def\be{\begin{equation}}
	\def\ee{\end{equation}}
\def\bea{\begin{eqnarray}}
	\def\eea{\end{eqnarray}}

\begin{document}

\title{SUBA-Jet: a new Model for Jets in Heavy Ion Collisions}

\author{Iurii Karpenko}
\affiliation{SUBATECH, Nantes University, IMT Atlantique, IN2P3/CNRS, 4 rue Alfred Kastler, 44307 Nantes cedex 3, France}
\affiliation{FNSPE, Czech Technical University in Prague, B\v{r}ehov\'a 7, Prague 11519, Czech Republic}
\author{Alexander Lind}
\affiliation{SUBATECH, Universit\'e de Nantes, IMT Atlantique, IN2P3/CNRS, 4 rue Alfred Kastler, 44307 Nantes cedex 3, France}
\author{Martin Rohrmoser}
\affiliation{ Institute of Physics, Cracow University of Technology, Podchor\k{a}\.{z}ych 1, PL-30-084 Krak\'ow, Poland}
\author{Joerg Aichelin}
\author{Pol-Bernard Gossiaux}
\affiliation{SUBATECH, Universit\'e de Nantes, IMT Atlantique, IN2P3/CNRS, 4 rue Alfred Kastler, 44307 Nantes cedex 3, France}

\date{\today}

\begin{abstract} 
\noindent
We present a new model for jet quenching in a quark gluon plasma (QGP). The jet energy loss has two steps. The initial jet parton with a high virtuality loses energy by a perturbative vacuum parton shower modified by medium interactions until it becomes on shell. Subsequent energy loss originates from elastic and radiative collisions with the medium constituents. Coherency of the radiative collisions is achieved by starting with virtual gluons that act as field dressing of the initial jet parton. These are formed according to a Gunion-Bertsch seed. The QCD version of the LPM effect is obtained by increasing the phase of the virtual gluons through elastic scatterings with the medium. Above a phase threshold, the virtual gluons will be formed and can produce coherent radiation themselves. The model has been implemented in a Monte Carlo code and is validated by successfully reproducing the BDMPS-Z prediction for the energy spectrum of radiated gluons in a static medium. Results for the more realistic case, in which the assumptions of the BDMPS-Z approach are released, are also shown. We investigate the influence of various parameters on the energy spectrum and the transverse momentum distribution, such as the in-medium quark masses, the energy transfer in the recoil process, and the phase accumulation criteria, especially for low and intermediate energy gluons.
\end{abstract}

\maketitle


\section{Introduction} \label{sec:introduction}
In the last decade, experimental results brought compelling evidence that in ultra-relativistic heavy-ion collisions, a plasma of quarks and gluons (QGP) is produced~\cite{STAR:2011fbd,PHENIX:2006dpn,CMS:2016xef,Matsui:1986dk}. Such a state of matter has earlier been  predicted by lattice gauge calculations (lQCD). Modern lQCD calculations robustly predict a crossover transition between the hadronic and the QGP phases, which happens (for vanishing chemical potential) approximately in the temperature range $T=154\pm9$~MeV~\cite{Borsanyi:2013bia,Bazavov:2014pvz}.
This new QGP state has, however, a very short lifetime of the order of a couple of fm$/c$ before it transforms into hadrons, which are finally observed in the detectors. The experimental and theoretical study of the properties of the QGP is presently one of the major challenges in hadron physics. Since quarks and gluons cannot be observed directly, it is necessary to study them by the analysis of hadron observables and it turns out that the multiplicity of the majority of hadrons, those which contain only up, down and strange quarks is well described by the assumption that the expanding system is at hadronization in statistical equilibrium~\cite{Andronic:2005yp}. More information can be obtained from the detailed analysis of the results of transport approaches, which describe the time evolution of the QGP. Employing the technique of Bayesian analysis constraints on the shape of the initial state, shear~\cite{Bernhard:2016tnd} and bulk~\cite{Bernhard:2019bmu} viscosity of the QGP medium and its equation of state~\cite{Pratt:2015zsa} can be established.

Another way to study the properties of the QGP before hadronization offer probes, which do not come to equilibrium with the expanding QGP. Besides leptonic probes, which suffer from low statistics, the studies concentrate on the one side on fast open or hidden heavy flavor mesons, and on the other side on jets. Heavy flavor quarks are produced in hard processes between projectile and target nucleons with a transverse momentum distribution which is quite different from that expected if being in equilibrium with the QGP. The expansion time of the QGP is too short to bring heavy quarks to a complete equilibrium with the QGP. Therefore, the change of the transverse momentum spectrum of heavy quarks is an image of their interaction with the QGP~\cite{Nahrgang:2014vza}. 

Jets also present a possibility to study the properties of the QGP during its expansion.  They are
produced in elementary collisions if the constituents of the projectile and target nucleons, quarks and gluons, collide with a large momentum transfer. Thanks to the factorization theorem, this elementary process can be calculated in the framework of  perturbative quantum chromo dynamics (pQCD) if one knows the initial momentum distribution of the partons in the nucleon. After being produced, these fast partons ($g, q$) with an energy $E$, being off-shell, radiate gluons of energy $\omega$ and transverse momentum $k_\perp$ with a distribution
\begin{equation}
    \frac{\mathrm{d}I^{\text{vac}}}{\mathrm{d}\omega \, \mathrm{d}^2 k_\perp^2} \propto \frac{\alpha_s}{E}P_{g,q}(z)\frac{1}{k_\perp^2} \,,
\end{equation}
where $z=\omega/E$ is the gluon energy fraction, $P_{g,q}(z)$ are the standard splitting functions, and ``$\perp$'' will be used throughout this manuscript to designate directions transverse to the jet whose direction is assumed to be along $Oz$ (while ``T'' will refer to directions transverse to some specific parton in the jet, like the radiated gluons). Finally, the fast and emitted partons hadronize into a bunch of hadrons whose momentum distribution is given by fragmentation functions. For an overview of jet production in elementary collisions we refer to refs.~\cite{Ellis:2007ib,Ellis}.

If a jet is produced in relativistic heavy-ion collisions the process is more complex~\cite{Wiedemann:2000za,Gyulassy:2000er,Wang:2001ifa} because:
\begin{itemize}

    \item The initial parton distribution in nucleons embedded in a nucleus is not the same as in a free nucleon.
    
    \item After production and being still off-shell, means carrying a momentum $p=\sqrt{E^2-Q^2}$ where $Q^2$ is its virtuality,  the parton may interact with the constituents of the QGP. This may modify the virtuality of the parton and change the color correlations~\cite{MehtarTani:2011tz}.
    
    \item After having lost its virtuality by emission of gluons, the fast parton becomes on-shell. Subsequently, the parton may have elastic and inelastic (radiative) collisions with the QGP partons in a time-ordered sequence.
    
    \item Gluons, emitted in inelastic (radiative) collisions, need a time, called formation time, before becoming formed and to get an on-shell gluon. If this time is shorter  than the time between two subsequent collisions of the fast particle in the medium, the inelastic collisions are independent. We are then in the (QCD-equivalent of the) Bethe-Heitler (BH) domain, where the gluon emission is described by the Gunion-Bertsch (GB) approach~\cite{Gunion:1981qs}. If, on the contrary, this time is longer, then we observe quantum interference between the inelastic processes, leading to the QCD equivalent of the Landau-Pomeranchuk-Migdal (LPM) effect in QED~\cite{Landau:1953um,Migdal:1956tc}. This process has been studied in a specific kinematic limit by Baier-Dokshitzer-Mueller-Peigne-Schiff and Zakharov and is known under the name BDMPS-Z~\cite{Baier:1996kr,Baier:1996sk,Zakharov:1996fv,Zakharov:1997uu}. In the fully coherent limit, when the formation time is large as  compared to the size of the system, on the average the medium-induced gluons remain virtual and are not converted into on-shell gluons. 
    
    During this formation time, the system, composed of the fast parton and the emitted gluon, may interact with further QGP partons. The transverse momentum transfer in these collisions modifies its formation time, which is given for a single collision as $t_{f}= \omega/k_{\operp}^2$ (where $\omega$ is the energy of the gluon and $k_{\operp}$ its transverse momentum with respect to the fast parton). Because the further collisions increase $k_{\operp}^2$,  its formation time becomes in fact a dynamical quantity. Therefore, there exist three different $\omega$ intervals:  For small $\omega$ we observe gluon production in independent inelastic collisions, for larger values of $\omega$ we observe coherent radiation. For even larger $\omega$, when the average formation time exceeds the size of the medium, only fluctuations of the formation time can lead to the production of real gluons.\footnote{Strictly speaking, the ``category'' to which each gluon belongs depends on its detailed evolution in the $(\omega,\vec{k}_{\operp})$ space, implying that these $\omega$-categories are somehow schematic.}
    
\end{itemize}

In order to analyze heavy ion collisions these already complex processes, which are usually addressed in a static QCD medium of constant length in which the density is kept constant, have to be formulated in a transport approach, which can be applied to a rapidly expanding QGP.

The first steps towards such a study have been paved by Wiedemann~\cite{Wiedemann:2000za},  who showed that the BDMPS-Z approach, which is defined on a Keldysh time contour and hence difficult to translate into a transport approach because it requires the knowledge of the future development of the system, could be transformed into an opacity expansion, which can serve subsequently~\cite{Zapp:2011ya} as a basis for a numerical study of the LPM effect and its consequences for the energy loss of a fast particle in a static medium.

The BDMPS-Z approach, which studies the medium induced gluon emission in the coherent regime, describes the  energy distribution of the emitted gluons under the condition that 
\begin{equation}
   E \gg \omega \gg \vert \bm{k}_\perp \vert, \vert \bm{l}_\perp \vert \ge \Lambda_{\text{QCD}} \,,
\end{equation}
where $\vert \bm{k}_\perp \vert$ and $\vert \bm{l}_\perp \vert$ are the transverse momenta of the emitted gluon and of the gluon exchanged with a medium whose sources are considered as static. Under these conditions the authors find a  gluon distribution $\mathrm{d}N/\mathrm{d}\omega \propto \omega^{-3/2}$. In the Bethe-Heitler limit the spectrum varies as $\mathrm{d}N/\mathrm{d}\omega \propto \omega^{-1}$, independent of the above mentioned kinematic constrains. 

Wiedemann and Zapp~\cite{Zapp:2011ya} have demonstrated for a static medium that for the first order opacity expansion the functional form of the gluon spectra agrees under the corresponding kinematic constrains with that of the BDMPS-Z approach. They showed, however, as well that these constraints are not realistic for fast partons passing a QGP. If one relaxes these conditions the energy distribution of the emitted gluons is modified substantially.

It is one of the goals of ultra-relativistic heavy-ion collisions to study how fast partons, created in the interaction of the incoming protons, pass a plasma of quark and gluons, which is created in such collisions. To achieve this goal, one has to embed the  results obtained for static QGP matter in a transport approach, which describes the time evolution of the heavy ion collisions. It is the purpose of this paper to outline the model, which we have developed for this. It is designed to describe the evolution of a parton in an expanding medium and we will present here all elements of this approach.  
This approach is a hybrid approach because the energy loss of the parton has two origins. Being produced in heavy ion collisions, the parton has initially a high virtuality, which decreases due to the emission of gluons. Such vacuum-like emissions take also place in the absence of the QGP but its presence modifies the energy loss of the leading parton. The branching is considered to be virtuality ordered, similar to JEWEL \cite{Zapp:2008zz,Zapp:2012ak} and MATTER\cite{Majumder:2010qh}. This allows to embed the algorithm in future easily in the space-time evolution of the medium through which the jet travels. Spin effects are not considered in the present version of the approach.
When the partons have reached low virtuality, the evolution switches towards a time-ordered scheme of the BDMPS-Z type, where the collisions with the QGP create an additional induced energy loss due to the scattering of the partons with the constituents of the QGP.  This two step scenario is similar to the decomposition adopted in \textsc{JETSCAPE}~\cite{JETSCAPE:2017eso}. 

In section~\ref{sec:outline} we present an outline of our approach, based, for its high virtuality component, on the ideas of the YaJEM~\cite{Renk:2008pp} approach, while its low virtuality component follows the same strategy as developed in \cite{Zapp:2011ya}. We consider, however, the influence of the phase space limitation in the gluon emission, what allows us to explore the whole $(\omega,k_T)$ space. This permits as well to calculate the recoil of the QGP parton, which may serve as a source term in hydrodynamical calculations of the expanding QGP.  We will show that for gluon energies larger than the critical energy $\omega_c$ we reproduce the GLV results \cite{Gyulassy:2000er}, what was not addressed in \cite{Zapp:2011ya}, which considered the harmonic approximation.  Our approach allows to go beyond the eikonal limit for the kinematical variables as well as for the phase accumulation and can therefore handle the kinematical region explored in heavy ion collisions.  We study furthermore the influence of finite masses of the QGP partons.  Coming to the details of our approach,  we start out with a discussion of how we describe the time evolution of highly virtual partons and its modification in a QGP in section~\ref{sec:numerical}. Then, in section \ref{sec:lowQ}, we present all the ingredients of the low virtuality component, including the model adopted for elastic scattering, the elementary radiation amplitude of low-virtuality partons, which scatter with the QGP partons, and how these cross sections can be modeled by a Monte Carlo approach, necessary for its use in a transport theory. The quantum interference of the radiation amplitudes, the LPM effect, and its implementation in a transport theory are discussed in subsection~\ref{sec:lpm}. 

In section~\ref{sec:results} we present the verification that  in a static medium of a given size, where analytical results of BDMPS-Z  are available,  our approach reproduces the quantum interference correctly, if we apply the condition under which these results are obtained. We show here as well the smooth transition to the fully coherent energy loss at ultra high gluon energies as well as to the Bethe-Heitler result at low gluon energy. In the fully coherent regime  in the average the gluons cannot be formed in the medium. Only fluctuations in the momentum transfer in the elastic collisions of the gluons or in the number of collisions can lead to the formation of such gluons. There we present as well how for a static QPG medium the energy loss is modified if the BDMPS-Z conditions are relaxed to a more realistic description of the passage of a fast parton through a QGP. 

In section~\ref{sec:conclusions} we conclude our results. The application of our model to heavy ion collisions, in which the QGP is expanding, as well as the comparison of our results with experimental data, we postpone to a future publication. Throughout this manuscript, we consider natural units with $\hbar = c = 1$.
     

\section{Outline of our hybrid approach} \label{sec:outline}
Hard scatterings between projectile and target nucleons can lead to a large momentum transfer either of the scattered partons or of the produced partons. Such partons are called jets or jet seeds and have a high virtuality, which is, due to the  lack of QCD based results, differently parameterized in different approaches. Subsequently, this parton loses its virtuality by emission of other partons. The splitting into several partons is described by the Sudakov form factor. In a medium, while losing its virtuality, the parton can also scatter with the QGP constituents. An effective way to describe the early jet evolution in the QGP medium is to assume that the scattering with the medium induces a gain of virtuality, as in YaJEM~\cite{Renk:2008pp}. Whenever the transverse-momentum broadening, acquired by a parton from the QGP during its formation time $t_f$, is larger then its own transverse momentum at "creation", one switches the evolution from the high virtuality cascade to the low virtuality evolution. This criterion is applied to each parton individually and adopted for instance in  \cite{Caucal:2018dla} when discussing the transition from vacuum-like emissions to induced BDMPS-Z gluons. Since the gluon formation time writes $t_f \simeq \omega/k_T^2$, the criteria to switch to the low virtuality evolution is $\hat{q} \times \frac{\omega}{k_T^2} \gtrsim k_T^2 \Leftrightarrow  \omega^3 \theta^4 \lesssim \hat{q}$, where $\theta=k_T/\omega$ is the gluon emission angle in the splitting. In \cite{Caucal:2018dla} the relation $\omega^3 \theta^4 \simeq \hat{q}$ indeed corresponds to the locus in the $(\omega,k_T)$ plane describing the end of the vacuum-like cascade. As $k_T^2 \approx z Q^2$ -- see eq.~(\ref{eq:kpdef1}) -- , where $Q$ is the virtuality of the parent parton while $z=\omega/E$ is the energy fraction of the daughter parton, this condition is equivalent to
\begin{equation}
\frac{z Q^4 }{E} \lesssim \hat{q} \,,
\label{criteriaswitch}
\end{equation}
which we adopt in our hybrid approach. This criterion differs slightly from the one advocated in JETSCAPE~\cite{JETSCAPE:2021ehl} where a switching virtuality $Q_{\rm sw}^2=\sqrt{E \hat{q}}$ is applied.\footnote{Although in practice, an average value is retained.} 
It is also equivalent to require that the transition takes place when the formation time of the gluon, emitted at high virtuality, 
exceeds the formation time of the gluon emitted in the low virtuality BMDPS-Z regime. This strategy of branching towards the process which has the smallest formation time is found in other MC approaches, like JEWEL \cite{Zapp:2012ak}. However, contrarily to JEWEL where the LPM effects are implemented in a modified virtuality-ordered DGLAP evolution equation incorporating formation time effects, our approach implements these LPM effects in a genuine time-ordered evolution, similarly to \cite{Zapp:2011ya}.
In the absence of any medium, we adopt the standard hypothesis that
splitting ends when the virtuality of the parton arrives at the lowest virtuality scale $2 Q_0$ with $Q_0 \simeq \Lambda_{\text{QCD}}$. Below $2 Q_0$ the parton enters the low-virtuality kinetic regime.

As soon as they enter the kinetic regime, the jet partons are considered as on-shell particles with a thermal mass and  scatter elastically with the partons (gluons and quarks) of the QGP, which is assumed to be in thermal equilibrium. These processes are modeled through a differential rate of transverse momentum exchange with the QGP particles evaluated by using standard QCD matrix elements where the exchanged gluon has an infrared regulator of $\mu^2 = \kappa m_D^2$ for the Debye mass $m_D \sim g T$ and sampled in a Monte Carlo procedure. This allows to assure momentum and energy conservation.
For the mass of the QGP particles two limiting cases are studied, $m_{\text{q}}=0$ and $m_{\text{q}}\to \infty$, the later choice being essentially considered for the purpose of benchmarking our results on theoretical calculations as BDMPS-Z.

As said, in vacuum a virtual gluon with an energy $\omega$ and a transverse momentum $k_\perp$, created in inelastic collisions, needs the time $t=\omega/k_\perp^2$  to be formed. In a strongly interacting medium the gluon may scatter with the medium while being formed. These collisions modify the transverse momentum of the gluon and cause a change of the formation time. To be able to describe gluon emission in a medium in a wide range of gluon energies and hence of formation times we start for the inelastic collisions from the Gunion-Bertsch approach~\cite{Gunion:1981qs, Aichelin:2013mra}, which describes the emission of a gluon with zero formation time. The finite formation time is then included by the rejection of the formation of a real gluon depending on the future path of the gluon in the medium. By this we can describe the Bethe Heitler regime as well as both regimes of BDMPS-Z. 
If the parton-gluon antenna resulting from this seed does not suffer an elastic scattering before its formation time, the gluon is assumed to be formed and both partons (the projectile parton and the formed gluon) are free to interact with the QGP independently. If, on the contrary, the projectile parton scatters inelastically with another QGP parton before the gluon is formed, the emission amplitudes interfere,  leading to the QCD equivalent of the Landau-Pomeranchuk-Migdal (LPM) effect. In practice we apply the heuristic BDMPS procedure to count the $N_s$ scattering centers, on which the virtual gluon scatters before being formed, and accept the formed gluon then with the probability 1/$N_s$, means effectively we replace the $N_s$ scattering centers a single one. If, finally, the formation time is even larger than the size of the system, the virtual gluon cannot be formed. As mentioned in the introduction, these three domains exist already when a fast parton traverses a brick of QCD matter and it is difficult to describe them with causal Monte Carlo approaches because the formation time is not known at the moment when the virtual gluon is formed. The approach, which we present here, is however more ambitious. We want to describe finally the gluon creation in a rapidly expanding QGP, as created in ultra-relativistic heavy-ion collisions, using the seed-rejection algorithm presented in details below.

In this article, we focus on the basic benchmark of the energy loss, which a high energy parton suffers in QGP matter. Therefore, even though the framework is constructed for an arbitrary space-time temperature profile, for the purpose of benchmarking and comparison to known semi-analytical results, we approximate the QGP medium by homogeneous bricks of temperature $T = 400$ MeV and fixed lengths, $L=1$, $2$, and $8$ fm. We perform the calculation with a mono-energetic stream of partons with an initial energy of $E=100$~GeV.

As it has been mentioned in section~\ref{sec:introduction}, the aim of each transport model for jets in ultra-relativistic heavy-ion collisions is to provide an integrated modeling of both, the bulk medium (using a hydrodynamic description) and jets, including the mutual interaction between the two.
The hydrodynamic description of the medium introduces a space-time evolution picture. This makes it necessary to develop as well a space-time picture of the jet shower evolution in the high virtuality regime, based on the purely momentum-space picture of the DGLAP equations.
With rare exceptions\footnote{Such as mode-by-mode hydrodynamics~\cite{Floerchinger:2013rya}.}, numerical solutions of the hydrodynamic expansion of a medium involves finite-difference or finite-volume schemes, which also discretize the time evolution into a series of finite time steps. In particular, the \texttt{vHLLE} code~\cite{Karpenko:2013wva}, which we plan to use in the upcoming studies of the jet quenching in relativistic heavy-ion collisions, is based on a finite-volume scheme with finite time steps. The low virtuality regime, where partons interact by cross sections, is naturally time-ordered and thus more suitable for a concurrent evolution.  Furthermore, to describe the interaction between the medium and jets, it is obvious to slice the time development of the jet shower into the same series of finite time steps $\Delta t$. We describe now in detail how we implement the jet evolution in our transport approach.


\section{Numerical realization of our approach: the high-virtuality component}
\label{sec:numerical}

In the high virtuality regime, jets evolve as virtuality-ordered parton cascades from an initial maximal virtuality scale $Q_\uparrow$ down to a minimal virtuality scale $Q_{\rm min}$. A virtuality ordering has been chosen here because 1/Q is the lifetime of the virtual state and therefore the algorithm can be easily embedded in the space time evolution of the environment. For future applications 
we consider to proceed to an angular ordering by limiting the kinematical range of the daughter partons \cite{Zapp:2008zz,Rohrmoser:2017vsa}.

While the maximal initial virtuality of the parton  is $Q_\uparrow=E$, the real initial virtuality $Q_1$ is obtained by sampling the Sudakov factor, between $Q_{\rm min}$ and $Q_\uparrow$. \footnote{Therefore our initial virtuality is different from the calculations of the JETSCAPE collaboration~\cite{JETSCAPE:2019udz} which start out with $Q^2=p_T^2/2$.} Quark jets, e.g., are assumed to evolve by emission of  gluons (bremsstrahlung)  as in the vacuum, however with modifications induced by the medium through which they travel. In the next subsection we first focus on the DGLAP algorithm adopted to describe such a jet evolution in the vacuum. We take  $Q_{\rm min}=2 Q_0$, with $Q_0 = 0.3$ GeV $\simeq \Lambda_{\text{QCD}}$. In the subsequent subsection we introduce the medium modifications. In order to maximize the effect of the medium on the high-virtuality component, we mostly keep the same value of $Q_{\rm min}$ as in the vacuum, disregarding for the time being the criteria given at eq.~(\ref{criteriaswitch}). For the ratio of the in-medium primary Lund plane and its equivalent in the vacuum, we provide, however, both results obtained by either applying this prescription -- Fig.~\ref{fig:highQ_50_diff_lund_plane} -- or by using the switch criteria expressed in eq.  eq.~(\ref{criteriaswitch}) -- Fig.\ref{fig:highQ_50_diff_lund_plane_switch} --, in order to illustrate the consequence of such criteria on the differential gluon-spectrum.


\subsection{Evolution in vacuum} 
\label{sec:highQvac}

In the vacuum, jets mainly evolve by emission of gluons as described by the DGLAP evolution equations. The jet particle follows a virtuality-ordered parton cascade with some additional phase space constraints (outlined below). Mathematically, such cascade can be seen as a tree, each branch being attributed some meta-index: $a\equiv \{i_a, {\rm t}_a\}$, $b\equiv \{i_b, {\rm t}_b\}$, $c\equiv \{i_c, {\rm t}_c\}$\ldots, where $i_{a/b/c/\ldots}$ unequivocally designates the position of the branch (e.g through some binary alphabet), while $t_{a/b/c/\ldots}$, designates the parton type (gluon or quark). For the sake of conciseness, $a$, $b$,\ldots will the generically used to designate each property $X_a$, $Y_a$,\ldots of the corresponding branch, while those of the first jet-parton will simply be denoted by $X_1$, $Y_1$,\ldots. $a\to b+c$ will indicate a generic parton splitting inside the shower.

The probability that a jet parton $a$  with a virtuality $Q_{a\uparrow}$ does not split in the virtuality interval $[Q_{a\uparrow},Q_a]$ into a parton and a gluon  is given by the Sudakov form factor~\cite{Ellis:1996mzs}
\begin{equation}
S_a(Q_{a\,\uparrow},Q_a)=\exp\left(-\int_{Q_a^2}^{Q_{a\,\uparrow}^2}\frac{\mathrm{d}Q^2}{Q^2}\int_{\chi_-(Q)}^{\chi_+(Q)}\mathrm{d}\chi\frac{\alpha_s(F(\chi,Q))}{2\pi}\sum_{a\rightarrow b\,,c}P_{a\rightarrow b\,,c}(\chi)\right)\,,\label{eq:Sudadef}
\end{equation}
where $P_{a\rightarrow b,c}$ is the DGLAP splitting function for a decay of parton $a$ into partons $b$ and $c$ at leading-order in QCD. For $Q_{a\,\uparrow}\gg Q_{\rm min}$, $S_a(Q_{a\,\uparrow},Q_{\rm min})\approx 0$, meaning that parton $a$ splits in this interval. The probability density $p\left(Q_a,\,\chi\right)$ that this jet parton $a$ splits exactly ``at'' a virtuality  $Q_a $  into partons  $b$ and $c$  is then given by 
\begin{equation}
p\left(Q_a,\,\chi\right)=S_a(Q_{a\,\uparrow}\,,Q_a)\left(\frac{\alpha_s(F(\chi,Q_a^2))}{2\pi}P_{a\rightarrow b,c}(\chi)\right)\,. \label{eq:partitionfQZ}
\end{equation}
In these expressions, $\chi$ represents the fraction of the leading energy-like quantity of parton $a$ carried away by parton $c$. The two common choices are $\chi=z=E_c/E_a$ (energy fraction) or $\chi=x=p_c^+/p_a^+$ (light-cone momentum fraction, where $Oz$ represents the direction of the propagation of $a$). $F(\chi,Q^2)= \chi(1-\chi)Q^2$~\cite{Ellis:1996mzs}.  The coupling constant is taken here in leading-order QCD,~\cite{Deur:2016tte}
\begin{equation}
    \alpha_s(Q^2)  = \frac{1}{b\ln \left ( \frac{Q^2}{\Lambda^2_{\text{QCD}}} \right )} \,, \hspace{5mm} \text{ with } \hspace{5mm} b=\frac{33-2N_f}{12\pi} \,.
\end{equation}
In the case $E_a \to \infty$ both choices for $\chi$ are equivalent. Distinction should be made for large but finite $E_a$, where NLO corrections $\propto 1/E_a$ are  important for the correct description of the splitting phase space. Neither choice is ideal in this respect, as the splitting probabilities $P$ are naturally expressed as a function of $z$ while the {\it absolute} lower and upper phase-space boundaries\footnote{Achieved for the lowest possible $Q_b$ and $Q_c$.}, $z_-$ and $z_+$ respectively, are given as
\begin{equation}
z_\pm(Q_a,Q_0,E_a)=\frac{1}{2}\left(1\pm\sqrt{\left(1-\frac{4Q_0^2}{Q_a^2}\right)\left(1-\frac{Q^2_a}{E_a^2}\right)}\right)\,,
\label{eq:xboundexact}
\end{equation}
whose dependence on $E_a$ --- the energy of the splitting particle --- makes the calculations cumbersome. It can be simplified by turning to the choice $\chi=x$:
\begin{equation}
x_\pm(Q_a,Q_0)= \frac{1}{2}\left(1\pm\sqrt{\left(1-\frac{4Q_0^2}{Q_a^2}\right)}\right),
\end{equation}
which does not depend on the energy $E_a$. However, the splitting probabilities $P$, expressed as a function of $x$, carry an extra dependence on the partons energies. For the purpose of establishing an efficient MC algorithm we have chosen to privilege $\chi=z$ --- the energy fraction --- in the sampling of the Sudakov form factor and to correct by a veto method for the too large phase space given by the boundary $x_-$ with respect to the correct boundary $z_-$. That will be expressed using the following identity:
\begin{equation}
\int_{z_-}^{z_+} \mathrm{d}x \cdots = r_{\int/\int} \times \int_{x_-}^{x_+} \mathrm{d}x \cdots \,,
\quad\text{where}\quad
r_{\int/\int}= \frac{\int_{z_-}^{z_+} \mathrm{d}x \cdots}{\int_{x_-}^{x_+} \mathrm{d}x \cdots} < 1 \,.
\end{equation}
Hence in the MC algorithm, the virtualities will be sampled according to cumulative distributions with $z$ taken between $x_-$ and $x_{+}$, with the restriction factor $r_{\int/\int}$ implemented through the veto method~\cite{Sjostrand:2006za}. 

As mentioned above, the jet evolution can be casted into a Monte Carlo algorithm.
The algorithm starts by randomly drawing the virtuality of the first particle of the cascade $Q_1$ from the  Sudakov factor $
S_1(Q_{\uparrow}\,,Q_1)$.  There the upper limit $Q_{\uparrow}$ is the maximal virtuality of the initial parton which is kinematically possible. The momentum of this parton defines the direction of the jet axis.

For any cascade particle $a$, the sampling of $Q_a$ proceeds as follow. The Sudakov $S_a(Q_{a\uparrow}\,,Q_{\rm min})$ is firstly calculated, and the eventuality for the particle not to split anymore is sampled with a probability $1-S_a(Q_{a\uparrow}\,,Q_{\rm min})$. If this is the outcome of the sampling, particle $a$ is considered as on-shell and a virtuality of $Q_0$ is assigned to it. It is then assumed that the further evolution of  that particle is given by the processes relevant for the regime of low virtuality, which are outlined later in section~\ref{sec:lowQ}.  

In the other case,  $Q_a>Q_{\rm min}$,   $Q_a$ is drawn according to eq.~(\ref{eq:Sudadef}) and particle $a$ may split into a new pair of particles, $b$ and $c$, see eq.~(\ref{eq:partitionfQZ}). The splitting is then repeated for the emitted partons $b$ and $c$ and their possible splitting products. The algorithm stops when all particles, produced in the splittings, have virtualities of $2Q_0$ or lower. The concrete steps in the algorithm for the splitting of a jet particle $a$ with virtuality $Q_a$ and a given momentum, assumed to be oriented along the $Oz$ axis, into a daughter particle with the momentum fraction $z_+$ are the following: 

\begin{enumerate}
\item If parton $a$ is a quark,  there is only the possibility of a splitting into a quark and a gluon, $q\to qg$. If parton $a$ is a gluon, a random number $\mathcal{R}'\in[0,1]$ is selected and compared with the following probability for a splitting into a gluon-pair
\begin{equation}\label{eq:ptype}
p_{g\rightarrow gg}(Q_a,E_a,z_+)=\frac{W_{g\rightarrow gg}(Q_a,E_a,z_+)}{W_{g\rightarrow q\bar{q}}(Q_a,E_a,z_+)+W_{g\rightarrow gg}(Q_a,E_a,z_+)}\,.
\end{equation}
There, the function $W_{a\rightarrow b,c}(Q_a,E_a,z)$ is the integrated distribution function for the branching of a gluon $a$ into the partons $b$ (with an energy fraction $z$) and c (with an energy fraction $1-z$). It is given by
\begin{equation}
W_{a\rightarrow b,c}(Q_a,E_a,z) \equiv {\int_{z_-}^{z}\mathrm{d}\tilde{z}\left(\frac{\alpha_s(F(\tilde{z},Q_a^2))}{2\pi}P_{a\rightarrow b,c}(\tilde{z})\right)}\,.
\label{eq:Wprob}
\end{equation}
If $\mathcal{R}'\leq p(g\rightarrow gg)$ the gluon $a$ splits into two gluons, otherwise into a quark and an antiquark. 

\item Both the virtualities $Q_b$ and $Q_c$ of the produced partons $b$ and $c$ are drawn randomly from the  partition functions   $S_b(Q_a,Q_b)$ and $S_c(Q_a,Q_c)$, respectively. In other terms, both $Q_{\uparrow c}$ and $Q_{\uparrow b}$ in eq.~(\ref{eq:Sudadef}) are taken as $Q_a$. This is done within a rejection loop: For a given value of $Q_a$, first $Q_b$ and then $Q_c$ are selected. If $Q_a^2 \geq Q_b^2+Q_c^2$, the values are accepted, otherwise both $Q_b$ and $Q_c$ are drawn again. The reason for this rejection loop is to guarantee values of $Q_b$ and $Q_c$ that allow for parton momenta that are kinematically possible.\footnote{ This rejection loop is not strictly necessary as the rejection can also be done at the very end of the selection of the splitting. However selecting already here virtualities that allow for kinematically possible parton momenta makes the whole algorithm more effective.}

\item The energy fraction $z$ between the momenta of one of the daughter partons (e.g.~$b$) with respect to mother parton $a$ is selected, using $W_{a\rightarrow b,c}(Q_a,z)$ as a cumulative distribution function.  $z$ is selected in between the boundaries, eq.~(\ref{eq:xboundexact}), imposed by energy and momentum conservation.

\item
For a splitting with a given energy fraction $z$, parton virtualities $Q_a$, $Q_b$, $Q_c$ and a parton energy $E_a$, the corresponding light cone energy fraction $x$ can be obtained as 
\begin{align}
x&=\frac{z\left(1+\tilde{t}_a\right)-\left(\tilde{t}_a+\tilde{t}_b-\tilde{t}_c\right)}{1-\tilde{t}_a}\,,
\label{eq:zx}
\end{align}
where $\tilde{t}_i$ is defined as $\tilde{t}_i=\frac{Q_i^2}{(p_a^+)^2}$ while the light-cone energy $p_a^+$ of particle $a$ is obtained from its energy $E_a$ as $p_a^+=E_a+\sqrt{E_a^2-Q_a^2}$. 
With the light-cone fraction $x$, the square of  the transverse momentum of particle $b$ and $c$ with respect to the momentum of particle $a$ is determined by
\begin{align}
{k}_T^2&=x(1-x)Q_a^2-(1-x)Q_b^2-xQ_c^2\,.\label{eq:kpdef1}
\end{align}
assuming that $p_L \gg Q_b, Q_c, p_T$.
\item The possible values for $Q_b$, $Q_c$ are constrained by the condition $k_T^2 \ge 0$.
If ${k}_T^2>0$ the kinematical constrains are fulfilled  and the splitting is accepted (see steps 5 and 6).  Otherwise, the selection of the splitting, starting from step 2, is repeated.\footnote{This latest rejection ensures that the correct PS boundaries are achieved, even if $x$ is sampled in a too large domain $[z_-,z_+]$.} 

\item Azimuthal angles $\varphi_b$ and $\varphi_c$ for the momenta of particles $b$ and $c$ relative to the momentum of particle $a$ are determined: $\varphi_b$ is selected from a uniform distribution in the range $[0,2\pi]$ and $\varphi_c$ is determined as $\varphi_c=\varphi_b +\pi$.

\item We reconstruct the energies and the momenta of particles $b$ and $c$ from the quadruplets $\{E_b=zE_a,Q_b,k_T,\varphi_b\}$ and $\{E_c=(1-z)E_a,Q_c,k_T,\varphi_c\}$.

\item We return to step 1 to pursue the splitting procedure for the daughter partons $b$ and $c$.
\end{enumerate}

\begin{figure}[htb]
\centering
\includegraphics[width=0.6\linewidth]{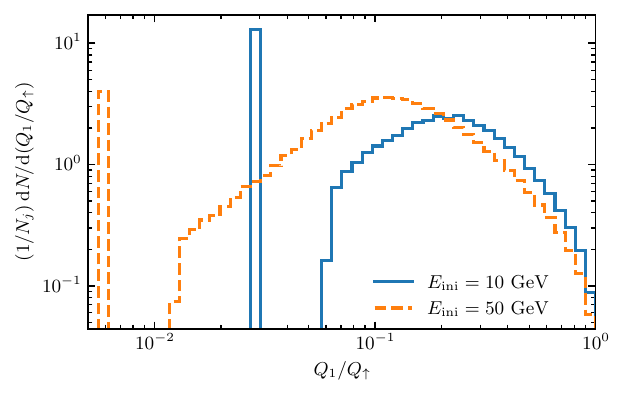}
\caption{Distribution of the ratio of the virtuality $Q_1$, the starting point of the jet evolution and the initial maximum virtuality $Q_{\uparrow} = E_{\text{ini}}$, averaged over all simulated parton cascades, for parton cascades with $E_{\text{ini}}=10$ and $50$~GeV.}
\label{fig:firstQ}
\end{figure}

We assume that between their generation (from a previous splitting) and their own splitting, the virtual partons travel as classical relativistic particles  and that the lifetime of such a state of virtuality $Q$ is given by $\tau = E/Q^2$.  This allows to translate the momentum space evolution into a phase space evolution. Positions of particles in space are obtained by using the assumption that both particles ($b$ and $c$), resulting from the splitting $a\to bc$ travel with a constant velocity 
\begin{equation}
\vec{v}_{b/c}=\frac{\vec{p}_{b/c}}{E_{b/c}}\,,
\end{equation}
until they split themselves. It turns out that the momenta are close to the speed of light, because already for the initial jet parton,  the starting point of our evolution, one has on the average $Q_1 \ll Q_\uparrow=E_{\rm ini}$ as seen in Fig.~\ref{fig:firstQ}.

Each of the jet partons above the virtuality scale $2Q_0$ can split into two partons with a lower virtuality, $Q_b$ and $Q_c$, according to the Sudakov form-factor. We can determine its equivalent life time, $\tau_a = E_a / Q_a^2$, by the Heisenberg uncertainty principle as well as its propagation in physical space, $\Delta\vec{ x}=\vec{p}_a/Q_a^2$. To model this process in an algorithm based on finite time steps, one assumes that the splitting takes place with a probability $\Delta t\cdot Q_a^2/E_a$ during a time step $\Delta t$.  This procedure ensures that the mean life-time of parton $a$ delineated by two successive splittings is indeed $\tau_a = E_a/Q_a^2$. 

\begin{figure}[htb]
    \centering
    \includegraphics[width=0.6\linewidth]{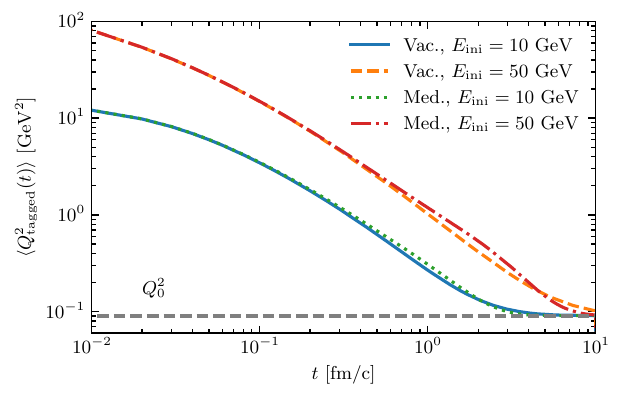}
    \caption{Mean squared virtuality of the initial quark tagged through further splittings and averaged over all simulated parton cascades as a function of time $t$ for parton cascades with $E_{\text{ini}}=10$ and $50$~GeV, in vacuum and with medium interactions. }
    \label{fig:Qvst}
\end{figure}

It is interesting to see the time evolution of the jets in this high virtuality regime. Fig.~\ref{fig:Qvst} shows the time evolution of the squared virtuality $\langle Q^2_{\text{tagged}}\rangle$ of the initially tagged quark for parton cascades with an initial quark and with initial energies of $E_{\text{ini}} = 10$ and $50$~GeV. For this we follow the virtuality of the fermion at each splitting. The virtuality is averaged over $10^6$ simulated cascades.
The average parton virtuality decreases fast, in agreement with the assumption of a time-like parton cascade with a strong ordering in the parton virtualities.  The higher the initial energy, the faster the initial virtuality decreases and, due to the increasing number of splittings, the longer it takes until the quark becomes on-shell. The mean virtuality for both energies decreases in the same pace for intermediate times.  At $t = 1$ fm$/c$ the virtuality is reduced by more than one order of magnitude, independent of the initial energy. At $t=5$~fm$/c$, particles with an initial virtuality of $50$ GeV  arrive finally at a virtuality of the order of $Q_0$.\footnote{This shows that the jet evolution in our modeling is not governed by a single time scale, even for its evolution in vacuum.} In the medium the decrease of the virtuality is slower at later times because the leading partons gain virtuality in collisions with the medium. 

\begin{figure}[htb]
    \centering
    \includegraphics[width=0.6\linewidth]{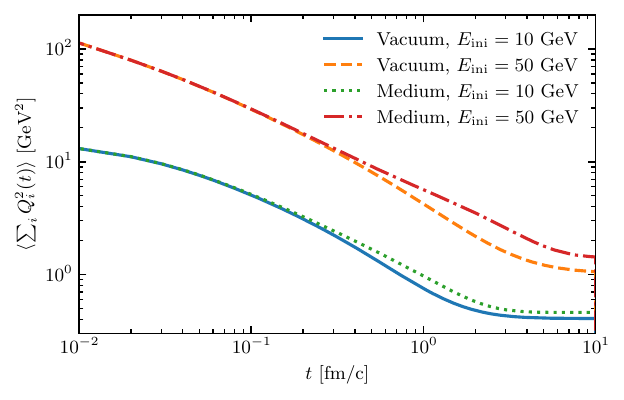}
    \caption{Mean of the sum of virtualities squared of all partons of a jet, averaged over all simulated parton cascades, as a function of time $t$ for parton cascades with $E_{\text{ini}}=10$ and $50$~GeV, with and without medium interactions. }
    \label{fig:Qsum}
\end{figure}

\begin{figure}[htb]
    \centering
    \includegraphics[width=0.6\linewidth]{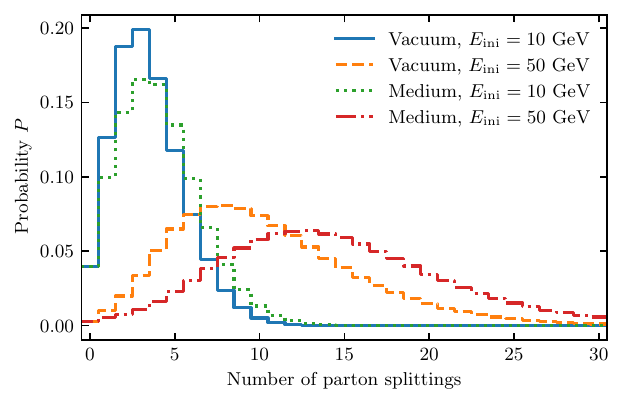}
    \caption{Distribution of the number of splittings, normalized to one, for $E_{\text{ini}}=10$ and $50$~GeV, and with and without medium interactions.}
    \label{fig:nsdist}
\end{figure}

Fig.~\ref{fig:Qsum} shows the average of the sum of virtualities squared of all partons in the cascade of a jet as a function of the time. The sum of all virtualities decreases much slower as a function of time than the virtuality of the leading parton because part of the virtuality of the initial parton is transferred to daughter partons with a lower virtuality, which live longer. Hence it takes considerably longer until these daughter partons are on the mass shell as compared to the time for the leading parton. The curves are observed to flatten out at large times, when all the partons in the cascade arrive at their final virtuality of the order of $Q_0$. The larger total virtuality at larger times for the more energetic jet shows that more splittings have occurred
and therefore more partons have been created. This is confirmed in Fig.~\ref{fig:nsdist} where the distribution of splittings (normalized to one) is shown.

\subsection{Medium modifications} 
\label{sec:highQmed}

In a medium, jet partons can scatter elastically or inelastically, where the latter results in medium-induced radiation. The consequence is a deflection of the jet particle momentum by a momentum transfer from the medium (and a modification of the momentum of the medium), which can lead to an additional branching of the jet particles. In order to account for these processes, an effective model, based on the YaJEM algorithm~\cite{Renk:2008pp}, is used:
The algorithm for the jet evolution in the vacuum, outlined in the previous subsection, 
is modified such that  every jet particle with virtuality $Q>2 Q_0$ changes its virtuality due to this medium interaction. This medium modifications are local in time and space. To allow for these medium modification, the vacuum evolution of the jet, which is given in momentum space, is extended to a phase space trajectory,  which allows to embed  the jet in the expanding QGP medium.

Following ref.~\cite{Renk:2008pp}, we effectively treat the interactions of the jet partons with the medium in the high-virtuality regime by a continuous increase of the virtuality of the jet parton,
\begin{equation} 
\frac{\mathrm{d}Q^2}{\mathrm{d}t} = \hat{q}(T,p) \,,
\label{eq:virt}
\end{equation}
which leads in the time interval $\Delta t$  to  a virtuality increase\footnote{In the code, the virtuality increase is translated into an increase in the energy of the parton, $\mathrm{d}E^2/\mathrm{d}t = \mathrm{d}Q^2/\mathrm{d}t = \hat{q}$. This ensures that the three-momentum of the parton is only changed by the parton splittings. The energy of the parton cascade is therefore not conserved, instead energy is effectively transferred from the medium to the jet parton. The energy loss due to the parton splittings will be however larger than the energy gain from $\hat{q}$, resulting in an overall energy loss of the leading parton during the parton cascade.} of $\Delta Q^2 = \hat{q}(T,p) \,\Delta t$.
For the calculation of the transport coefficient $\hat{q}(T,p)$, we could e.g.\ adapt the model E of ref.~\cite{Gossiaux:2008jv} for heavy quarks, to be used for light quarks. For a ``large'' momentum of $20$~GeV$/c$, $\hat q$ can be parameterized in this model as
\begin{equation}
\frac{\hat{q}_{\rm GA}(T,p=20\text{ GeV})}{T^3}\approx \frac{210}{1+8\, \frac{T}{T_c}}\,,\quad\text{with}\quad
T_c=0.15 \text{ GeV}\,.
\label{eq:qhatGA}
\end{equation}
At smaller (or larger) $p$, a good parameterization of $\hat{q}_{\rm GA}(T,p)$ for quarks is provided by the following fitting function:
\begin{equation}
\hat{q}_{\rm GA}(T,p) =  
\hat{q}_{\rm GA}(T,p=20\text{ GeV}) \times 
\hat{q}_{\rm cof}(p)\,,
\quad\text{where}\quad
\hat{q}_{\rm cof}(p)=  \frac{1.69 + 1.25\,p}{4.07 + p + 0.85 \ln \left (p + 1 \right )} \,,
\label{eq:qhatofpGA}
\end{equation}
satisfying $\hat{q}_{\rm cof}(20\text{ GeV})=1$ (dimensionless). In this definition, $p$ is the dimensionless numerical value of the momentum expressed in $\text{GeV}$. For gluons, a color factor of $C_A/C_F$ is applied.
This $p$-dependent behavior of $\hat{q}_{\rm GA}$ in Eq.~(\ref{eq:qhatGA}) is similar to the model employed in ~\cite{JETSCAPE:2023hqn}, where $\hat{q}$ monotonously rises with the jet-parton energy. In addition, in ~\cite{JETSCAPE:2023hqn} coherence effects at high-virtualities are described via a $\hat{q}$ dependence on the jet-parton virtuality $Q$, which is monotonously falling. We leave the treatment of coherence effects at high virtualities for a later adaptation of our proposed algorithm and note that the approach, proposed in this section, does not depend on the explicit form of $\hat{q}$.

\begin{figure}[htb]
    \centering
    \includegraphics[width=0.6\linewidth]{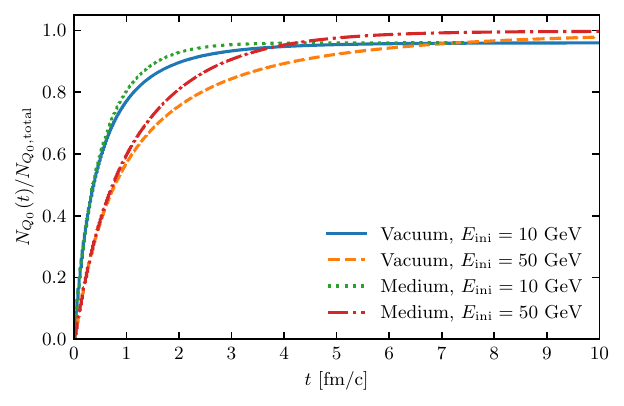}
    \caption{Fraction of the initial tagged quarks that have reached the minimal virtuality $2 Q_0$ as a function of time, for different initial energies and with/without medium modifications.}
    \label{fig:tq0ratio}
\end{figure}

A while ago, the JET collaboration performed a first extraction of the $\hat{q}$ coefficient by comparing various jet models to experimental data~\cite{JET:2013cls}. This leads to
\begin{equation} \frac{\hat{q}_{\rm JET}}{T^3}\approx 5.5\times \frac{2}{1+\frac{T}{T_c}}\,,
\label{eq:qhatJEF}
\end{equation}
without any momentum dependence. In our approach, we combine both forms, eq.~(\ref{eq:qhatofpGA}) and eq.~(\ref{eq:qhatJEF}), by rescaling $\hat{q}_{\rm GA}$ to $\hat{q}_{\rm JET}$ for large momenta and conserving its momentum dependence. The $\hat{q}$ law adopted in our calculations of eq.~(\ref{eq:virt}) is thus
\begin{equation} 
\hat q_{\text{JET}}(T,p) = \hat q_{\rm JET}(T) \times \hat q_{\rm cof}(p)\,.
\end{equation}
The consequences of the in-medium induced virtuality increase is shown in Figs.~\ref{fig:Qvst} and \ref{fig:Qsum} for initial parton energies of $10$ and $50$ GeV. The jet partons lose a large fraction of their initial virtuality without being influenced by the medium. Only later, when their virtuality is already considerably reduced and when they have traveled more than a mean free path in the medium, the contribution of the medium to their virtuality becomes visible. At this late times the medium contribution to the sum of the squared virtualities of all jet partons may reach $50\%$ of  as shown on Fig.~\ref{fig:Qsum}. For the individual tagged partons the medium contributions still may amount to more than $25\%$ of the parton virtuality at a time of $2$ to $3$~fm/c as can be seen in Fig.~\ref{fig:Qvst}, while at later times virtualities of the tagged particles drop sharply, due to a larger amount of medium induced splittings.  
The aforementioned contributions depend, of course, on the choice of $\hat q$.  A higher virtuality leads to more splittings in which the virtuality is reduced, as shown in Fig.~\ref{fig:nsdist}.  In Fig.~\ref{fig:tq0ratio} the fraction of the initial tagged partons, which have reached the minimum virtuality of $2 Q_0$, is shown for different initial energies as a function of time. The consequences of the presence of a medium (Eq.~(\ref{eq:virt}) ) are also shown. The medium-induced virtuality increase  (Eq.~(\ref{eq:virt}) has the consequence that the tagged initial quark reaches the minimum virtuality earlier due to more virtuality reducing splittings per time.  Figs. \ref{fig:Qsum} - \ref{fig:tq0ratio} show that for larger jet energies the effect of the medium modifications plays a bigger role.  On the one side $\hat{q}$ increases with the jet energy in our model; on the other side, initial jets with larger energies need more time to reach a low virtuality close to $2 Q_0$. Hence the accumulated virtuality contribution from the medium ultimately becomes larger.  For an initial energy $E_{\text{ini}} = 10$ GeV  in about $80\%$ of the jets the initial quark has reached after around $1$ fm$/c$ the minimal virtuality and has entered the low virtuality regime. For $E_{\text{ini}} = 50$ GeV this happens after about $2$ fm$/c$.

\begin{figure}[H]
\centering
\begin{minipage}[t]{.49\textwidth}
\centering
\includegraphics[width=\linewidth]{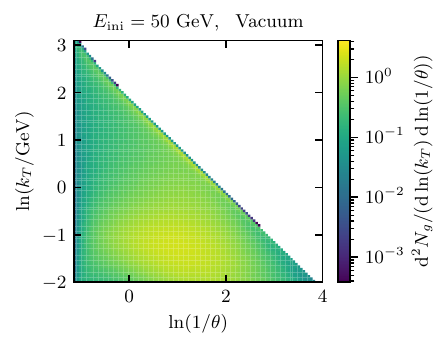}
\caption{Primary Lund jet plane for the high virtuality shower for $E_{\rm ini} = 50$ GeV. 
}
\label{fig:highQ_50_vacuum_lund_plane}
\end{minipage}\hfill%
\begin{minipage}[t]{.49\textwidth}
\centering
\includegraphics[width=\linewidth]{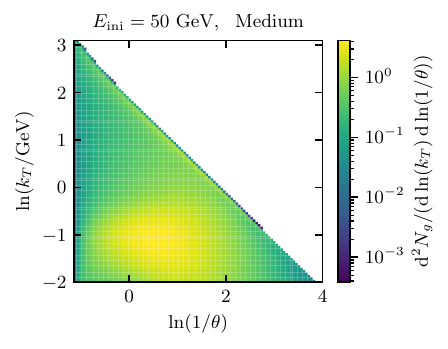}
\caption{Same as Fig~\ref{fig:highQ_50_vacuum_lund_plane} but with medium modifications. 
}
\label{fig:highQ_50_medium_lund_plane}
\end{minipage}
\end{figure}

\begin{figure}[H]
\centering
\begin{minipage}[t]{.49\textwidth}
\centering
\includegraphics[width=\textwidth]{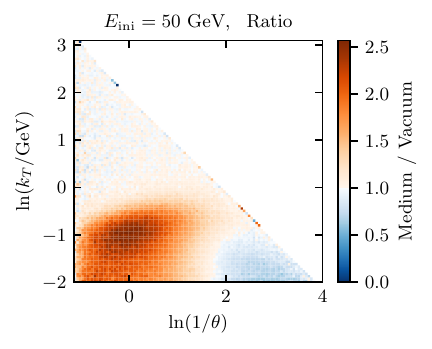}
\caption{The ratio between the primary Lund jet planes in Figs.~\ref{fig:highQ_50_medium_lund_plane} and \ref{fig:highQ_50_vacuum_lund_plane}, highlighting the effect of the medium modifications. 
}
\label{fig:highQ_50_diff_lund_plane}
\end{minipage}\hfill%
\begin{minipage}[t]{.49\textwidth}
\centering
\includegraphics[width=\textwidth]{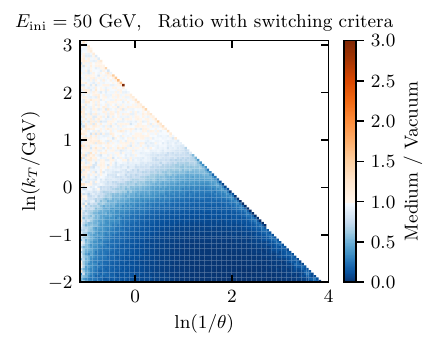}
\caption{Same as Fig~\ref{fig:highQ_50_diff_lund_plane}, but terminating the high virtuality according to the switch criterion expressed in eq.~(\ref{criteriaswitch})}
\label{fig:highQ_50_diff_lund_plane_switch}
\end{minipage}
\end{figure}

In Figs. \ref{fig:highQ_50_vacuum_lund_plane} and \ref{fig:highQ_50_medium_lund_plane}, we display the primary Lund plane for an initial quark-jet of $E_{\rm ini}=50\,{\rm GeV}$ evolving in the vacuum and in a medium, respectively. The upper boundary of the filled region corresponds to a gluon energy of 25 GeV and is only achievable  in the first splitting. The "bulk" of the parton emission, observed for $\theta \approx 1/e$ and $k_T \approx 1\,{\rm GeV}/e$, corresponds to much lower values of parton energies $\approx 1\,{\rm GeV}$. The variations in the Lund plane density is due to the running of the coupling constant adopted in our model but also due to the boundaries imposed by eq.~(\ref{eq:kpdef1}), which narrows the acceptable $x$-interval. On the right panel, the medium  increases the overall density of emitted partons, while letting the part of the diagram located at $k_T\gtrsim 1\,{\rm GeV}$ unaffected. The medium-effect can be better scrutinized on Fig.~\ref{fig:highQ_50_diff_lund_plane}, where the ratio of medium and vacuum emission is displayed. On this figure one distinguishes more clearly a strong enhancement of the late-time soft radiation at large angles, accompanied by a slight reduction of the soft collinear emission, resulting from the effective virtuality increase with time encoded in eq.~(\ref{eq:virt}).

In Fig.~\ref{fig:highQ_50_diff_lund_plane}, we show the same ratio but terminate the high virtuality according to the switch criterion of eq.~(\ref{criteriaswitch}). With this criterion and for this specific choice of parameters, the high-$Q$ stage evolves on the average down to $Q\approx 2-3\, {\rm GeV}$. Therefore, the soft collinear gluons emitted at late times in the vacuum are excluded by the switch criterion and consequently strongly depleted. For the present choice of parameters, gluons emitted in the high-$Q$ stage are mostly radiated for $t\lesssim 1\,{\rm fm}/c$, and do not suffer from the medium effect modeled with eq.~(\ref{eq:virt}) -- see as well Fig.~\ref{fig:Qvst} --. This explains why the enhancement observed in Fig. \ref{fig:highQ_50_diff_lund_plane} is not seen in Fig.~\ref{fig:highQ_50_diff_lund_plane_switch}.


\section{Numerical realization of our approach: the low-virtuality component} 
\label{sec:lowQ}

\subsection{Thermal masses}

\begin{figure}[htb]
\centering
\includegraphics[width=0.5\linewidth]{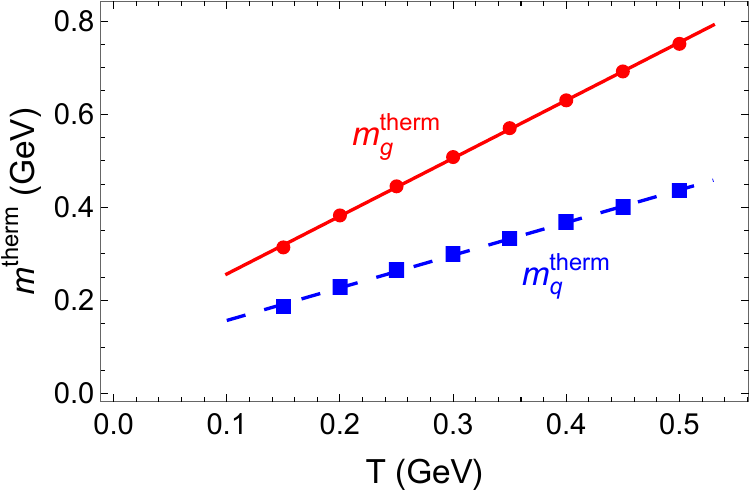}
\caption{The thermal masses of quarks (squares) and gluons (circles), as well as the fits (solid and dashed lines), used in the calculation.}
\label{fig:mtherm}
\end{figure}

After the virtuality of the partons, which may be a quark, an antiquark, or a gluon, is reduced to $2 Q_0$ or below, the partons are considered as an on-shell particle with its thermal mass. For gluons it is taken as the pole mass of the gluon HTL propagator~\cite{Blaizot:2001nr} for $p\approx 0$: $m_g^{\rm therm}=m_D/\sqrt{3}$, with
\begin{equation}
m_D^2 = \left ( 1 + \frac{N_f}{2 N_c} \right ) 4\pi  \alpha_{s,{\rm eff}}  T^2 \,,
\label{eq:md}
\end{equation}
taking $N_f=N_c=3$ as well as the effective fixed coupling constant $\alpha_{s,{\rm eff}}$ discussed below in eq.~(\ref{eq.alphaeffGA}). The quark thermal mass is obtained in a similar way from the HTL expression $m_q^{\rm therm} = \sqrt{\frac{\pi C_F \alpha_{s,{\rm eff}}}{2}}\,T$. These thermal masses are illustrated as a function of $T$ on Fig.~\ref{fig:mtherm}. One notices that they are well fitted by the following linear form:
\begin{equation}
m_g^{\text{therm}}(T) = 0.13 \text{ GeV} + 1.24\times T \,,
\label{eq:mthg}
\end{equation} 
and
\begin{equation}
 m_q^{\text{therm}}(T)= 0.087 \text{ GeV} + 0.7\times T \,
\label{eq:mthq}
\end{equation}
with T in GeV.
In the code we use these linear fit functions.

The jet partons scatter elastically or inelastically with the QGP partons, the implementation will be described in the next two subsections.

\subsection{Coupling constant}
\begin{figure}[htb]
    \centering
    \includegraphics[width=0.6\linewidth]{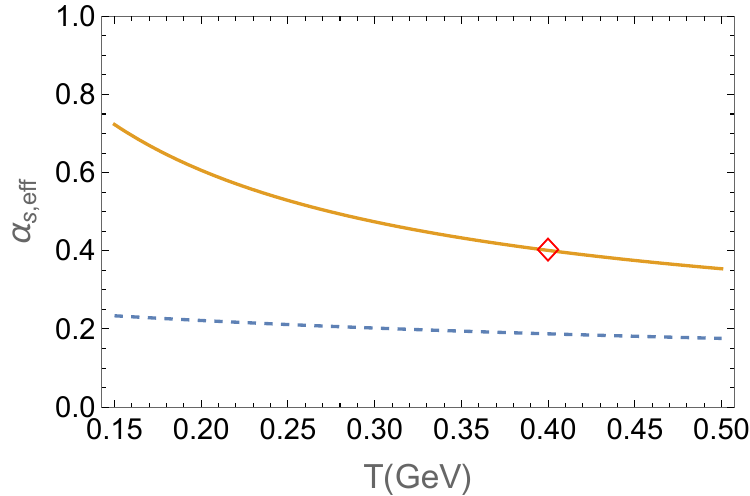}
    \caption{Effective $\alpha_s(T)$ used to reproduce some benchmark $\hat{q}$, either the one associated to the model (E) of ref.~\cite{Gossiaux:2008jv} (full red line) or the one from the JET collaboration (dashed blue line). The open diamond illustrates the value taken for the numerical applications in this work for the low-virtuality component, discussed in section \ref{sec:results}.}
    \label{fig:alphaseff}
\end{figure}

During the evolution of the QGP, the elastic collisions will be probed especially at small momentum transfer. For this reason, we neglect the dependence of $\alpha_s$ on $Q^2$ and consider only an effective temperature dependence of $\alpha_s$ in eqs.~(\ref{eq:gammael}) to  (\ref{eq:qhatsimple}). Our method relies on the models developed in ref.~\cite{Gossiaux:2008jv} that have led to a good agreement with experimental data in the heavy flavor (HF) sector. This is in particular true for the running model (E). We have evaluated the $\hat{q}(T,p)$ for model (E) --- adapted to the light-flavor case --- and then fitted $\alpha_s(T)$ of model (C) to reproduce at best these $\hat{q}$ values. This method leads to 
\begin{equation}
\alpha_{s,{\rm eff}}(T)\approx
\frac{0.42}{\ln\left(1.15+0.64\frac{T}{T_c}\right)}\,, 
\quad\text{with $T_c=0.15\text{ GeV}$}.
\label{eq.alphaeffGA}
\end{equation}
It is illustrated by the full curve in Fig.~\ref{fig:alphaseff}.  It has been used for the determination of the thermal masses, eqs.~(\ref{eq:mthg}) and (\ref{eq:mthq}), and will be used in the rest of the manuscript if not stated differently. There exists an alternative way to determine
$\alpha_{s,{\rm eff}}(T)$ based on the differential rate. We come back to this after eq.~(\ref{eq:qhatsimple}).

\subsection{Elastic collisions}

Within a time step $\Delta t$ each jet parton has a chance to scatter elastically with the partons from the QGP with a probability $\Gamma_{\rm el} \Delta t$, according to the differential elastic rate $\mathrm{d}\Gamma_{\rm el}^{q,g} / \mathrm{d}^2 q_T$. Focusing on the quarks inside the jet, this rate is taken in our approach as the weighted average of the $qq(\bar{q})$ and $qg$  scattering cross section  $\mathrm{d}\sigma_{\rm el}^{qq(\bar{q})}$ and $\mathrm{d}\sigma_{\rm el}^{qg}$ with the light quark and antiquark density  $n_{q+\bar{q}}$ and the  gluon density  $n_g$ in the QGP, respectively. Here, $\mathrm{d}\sigma_{\rm el}^{qq(\bar{q})}$ and $\mathrm{d}\sigma_{\rm el}^{qg}$ are the pQCD cross sections limited to the $t$-channel exchanges and are IR regulated:
\begin{equation}
\frac{\mathrm{d}^2\sigma_{\rm el}^{qq(\bar{q})}}{\mathrm{d}^2q_T} = \frac{2 C_F}{N_c} \frac{\alpha_s^2}{(q_T^2+\mu^2)^2}
    \quad\text{and}\quad
    \frac{\mathrm{d}^2\sigma_{\rm el}^{qg}}{\mathrm{d}^2q_T} = \frac{C_A}{C_F} \frac{\mathrm{d}^2\sigma_{\rm el}^{qq}}{\mathrm{d}^2q_T} \,, 
\end{equation}
where $N_c$ is the number of colors and $C_{F/A}$ are the usual Casimirs. For the parton densities, we consider classical relativistic distributions for massless particles and get $n_{q+\bar{q}}=\frac{4 N N_f T^3}{\pi^2}$ and $n_g=\frac{2(N^2-1) T^3}{\pi^2}$, where $N_f$ is the number of flavors (in our case $N_f=3$), and $T$ is the local QGP temperature. The differential elastic rate for a quark is then
\begin{equation}
\frac{\mathrm{d}\Gamma_{\rm el}^q}{\mathrm{d}^2 q_T} = n_{q+\bar{q}} \frac{\mathrm{d}^2\sigma_{\rm el}^{qq}}{\mathrm{d}^2q_T} + n_g \frac{\mathrm{d}^2\sigma_{\rm el}^{qg}}{\mathrm{d}^2q_T} = \left ( n_{q+\bar{q}} + n_g \frac{C_A}{C_F} \right ) \frac{\mathrm{d}^2\sigma_{\rm el}^{qq}}{\mathrm{d}^2q_T} \,,
\end{equation}
resulting in
\begin{equation}
\frac{\mathrm{d}\Gamma_{\rm el}^q}{\mathrm{d}^2 q_T}=\left(1+\frac{N_f}{N_c}\right)\frac{4(N_c^2-1)T^3}{\pi^2}\frac{\alpha_s^2}{(q_T^2+\mu^2)^2} \,.
\label{eq:gammael}
\end{equation}
After $q_T$ integration, the resulting total elastic rate for the quarks is then
\begin{equation}
\quad \Gamma_{\rm el} =\left(1+\frac{N_f}{N_c}\right)\frac{4(N_c^2-1)T^3}{\pi}\frac{\alpha_s^2}{\mu^2} \,.
\label{eq:Gammaeltot}   
\end{equation}
For a gluon we apply the same rate multiplied by the color factor $C_A / C_F$. The elastic scattering, eq.~(\ref{eq:gammael}), requires an IR regulator $\mu$. Here we follow ref.\cite{Gossiaux:2008jv} where $\mu$ is given in terms of the Debye mass $m_D$,
\begin{equation}
\mu^2 = \kappa m_D^2\,,
\label{eq:mu}
\end{equation}
with a momentum independent factor $\kappa = 0.16$. This value of $\kappa$ is obtained, if one requires that also in QCD the transition of elastic collisions with a large momentum transfer, described by a Born type propagator, joins smoothly that with a small momentum transfer, described by a hard thermal loop (HTL) propagator, as discussed in ref.~\cite{Gossiaux:2008jv}.

Thanks to the IR regulator, the total elastic collision rate does not diverge, however its second moment - the transport coefficient $\hat{q} = \int q_T^2\frac{\mathrm{d}\Gamma_{\rm el}}{\mathrm{d}^2 q_T}/\int \frac{\mathrm{d}\Gamma_{\rm el}}{\mathrm{d}^2 q_T}$ does if $q_T$ is not limited from above. Therefore, to match the effective $\hat{q}$ of a more exact approach, model C of ref.~\cite{Gossiaux:2008jv}, which uses HTL propagators, we limit the transverse momentum exchange to $q_T^2 < q_{T,{\rm max}}^2=2ET$,  where $E$ is the energy of the incoming parton\footnote{Not to be confused with the initial jet parton.}. 
With this upper boundary one obtains a finite transport coefficient for jet quarks:
\begin{equation}
\hat{q}\approx  \left(1 + \frac{N_f}{N}\right) \frac{4\alpha_s^2(N^2-1)T^3}{\pi} \left[\ln \left ( 1+\frac{q_{T,{\rm max}}^2}{\mu^2} \right ) - 1 + \frac{1}{1+\frac{q_{T,{\rm max}}^2}{\mu^2}}\right] \,.
\label{eq:qhatsimple}
\end{equation}
This expression is quite close to the one found in MARTINI based as well on the HTL approach (see eq.~(18) in ref.~\cite{JET:2013cls}). In the actual calculation we do not use $\hat q$ but a Monte Carlo simulation of the total elastic rate, eq.~(\ref{eq:Gammaeltot}). When an elastic collision occurs, the momentum exchange is sampled according to eq.~(\ref{eq:gammael}).  For the elastic scattering, once the transverse momentum exchange is fixed, the other kinematic variables -- longitudinal momentum and energy exchange -- are fixed using three possible prescriptions, described in subsection~\ref{subsubsect:phaseaccumalgo}.

An alternate method to determine $\alpha_{s,{\rm eff}}(T)$ is to use  $\alpha_s$ in eq.~(\ref{eq:qhatsimple}) as a free parameter.  We can then determine, for a given T, 
$\alpha_{s,{\rm eff}}(T)$ by the requirement that eq.~(\ref{eq:qhatsimple}) 
reproduces $\hat{q}_{\rm JET}$~parameterized by eq.~(\ref{eq:qhatJEF}). 
The $\alpha_{s,{\rm eff}}(T)$ resulting from this method is illustrated by a dashed curve in Fig.\ref{fig:alphaseff}. It appears to be at most half as large as eq.~(\ref{eq.alphaeffGA}). Future comparison  with experimental data will allow to constrain better this effective coupling constant.


\subsection{Inelastic collisions}

The elementary inelastic collision, which we employ in our approach, is the radiation of a virtual gluon from a jet parton. The formalism, which we apply, is the QCD extension of the QED Bremsstrahlung and has been studied by Gunion and Bertsch (GB)~\cite{Gunion:1981qs}. As discussed, the emitted gluon with a four-vector $(\omega,k_{\operp},k_z)$ 
needs a time $t_f=\omega/k_{\operp}^2$,\footnote{In the numerical code, gluons are also radiated from secondary partons; in this case, the notation ``$\perp$" is slightly abusive, although still pretty clear.} called formation time, until it becomes real. 
If the formation time is shorter than the mean free path $\lambda$ of the incoming parton in the medium, gluon emission from the different vertices is independent and we are in the Bethe Heitler (BH) regime. If, however, the formation time is larger than the mean free path, which depends on the temperature via the density ($\tau_f \propto T$), the gluon emission amplitudes interfere destructively, a process which has been studied in the BDMPS-Z approach~\cite{Baier:2001yt,Baier:1998yf,Baier:1996sk}. In this regime the gluon emission $\omega \mathrm{d}N/\mathrm{d}\omega$ is suppressed with respect to the BH regime. If the formation length exceeds
the length $L$ of the QGP (assumed to be represented by a box of length L), which is the case if $\omega$ exceeds the critical value $\omega_c\sim \hat{q} L^2$, gluons will, on the average, not be formed anymore and will only be emitted by fluctuations. 
The three $\omega$ regions are schematically shown in Fig.~\ref{fig:energyspectrum}. They are characterized by a different dependence of the power spectrum $\omega\frac{\mathrm{d}N}{\mathrm{d}\omega}$ on $\omega$. We see that  with respect to the BH regime, the other two regimes show a suppression of the gluon emission. We discuss here first the BH regime. In our MC approach, the two other regimes are implemented as a partial rejection of the BH cross section as described in the next subsection.

\begin{figure}[htb]
\centering
\begin{tikzpicture}[domain=0:3]
\draw[->] (0,0) -- (6,0) node[right] {$\omega$};
\draw[->] (0,0) -- (0,3) node[above] {$\displaystyle \omega\frac{\mathrm{d}N}{\mathrm{d}\omega}$};
\draw[color=blue] (-0.1,2.5) node[left,color=black] {$\sim L/\lambda$} -- node[above] {GB} (1.5,2.5);
\draw[color=green!50!black!80] (1.5,2.5) -- node[below=0.4cm] {BDMPS-Z} (4,1.5);
\draw[color=orange] (4,1.5) -- node[right=0.5cm] {GLV} (5,0.4);
\draw[dashed] (1.5,2.5) -- (1.5,0) node[below] {$\omega_{\text{BH}}$};
\draw[dashed] (4,1.5) -- (4,0) node[below] {$\omega_c$};
\draw[dashed] (1.5,2.5) -- (5.5,2.5);
    \draw[->] (3.4,2.2) -- node[right=0.1cm] {$\displaystyle \frac{1}{\sqrt{\omega}}$} (3.4,1.9);
    \draw[->] (4.8,2.3) -- node[right=0.1cm] {$\displaystyle \frac{1}{\omega}$} (4.8,1.0);
    \end{tikzpicture}
    \caption{Schematic diagram of the energy radiation spectrum in the different theoretical regimes.}
    \label{fig:energyspectrum}
\end{figure}

The GB cross section $\sigma_{\text{inel}}$ depends (after integrating out energy and momentum conservation) on five kinematic variables: $\vec{k}_{\operp}$, the transverse momentum of the virtual gluon, $\vec{l}_{\operp}$, the momentum transfer from the QGP parton to the projectile and $x$, the light-cone momentum (LCM) fraction of the virtual gluon with respect to the one of the projectile.\footnote{These definitions strictly apply in the cm frame where both incoming partons are colinear.} It depends also on the thermal mass of the time like partons, eq.~(\ref{eq:mthg}) or eq.~(\ref{eq:mthq}).
The propagator of the exchanged gluon is $\propto 1 / (l_T^2+\mu^2)$ with $\mu$ given by eq.~(\ref{eq:mu}). We perform the calculations for two extreme assumptions on the mass of the QGP partons $m_q$ (and $m_g$): Either they are set to 0 or to $\infty$,  what will in the following  be generically indicated by ``$m_q=0$'' or ``$m_q=\infty$'', implicitly assuming that $m_g=m_q$ for the QGP partons. The latter case will be used for calibration purposes, when comparing to the BDMPS-Z calculation, where static scattering centers are assumed. 

In the high energy limit, the inelastic Gunion-Bertsch cross section is given by the factorized form 
\begin{equation}
\frac{\mathrm{d}^5\sigma_{\rm rad}}{(\mathrm{d}x \, \mathrm{d}^2 l_\perp \, \mathrm{d}^2 k_\perp)} \sim \left \vert \mathcal{M}_{\rm el}\right \vert^2 \times P_g \times \Theta \,, \label{eq:inelxsec}
\end{equation}
which is used in our approach. $\mathcal{M}_{\rm el}$ is the matrix element for the elastic collision, $P_g$ the conditional probability of emission, and $\Theta$ is the phase space constraint. The exact forms of $\mathcal{M}_{\rm el}$ and $\Theta$ depend on the choice of $m_q = 0$ or $m_q \to \infty$. The exact expressions and the details of their derivation are given in eqs.~(\ref{eq:mq0xsec}) and (\ref{eq:mqinfxsec})  of appendices~\ref{app:gbmq0sampling} and \ref{app:gbmqinfsampling}, respectively, which also explain the MC procedure used for the sampling. The $\alpha_s$ in $P_g$ is fixed to $0.4$, whereas $\mathcal{M}_{\rm el}$ and its $\alpha_s$ dependence are taken as discussed in the previous subsection.  The gluon spectrum in the GB approach falls off like $1/\omega$, as shown in Fig.~\ref{fig:energyspectrum},
\begin{equation}
\omega\frac{\mathrm{d}N^{\text{GB}}}{\mathrm{d}\omega} \simeq \frac{\alpha_s C_R}{\pi} \frac{L}{\lambda_q}\,
\label{eq:gb}
\end{equation}
with $\lambda_q$ being the mean free path. This is the result for the massless partons, a case which is plagued by soft and colinear divergences. The generalization to finite masses -- that cure the divergences -- leads to some deviations with respect to this law (see eq.~(\ref{eq:dsigmaradsup3dx}), where an extra dependence on $x$ stems from $\tilde{m}_g$ defined through eq.~(\ref{eq:mgtilde}). 

In order to prepare in our approach for the calculation of the formation of on-shell  gluons, what will be discussed in the next section, we perform for each inelastic collision the following steps in the Monte Carlo generation of the kinetic variables:
\begin{itemize}
\item The sampled gluon is added to the list of the virtual gluons for a given projectile. The emission amplitude has an initial phase $\varphi=0$ and its collision counter is set to $N_s=1$.  

\item The initial energy-momentum of the virtual gluon is stored to be latter updated after further collisions.

\item The energy-momentum of the jet parton  is possibly corrected for the energy momentum carried by the virtual gluon and the transverse momentum it receives from the collision with the QGP parton.\footnote{See discussion in subsection \ref{subsubsect:phaseaccumalgo}}.

\item The color configuration is updated for the gluon emission, i.e.~the quark projectile with initial color $c$ gets a new color $c+1$, whereas the radiated gluon gets color $c$ and anti-color $c+1$. Collisions with the QGP partons are presently considered here as colorless. We will improve on this in future.

\end{itemize}

A flow diagram of the full Monte Carlo algorithm is given in Fig.~\ref{fig:flowdiag}.
\begin{figure*}[htb]
    \tikzstyle{block} = [draw, rounded corners, minimum height=2.0em, minimum width=4em]
    \tikzstyle{fitblock} = [draw, rounded corners, minimum height=6.0em, minimum width=4em]
    \tikzstyle{line} = [-{Latex[scale=1.0]}, solid]
    \begin{tikzpicture}[auto, node distance=1.0cm]
        \node [block] (rad) {virtual incoherent gluon formation according to GB seed};
        \node [block, below of=rad] (ns) {$N_s = 1$, $\varphi = 0$};
        \node [block, below of=ns] (evolve) {Evolve one timestep $\Delta t$};
        \node [block, below of=evolve] (phase)
            {\begin{tabular}{c}
                Gluon phase accumulation\\
                $\varphi \to \varphi + \Delta \varphi$
            \end{tabular}};
        \node [block, below of=phase] (inmedium) {Still in medium ($T > T_c$)?};
        \node [block, below=0.7cm,left of=inmedium] (inmediumN) {No};
        \node [block, below=0.7cm,right of=inmedium] (inmediumY) {Yes};
        \node [block, below of=inmediumY, text width=2.0cm, align=center] (phasecrit) {$\varphi > \varphi_c$?};
        \node [block, below=0.7cm,left of=phasecrit] (phasecritY) {Yes};
        \node [block, below=0.7cm,right of=phasecrit] (phasecritN) {No};
        \node [block, below of=phasecritN] (el) {Elastic rescattering?};
        \node [block, below=0.7cm,left of=el] (elY) {Yes};
        \node [block, below=0.7cm,right of=el] (elN) {No};
        \node [block, below=1.2cm of el] (increment) {$N_s \to N_s + 1$, update $k$};
        \node [block, below of=elY, draw=none] (dummyincrement) {};
        \coordinate [right=0.9cm of increment, draw=none] (dummy1) {};
        \coordinate [right=1.1cm of elN, draw=none] (dummy2) {};
        \coordinate [right=3.0cm of evolve, draw=none] (dummy3) {};
        \node [block, below=3.0cm of phasecritY, draw=none] (dummyform) {};
        \node [block, below=3.7cm of phasecrit] (form)
            {\begin{tabular}{c}
                Accept gluon with\\
                probability $1/N_s$
            \end{tabular}};
        \node [block, below=0.9cm,left of=form] (formN) {No};
        \node [block, below=0.9cm,right of=form] (formY) {Yes};
        \node [block, right=0.8cm of formY, fill=green!20] (add) {Add virtual gluon as radiated/realised};
        \node [block, below=6.8cm of inmediumN, fill=red!20] (discard) {Discard virtual gluon};
        \node [block, below of=formN, draw=none] (dummydiscard) {};
        \node [block, below of=add] (adjust) {Adjust projectile kinematics accordingly};
        \draw [line] (rad) -- (ns);
        \draw [line] (ns) -- (evolve);
        \draw [line] (evolve) -- (phase);
        \draw [line] (phase) -- (inmedium);
        \draw [line] (inmediumY) -- (phasecrit);
        \draw [line] (phasecritN) -- (el);
        \draw [line] (elY) -- (dummyincrement);
        \draw [line] (increment) -- (dummy1);
        \draw [line] (dummy1) -- (dummy3);
        \draw [line] (elN) -- (dummy2);
        \draw [line] (dummy3) -- (evolve);
        \draw [line] (phasecritY) -- (dummyform);
        \draw [line] (formY) -- (add);
        \draw [line] (inmediumN) -- (discard);
        \draw [line] (formN) -- (dummydiscard);
        \draw [line] (add) -- (adjust);
        \draw [line] (discard) -- (adjust);
    \end{tikzpicture}
    \caption{A diagram summarizing the flow of the Monte Carlo algorithm for the coherent medium-induced gluon radiation in our model.}
    \label{fig:flowdiag}
\end{figure*}


\subsection{The BDMPS-Z regime} 
\label{sec:lpm}
\subsubsection{General considerations}

\begin{figure}[htb]
\centering
\begin{tikzpicture}
			\node[anchor=south west,inner sep=0] at (0,0) {\includegraphics[angle=-90,width=0.40\textwidth]{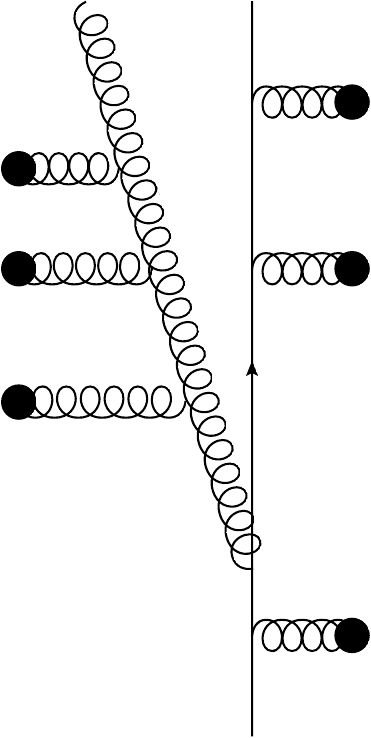}};
			\node at (0.3,1.5) {$E$};
			\node at (6.8,3.1) {$\omega$};
			\node at (6.8,2.1) {$k_\perp$};
            \node at (2.2,1.9) {$k_{\perp}^{\mathrm{pre}}$};
			\node at (4.43,4.2) {$N_s$};
			\node[rotate=180] at (4.41,3.8) {{\makebox[20ex]{\upbracefill}}};
			\node[circle,fill,inner sep=0.8pt] at (3.72,2.8) {};
			\node[circle,fill,inner sep=0.8pt] at (3.92,2.8) {};
			\node[circle,fill,inner sep=0.8pt] at (4.12,2.8) {};
		\end{tikzpicture}
    \caption{Illustration of the LPM effect for coherent gluon radiation and some of the relevant variables.}
    \label{fig:lpm}
\end{figure}

The virtual gluons, created in inelastic collisions of the incoming parton, may scatter
with the partons of the QGP, as shown schematically in Fig.\ref{fig:lpm} .
For soft gluons this has been studied by GW(P) in~\cite{Gyulassy:1993hr,Wang:1994fx} and in the series of BDMPS-Z papers~\cite{Baier:1996kr,Baier:1996sk,Zakharov:1996fv,Zakharov:1997uu}. There it has been shown that, in analogy to the LPM effect in QED, the gluon radiation amplitudes, created in subsequent scatterings of a colored projectile parton in the medium, interfere destructively. It is a distinctive new feature of QCD radiation  that the virtual gluons themselves can scatter with the partons of the QGP. These interactions modify the formation time of the gluon.  As discussed, in vacuum for  a  gluon with four-momentum $k\equiv(\omega,\vec{k}_\perp, k_z)$, the  formation time $t_{f}$ can be expressed as~\cite{Dokshitzer:1991wu} 
\begin{equation}
t_{f}=\lambda_{\text{coh}} \simeq \frac{\omega}{k_\perp^2} \,.
\end{equation} 
An on-shell gluon may scatter with the QGP partons. Elastic collisions of a gluon  with a QGP parton increase its transverse momentum $k_{\operp}$ which is originally equal $k_\perp^{\rm pre}$. In an medium such a momentum transfer can be quantified by the transport coefficient  $\hat q$, the average momentum transfer squared per unit time
\begin{equation}
\hat q = \frac{\mathrm{d}\langle k_\perp ^2\rangle}{\mathrm{d}t} \,.
\end{equation}
Following the heuristic arguments given by BDMPS-Z, the average transverse momentum transfer per unit length, eq.~(\ref{eq:qhatsimple}), applies also to virtual gluons, which scatter elastically with QGP partons. These collisions with the QGP result in a shortening of the in-medium formation time, deduced from the implicit equation 
\begin{equation}
t_{f}  \simeq \frac{\omega}{k_\perp^2(t_f)} \simeq \frac{\omega}{\hat q t_{f}} \hspace{5mm} \to \hspace{5mm} t_{f}^{\text{BDMPS-Z}}\simeq\sqrt{\frac{\omega}{\hat q}}\,,
\label{eq:tfsimple}
\end{equation}
where the expression is derived assuming that $\hat{q} t_f \gg (k_{\operp}^{\rm pre})^2$.\footnote{Since the typical $k_{\perp}\approx \mu\approx m_g^{\rm therm}$, this condition writes $\sqrt{\omega \hat{q}}\gg \mu^2$, which is satisfied provided $\omega\gg \omega_{\rm BH}$. }
If the parton mean free path is larger than $t_f$, the amplitudes for gluon emission on the trajectory of the parton do not interfere. If, however, $t_f$ is larger than the mean free path, interference becomes important and the relative phases of the amplitudes, $\Delta \varphi = k_\mu(x_1-x_2)^\mu \approx \omega \left \vert \bm{x}_1-\bm{x}_2 \right \vert (1-\cos\theta)$ have to be added. This is the LPM (BDMPS-Z) domain. 
For a medium of length $L$ the gluon is formed in the medium if $t_f \lesssim L$. The largest momentum a gluon can get on average, while traversing the medium, is $k_{\perp, \max}^{2} \simeq \hat q L$. We find therefore that above a limiting gluon energy of $\omega_c \sim \hat q L^2$, 
gluons can only be formed by fluctuations and are therefore strongly suppressed.
 
One of the essential results of the BDMPS-Z multiple soft scattering approximation is the energy spectrum of gluons emitted by an energetic projectile. This distribution is modified by the interference as compared to independent emissions in the Bethe-Heitler regime. It shows a $1/\sqrt{\omega}$ dependence, which is characteristic for the non-Abelian LPM effect. 

Coming to more precise estimates, the radiation spectrum of gluons radiated inside a medium of finite length $L$ with intermediate energies, $\omega_{\text{BH}} < \omega < \omega_c$, is given by~\cite{Caron-Huot:2010qjx,Mehtar-Tani:2019tvy}
\begin{equation}
    \omega \frac{\mathrm{d}N^{\text{BDMPS-Z}}}{\mathrm{d}\omega} \simeq \frac{2 \alpha_s C_R}{\pi} \ln \left \vert \cos(\Omega L) \right \vert \,, \label{eq:bdmpsz}
\end{equation}
where $C_R$ is the Casimir factor corresponding to the representation of the incoming parton and $\Omega$ is the complex frequency
\begin{equation}
    \Omega \equiv \frac{(1-i)}{2} \sqrt{\frac{\hat{q}_g}{\omega}} \,, \label{eq:complexfreq}
\end{equation}
where $\hat{q}_g$ is the gluonic transport coefficient, related to that of the quark --- e.g.\ eq.~(\ref{eq:qhatsimple})  --- through
\begin{equation}
    \hat{q}_g = \frac{C_A}{C_F} \hat{q} \,.
\end{equation}
In the limit of  a large path length, $L \to \infty$, we can approximate
\begin{equation}
    \ln \left \vert \cos \left ( \frac{(1-i)}{2} \sqrt{\frac{\hat{q}_g}{\omega}} L \right ) \right \vert \overset{L \to \infty}{\approx} \frac{1}{2} \sqrt{\frac{\hat{q}_g}{\omega}} L = \sqrt{\frac{\omega_c}{2 \omega}} \, .
    \label{eq:BDMPSLL}
\end{equation}
The BDMPS-Z spectrum becomes then
\begin{eqnarray}
\omega \frac{\mathrm{d}N^{\text{BDMPS-Z}}}{\mathrm{d}\omega} \overset{L \to \infty}{\approx} \frac{\sqrt{2} \alpha_s C_R}{\pi} \sqrt{\frac{\omega_c}{\omega}} \,, \label{eq:bdmpszLinf}
\end{eqnarray}
which shows the characteristic $1/\sqrt{\omega}$ behavior.\footnote{Stricto sensu, the BDMPS-Z regime is obtained by choosing self-consistently the hard scale $q_{\perp,{\rm max}}^2$ appearing in the expression of $\hat{q}(q_{\perp,{\rm max}}^2)$, a hard scale  such that $q_{\perp,{\rm max}}^2=\sqrt{\omega \hat{q}(q_{\perp,{\rm max}}^2)}$ --- see the discussion in ref.~\cite{Mehtar-Tani:2019tvy} --- leading to log corrections wrt the $\omega^{-1/2}$ behavior.} The energy limits are defined as
\begin{equation}
\omega_{\text{BH}} = \lambda_q \mu^2 \,, \hspace{5mm} \text{ and } \hspace{5mm} \omega_c = \frac{\hat{q}_g L^2}{2} \simeq \frac{\mu^2 L^2}{2 \lambda_g} \,, \label{eq:limits}
\end{equation}
with $\lambda_q = \Gamma_{\text{el}, q}^{-1} \sim (\alpha_s T)^{-1}$ being the quark Coulomb like mean free path, and $\lambda_g = (C_F/C_A)\lambda_q$ that for the gluon. The IR regulator $\mu^2$ is given by eq.~(\ref{eq:mu}). The gluon spectrum  eqs.~(\ref{eq:bdmpszLinf}) is shown as well schematically in Fig.~\ref{fig:energyspectrum}. 

In the fully coherent regime, i.e.\ for  a large gluon energy, $\omega > \omega_c$, the radiation spectrum is given by the first-order opacity expansion of GLV~\cite{Gyulassy:2000er,Mehtar-Tani:2019tvy}\footnote{We mention that the scales, at which $\hat{q}_g$ should be taken in the BDMPS-Z and in the GLV regimes, differ. The first relies on a self-consistent estimation of the total transverse momentum transfer during the formation time, while the latter corresponds to $\hat{q}_0$, i.e.\ eq.~(\ref{eq:qhatsimple}) stripped from its Coulomb logarithm.}
\begin{equation}
\omega \frac{\mathrm{d}N^{\text{GLV}}}{\mathrm{d}\omega} \simeq \frac{\alpha_s C_R}{8} \frac{\hat{q}_g L^2}{\omega} \,.
\label{eq:glv}
\end{equation}

This spectrum is also  shown schematically in Fig.~\ref{fig:energyspectrum}.  
Recently the opacity approach has been further advanced~\cite{Mehtar-Tani:2019tvy,Barata:2021wuf} and offers now a better transition around $\omega=\omega_c$. This result will be used to calibrate our Monte Carlo approach.
If one wants to model the passage of a fast parton through matter,  these three domains of energy loss have to be considered. This  is a big challenge for transport approaches, which are based on finite time step methods and the causality assumption, i.e.\ that the development in the next time step is exclusively based on the information available at the present and past time steps.  Finite formation times, where only the further development of the system decides whether a virtual gluon becomes reabsorbed or whether  it is finally converted into an on-shell gluon, are difficult to formulate in such approaches. This is already true for static QGP matter at a fixed temperature and an even bigger challenge for approaches, which want to simulate the energy loss of a fast particle in an expanding QGP, created in heavy ion collisions. The transition of the analytical static matter results to such transport approaches is not unique and therefore it is necessary to verify that the essential features are reproduced independent of the different possible implementations.

\subsubsection{Phase accumulation algorithm}
\label{subsubsect:phaseaccumalgo}
In the original BDMPS-Z work, the $\omega$ power spectrum is created by the interference between the forward going and the backward going amplitudes for the gluon radiation. Interference terms cannot be easily formulated in usual Monte Carlo based transport theories, which propagate particles and where the collisions among these particles are independent. The goal of this subsection is to explain how such interference terms can be mocked up in transport approaches resorting to an algorithm based on phase accumulation.
As seen in ref.~\cite{Baier:1996kr}, the gluon radiation probability involves phases 
which are accumulated between the radiation vertex $i$ at time $t$ (in the direct amplitude) and the radiation vertex $j$ at time $t'$ (in the conjugate amplitude). Each length element $\Delta_l$ along the path from $i \to j$ leads to an associated contribution $\frac{\kappa}{\lambda} U_l^2 \Delta_l$, where $l$ runs from $i$ to $j-1$ and $\vec{U}_l$ is --- up to a factor $\mu^{-1}$ --- the difference between the final $\vec{k}_\perp$ and the accumulated transverse momentum after $l-i$ scatterings, which can be rewritten in terms of the ``local'' $\vec{k}_\perp$ of the gluon.  If a gluon is
emitted at a vertex $i$ in the amplitude and at a vertex $j$ in the conjugate amplitude
we obtain a global phase
\begin{equation}
\Phi_{j,i} \approx \sum_{i\le l\le j} \frac{k_{\perp,l}^2}{2 \omega}\times\lambda \,,
\end{equation}
where  $\lambda$ is the elastic mean free path. Such phase increases with the ``distance'' $j-i$. If this distance and the equivalent phase $\Phi_{j,i}$ becomes too large, the interference disappears and the gluon emissions are considered to be independent. Such a criteria allows to define an ensemble of $N_s$ scattering centers that radiate coherently.

In our algorithm, we take over this expression and assume that a gluon is formed and can be considered as an on-shell gluon (which may undergo further collisions and emissions) if the accumulated phase exceeds a critical value $\varphi_c$. If one assumes, in a MC algorithm, that  on the average $k_{\perp,l}^2 \simeq (l-i)\lambda \times \hat{q}$  and if one imposes $\Phi_{j,i} < \varphi_c$, one gets, as a formation criteria:
\begin{equation}
\frac{\hat{q}\lambda^2}{2\omega}  \times \sum_{i\le l \le j} (l-i) \approx \frac{\hat{q}\lambda^2 (j-i)^2}{4\omega}\approx \varphi_c \Leftrightarrow (j-i) \approx \frac{2}{\lambda} \sqrt{\frac{\omega}{\hat{q}}\,\varphi_c} \,.
\end{equation}
Hence, the number of inelastic collision amplitudes that contribute to a coherent gluon emission scales like 
\begin{equation}
N_s \simeq \frac{2}{\lambda} \sqrt{\frac{\omega}{\hat{q}}\,\varphi_c} \,.
\label{eq:NsMC}
\end{equation}
As the associated power spectrum will be reduced by a factor $1/N_s$ as compared to the GB one, it will acquire a $\omega^{-1/2}$ slope, as predicted by the BDMPS-Z approach.  The threshold $\varphi_c$ cannot be calculated from first principle. It may be used in order to tune the radiation spectrum to the BDMPS-Z reference given by eq.~(\ref{eq:bdmpszLinf}).

\begin{figure}[htb]
\centering
\begin{subfigure}{.3\textwidth}
\centering
\begin{tikzpicture}
        \node[anchor=south west,inner sep=0] at (0,0) {\includegraphics[width=\linewidth]{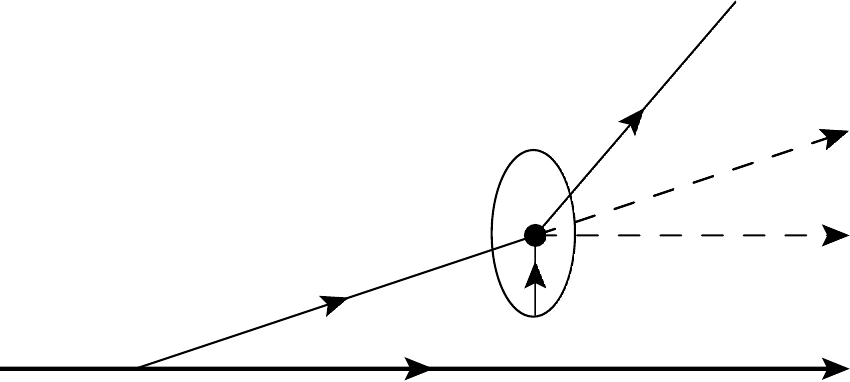}};
        \node at (5.1,-0.2) {$+z$};
        \node at (1.0,-0.2) {Jet projectile};
        \node at (3.97,0.3) {$\vec{q}_{\perp} = (q_{\perp, x}, q_{\perp, y}, 0)$};
        \node at (1.5,1.0) {$k = (k^{+}, k^{-}, k_x, k_y)$};
        \node at (2.3,2.1) {$k' = (k^{+\prime} = k^{+}, k^{-\prime}, k_x', k_y')$};
      \end{tikzpicture}
      \caption{$k^{+}$ conservation, with $k^{+\prime} = k^{+}$ where $k^{+} = \omega + k_L$ and transverse spatial kick in the $x$-$y$ plane only.}
\label{fig:elasticprescriptions1}
\end{subfigure}\hfill%
\begin{subfigure}{.3\textwidth}
\centering
\begin{tikzpicture}
\node[anchor=south west,inner sep=0] at (0,0) {\includegraphics[width=\linewidth]{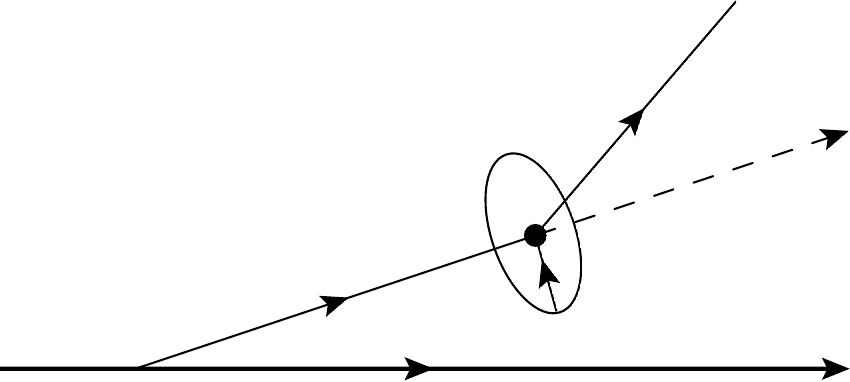}};
        \node at (5.1,-0.2) {$+z$};
        \node at (1.0,-0.2) {Jet projectile};
        \node at (3.76,0.3) {$\vec{q}_{T} = (q_{T, x}, q_{T, y}, q_{T, z})$};
        \node at (1.5,0.9) {$k = (\omega, \vec{k})$};
        \node at (2.9,2.1) {$k' = (\omega' = \omega, \vec{k}')$};
\end{tikzpicture}
\caption{Energy conservation, with $\omega' = \omega$ and transverse spatial kick perpendicular to initial direction $\vec{k}$.}
\label{fig:elasticprescriptions2}
\end{subfigure}\hfill%
\begin{subfigure}{.3\textwidth}
\centering
\begin{tikzpicture}
\node[anchor=south west,inner sep=0] at (0,0) {\includegraphics[width=\linewidth]{plots/ElasticPrescription.pdf}};
\node at (5.1,-0.2) {$+z$};
\node at (1.0,-0.2) {Jet projectile};
        \node at (3.76,0.3) {$\vec{q}_{T} = (q_{T, x}, q_{T, y}, q_{T, z})$};
        \node at (1.5,0.9) {$k = (\omega, \vec{k})$};
        \node at (2.9,2.2) {$k' = (\omega', \vec{k}')$};
        \node at (2.0,1.7) {\tiny $\omega' = \omega - \left (\sqrt{(m_{\text{th}}^q)^2+q_T^2}-m_{\text{th}}^q \right )$};
\end{tikzpicture}
\caption{Energy reduction, with $\omega' < \omega$ and transverse spatial kick perpendicular to initial direction $\vec{k}$.}
\label{fig:elasticprescriptions3}
\end{subfigure}
\caption{Illustrations of the three different prescriptions for the elastic scatterings, in the jet frame, assumed to be aligned on $Oz$.}
\label{fig:elasticprescriptions}
\end{figure}

Let us now come to the concrete algorithm adopted in our approach.  A virtual gluon, formed in a collision during the time step $\Delta t_i\equiv [t_{i},t_{i+1}]$  with a Gunion-Bertsch seed, can scatter elastically in later time steps, $l > i$ according to eq.~(\ref{eq:Gammaeltot}). If such a virtual gluon suffers an elastic collisions in the time step $\Delta t_l$ our algorithm for the LPM effect for such a virtual gluon proceeds as follows:
\begin{itemize}

\item The norm $q_{T,l}$ of the transverse momentum transfer from the QGP parton to the virtual gluon is randomly drawn from the differential scattering rate 
(\ref{eq:gammael}), where $\mu$ depends on the local temperature, e.g. through eq.  (\ref{eq:mu}).

\item  Since previous considerations on the phase-accumulating algorithm do not allow us to fix the other kinematic variables unambiguously, we have chosen, as a first guiding principle, to constrain the virtual gluon to stay on-shell during its formation. This choice appears to be natural for a MC algorithm based on particle degrees of freedom.  For the other kinematic variables we employ one of the three following prescriptions,  illustrated in Fig.~\ref{fig:elasticprescriptions}):
    
\begin{itemize}
\item Option 1: \emph{$k^{+}$ conservation}; $k^{+'} = k^+$, means the light cone momentum is conserved. Then, direction of the transverse momentum exchange is chosen
$\perp$ to the jet axis (with $\vec{q}_\perp=q_{T,l}$) and $k^- $ is adjusted to keep the gluon on the mass shell. This description conserves neither energy nor momentum but follows the Zakharov~\cite{Zakharov:1996fv} approach.

\item
Option 2: \emph{Energy conservation}; $\omega' = \omega $, means the energy of the gluon is conserved. The transverse momentum exchange is chosen $\perp$ to the local direction of the virtual gluon and the longitudinal momentum is then
adjusted to keep the gluon on the mass shell. This is the correct formula for scattering on an infinite massive QGP parton as assumed in BDMPS calculations~\cite{Baier:1996sk}.\footnote{Notice that it is not always possible to satisfy both, the on-shellness condition and the longitudinal momentum conservation,  If this is not possible for a given momentum exchange the scattering is rejected.}

\item Option 3: 
\emph{Energy reduction};
\begin{equation}
\omega'=\omega-\left (\sqrt{(m_q^{\text{therm}})^2+q_{T,l}^2}-m_q^{\text{therm}} \right ) \,,
\label{eq:energyreduction}
\end{equation}
assuming that the QGP parton is initially at rest with its thermal mass given in eq.~(\ref{eq:mthq})\footnote{Independent of the prescription made for the scattering centers (infinitely massive ($m_q = \infty$) or zero mass ($m_q = 0$)) to evaluate the inelastic cross section.}. Then the longitudinal gluon momentum is adjusted to satisfy the on-shell condition. If this is not possible the event is discarded. This option is the most suitable for the physical case of QGP partons with finite thermal mass.
\end{itemize}
\item If the elastic collision is not vetoed by the energy-momentum conservation condition, the collision counter $N_s$ for the virtual gluon is incremented by 1.
 
\end{itemize}

In the time steps after its formation we follow the virtual gluon and perform the following steps:
\begin{itemize}    
\item The phase $\varphi$ of the virtual gluon is increased in each time step $l$  after the emission by
\begin{equation}
\Delta\varphi_l=\frac{2P\cdot k'}{E}\cdot \frac{\Delta t}{\hbar c}\,,
\end{equation}
where $P$ and $E$ are the four-momentum and energy of the projectile which has emitted the gluon, and $k'$ is the gluon momentum at the beginning of the time step.\footnote{In practice, the time steps are small enough that this prescription is irrelevant.} Hence at the end of the time step $j$ the total phase is $\sum_{l=i}^j \Delta \varphi_l$. In the limit of an energetic projectile $E\gg m_q, \omega \gg m_g$, $(P\cdot k)/E$ reduces to $(m_g^2+k_\perp(t')^2)/k^+ \approx (m_g^2+k_\perp(t')^2)/2\omega$ In the $E\gg m_q, \omega \gg m_g$ limit the expression  coincides with the one used in \cite{Zapp:2011ya}. Note the extra factor 2 as compared to~\cite{Zapp:2011ya} , for both, for the phase accumulation formula and for the formation phase.  $k_\perp(t')$ is the instantaneous component of the virtual gluon transverse to the jet direction.

\item If the virtual gluon scatters again elastically we apply the procedure as described before.

\item A formation check is performed: Once the gluon phase $\varphi$ reaches a threshold value $\varphi_c$ while the local temperature of the medium is above the critical temperature, $T_c = 150$ MeV, the virtual gluon is accepted as formed with a probability $1/N_s$ or discarded with a probability $1-1/N_s$. As discussed above, this allows to achieve the $1/\sqrt{\omega}$ suppression of the energy spectrum predicted by BDMPS-Z, since $N_s \sim t_f \sim \sqrt{\omega}$. If the local temperature falls below $T_c$ and the gluon is not formed yet, the virtual gluon is discarded as unformed.

\item Once a gluon is formed, it is considered as an on-shell particle, which can undergo elastic and inelastic collisions. The four-momentum of the projectile parton is adapted accordingly to the outcome of the gluon formation. The two possible schemes are: a) subtract  $k-l$ from the projectile parton at the moment when the virtual gluon is emitted and give it back to the projectile if and when the virtual gluon is discarded or b) subtract $k-l$ from the projectile only if and when the virtual gluon is realized. In the present study, we will place ourselves in the eikonal limit where the gluon emission leaves the momentum of the projectile parton  unchanged. Therefore this correction does not play a role for the observables.
\end{itemize}

A flow diagram of the implementation of the gluon formation in our approach is presented Fig.~\ref{fig:flowdiag}. The algorithm resembles the one described in \cite{Zapp:2011ya}, where successive elastic scatterings of the radiation candidates increase their $k_\perp$, and hence their chance to become a real gluon in the medium. We want to mention that this algorithm is adapted to an expanding medium because it takes into account the local temperature of the QGP at which the elastic scattering of the virtual gluon takes place (and allows to stop the evolution if the QGP temperature falls below the critical temperature $T_c$).

Besides, in a sufficiently large and dilute medium a low-energy on-shell gluon can also be formed for $N_s=1$. In this case we reproduce effectively the Gunion-Bertsch cross section (apart from possible momentum changes of the projectile parton due to later interactions with the QGP). Our algorithm smoothly interpolates between these two regimes.

The question that naturally arises is whether one can also reproduce the GLV spectrum in such a MC approach with a phase accumulating algorithm. If one considers the case $\lambda \ll L \ll L^\star$, where $L^\star\approx \sqrt{\frac{\omega}{\hat{q}}}$ is the typical time for in-medium gluon formation, the total number of scattering centers $N=L/\lambda$ is too small to ensure the full in-medium gluon production. It is argued in ref.~\cite{Peigne:2008wu} that in this case the induced gluon production is dominated by a single hard rescattering --- the GLV regime --- 
leading to the simplified expression 
\begin{equation}
\frac{\mathrm{d}I^{\rm GLV}}{\mathrm{d}\omega}\simeq \frac{\alpha_s L}{\lambda} \int \frac{\mathrm{d}^2 k_\perp}{k_\perp^2} \times \frac{\mu^2}{k_\perp^2+\mu^2} \left(1-\cos \frac{L k_\perp^2}{2 \omega} \right) \,,
\end{equation}
where the last factor represents the coherence between the vertex where the gluon is emitted from the projectile parton and the single elastic rescattering of that gluon.
Due to the last factor, the incoherent radiation is suppressed if $\frac{L k_\perp^2}{2\omega}$ is significantly smaller then 1.\footnote{When  $\frac{L k_\perp^2}{2\omega}\gtrsim 1$, the cos starts to oscillate fast and averages to a subleading contribution.} Then
\begin{equation}
\frac{\mathrm{d}I^{\rm GLV}}{\mathrm{d}\omega}\simeq \frac{\pi \alpha_s L}{\lambda} \int_ {\frac{2\omega}{L}}^{+\infty}\frac{\mathrm{d}k_\perp^2}{k_\perp^2} \times \frac{\mu^2}{k_\perp^2+\mu^2}\approx \frac{\pi \alpha_s L \mu^2}{\lambda} \int_{\frac{2\omega}{L}}^{+\infty}\frac{\mathrm{d}k_\perp^2}{k_\perp^4} \sim
\frac{\pi \alpha_s L^2 \mu^2}{\lambda \omega} \sim \frac{\alpha_s \omega_c}{\omega} \propto L^2 \,.
\end{equation} 
If $L\ll L^\star$, the number of scattering centers $N\ll N_s$ and thus on the average the virtual gluons are not converted into real ones. This can only be achieved by a statistical fluctuation of the transverse momentum. Starting from a GB spectrum $\propto k_\perp^{-4}$, the probability for the associated phase $\varphi = \frac{k_\perp^2 L}{\omega}$ to exceed $\varphi_c$ is given by the proportion of initial virtual gluons with $k_\perp^2> \frac{\omega \varphi_c}{L}$, which scales like $\frac{\mu^2 L}{\omega \varphi_c}$. Since the radiative cross section is sampled $\frac{L \alpha_s}{\lambda}$ times over path length $L$, the ensuing MC power spectrum writes
\begin{equation}
\frac{\mathrm{d}I^{\rm MC}}{\mathrm{d}\omega}\propto \frac{\hat{q}L^2}{\omega\varphi_c} \,,
\end{equation}
which has the correct GLV scaling (of eq.~(\ref{eq:glv})). 
While the phase accumulation method allows to reproduce the $\omega$ dependence 
in the BDMPS-Z as well as in the GLV distribution, one realizes that the numerical parameter $\varphi_c$ enters both expressions with a different power. It can hence be used to adjust the typical value of $\omega$ for which the slope of the power spectrum passes from $\omega^{-1/2}\to \omega^{-1}$ in the numerical simulations\footnote{Which is of course $\simeq \omega_c$.}.

For the parameter set used in the present version of our model ($\alpha_{s,{\rm eff}}$, $\kappa$,\ldots), the value $\varphi_c=6$ is chosen to match the transition between the BDMPS-Z regime (eq.~(\ref{eq:bdmpsz})) and the fully coherent GLV regime (eq.~(\ref{eq:glv})), as documented in appendix~\ref{app:calibration}.
In addition, a second numerical parameter is mandatory to adjust the absolute level of the $\mathrm{d}N/\mathrm{d}\omega$ MC-spectrum in the full BDPMS-Z-GLV regime. This is achieved by tuning, for large $k^+$, the gluon mass in the GB-seed algorithm to an effective gluon mass $m_g^{\rm reg}$. While the gluon mass has little influence on $\mathrm{d}N/\mathrm{d}\omega$ in the BDMPS-Z regime, being a subleading scale, it effectively affects the strength of the radiation from the GB seed --- as seen  in eq.~(\ref{eq:dsigmaradsup3dx}) --- and can be used as an adjustable parameter. The optimal value for $m_g^{\rm reg}$ is found to be around  15\% of $T$, as documented in appendix~\ref{app:calibration}.

As a last feature of our code, let us mention that the inelastic cross-section of any projectile parton with the QGP partons is sampled independently of existing  virtual gluons possibly emitted earlier and still under formation. In other words, several  virtual gluons can coexist for a given projectile.

\subsubsection{Gluon-initiated jets}

Although the presentation of the gluon-induced radiation as well as the discussion of the numerical results in the next section is made in the context of quark-initiated jets, SUBA-Jet also includes the medium evolution of gluon-initiated jets. In this case, the $\hat{q}$ parameterization used in eq.~(\ref{eq:virt}), the rate of elastic scatterings in eq.~(\ref{eq:Gammaeltot}), and the rate of inelastic scattering in eq.~(\ref{eq:inelxsec}) --- or more specifically eqs.~(\ref{eq:mq0xsec}) and (\ref{eq:mqinfxsec}) --- are multiplied by the appropriate color factor $C_A / C_F$.

\section{Results from the low virtuality component} \label{sec:results}

\begin{figure}[H]
\centering
\begin{minipage}[t]{.48\textwidth}
\centering
        \includegraphics[width=0.8\linewidth]{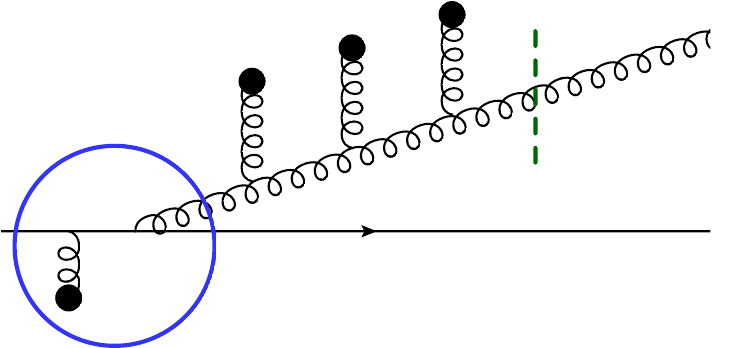}
        \captionof{figure}{Parton cascade with medium modifications for the BDMPS-Z conditions.The blue circle indicates an inelastic scattering resulting in a gluon emission. The green dashed line is the moment when the  virtual gluon becomes a formed gluon.}
        \label{fig:rescatter1}
    \end{minipage}\hfill%
    \begin{minipage}[t]{.48\textwidth}
        \centering
        \includegraphics[width=0.8\linewidth]{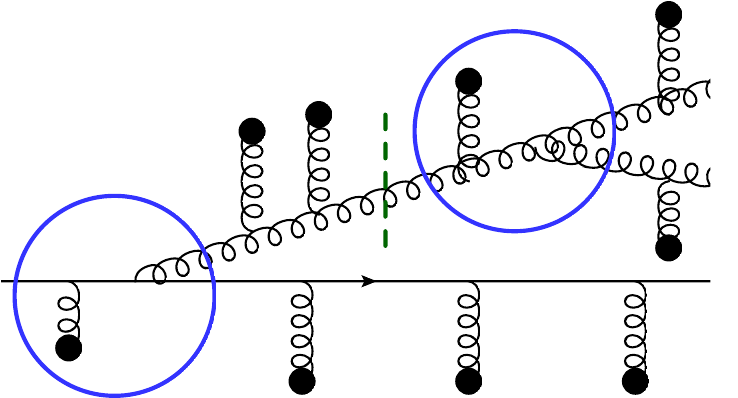}
        \captionof{figure}{Realistic parton cascade with medium modifications. The blue circle indicates an inelastic scattering resulting in a gluon emission. The green dashed line is the moment when the  virtual gluon becomes a formed gluon.}
        \label{fig:rescatter2}
    \end{minipage}
\end{figure}

In this section we present the results from the Monte Carlo implementation of the low virtuality component of our model. We start out with those for the GB radiation of virtual gluons, which are the seed of the later evolution. We show next that our Monte Carlo calculation of the $\omega$ spectrum reproduces the BDMPS-Z and GLV limits in their respective range of validity.  The processes we take into account for this study are displayed in Fig.~\ref{fig:rescatter1}.
Finally we relax some of the assumptions and show the role of the different model parameters and options, before in the last subsection the most ``realistic'' implementation of our model is considered, which corresponds to the following scenario: $m_q = 0$ in the initial GB seed, energy reduction prescription for the elastic scattering of virtual gluons and  elastic as well as inelastic scattering of on-shell gluons and partons, as displayed in Fig.~\ref{fig:rescatter2}.  This parameter choice will be the basis for future phenomenological studies with our model.

For all simulations, we disable the change of the longitudinal momentum of the incoming projectile due to elastic collisions or inelastic collisions, so effectively we work in the eikonal approximation. The medium temperature is $T=400$~MeV and the strong coupling constant is $\alpha_s = 0.4$, a value which corresponds to eq.~(\ref{eq.alphaeffGA}). $N_f$ is set to $3$. For these parameters eq.~(\ref{eq:mu}) gives $\mu\approx 0.44$~GeV, as well as $m_g^{\rm therm}=0.626$~GeV, $m_q^{\rm therm}=0.367$~GeV, $\lambda_{\rm el}^{q}=0.18$~fm, and $\lambda_{\rm el}^{g}=0.08$~fm. Eq.~(\ref{eq:qhatsimple}) yields then $\hat{q}_q(p=10\text{ GeV}/c)=2.9$~GeV$^2/$fm while the corresponding transport coefficient stripped of its Coulomb logarithm, is $\hat{q}_{q,0}=1.06$~GeV$^2/$fm.


\subsection{Benchmark of the Gunion-Bertsch radiation seed}

To benchmark the virtual gluon radiation procedure, we first produce the energy spectrum $\mathrm{d}N/\mathrm{d}\omega$ of the radiated gluons in a box of length L filled with QGP matter. We assume a mono-energetic stream of partons with an energy of $E=p=100$~GeV. The virtuality at the beginning of the low virtuality stage is set to $Q = m_Q= m_q^{\text{therm}}(T) $. 
\begin{figure}[H]
\centering
\includegraphics[width=0.6\textwidth]{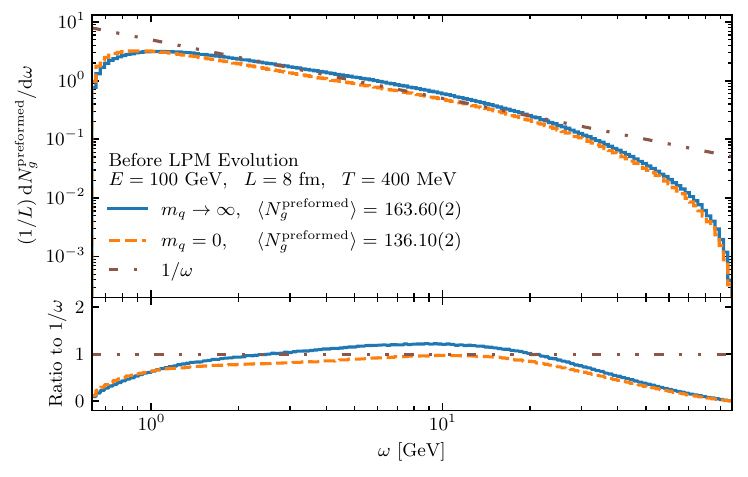}
\caption{Top: Radiation spectrum of virtual  gluons immediately after their creation, i.e.\ before the LPM evolution,  in our Monte Carlo approach. We display the result  for different choices of the mass of the QGP partons, $m_{q}$, as compared to the superficial $\omega^{-1}$ dependence of the Gunion Bertsch radiation formula with an arbitrary normalization (see text for details). $N_g^{\rm preformed}$ is the number of preformed gluons.
Bottom: Ratio of the spectrum for different choices of $m_q$ and the Gunion-Bertsch $1/\omega$ expectation. } 
\label{fig:dNdomega-beforeLPM}
\end{figure}
In Fig.~\ref{fig:dNdomega-beforeLPM} we present the energy distribution of the virtual gluons immediately after the generating inelastic collision, corresponding to the integration of eq.~(\ref{eq:inelxsec}) over $\mathrm{d}^2k_{\operp} \, \mathrm{d}^2q_{\operp}$. We display the result of our Monte Carlo sampling for two cases: For a mass of the QGP parton of $m_{\text{q}} = \infty$ (solid blue line) and of $m_{\text{q}} = 0$ (dashed orange line). These curves are compared to the generic spectrum of the GB approach, which falls off $\propto \omega^{-1}$. At intermediate $\omega$,  sufficiently large that masses and the infrared cutoff $\mu$ do not play a role and sufficient small that the energy/momentum conservation can be neglected, our sampled spectrum agrees with that from the GB approach. The finite gluon mass enters our cross section formula via $\tilde{m}_g^2(x)=(1-x) m_g^2 + x^2 m_Q^2$. \footnote{ In particular,  $m_g$ is taken as $m_g^{\rm reg}$ for $\omega \gtrsim T$, which is quite small, resulting in $\tilde{m}_g^2(x)\approx x^2 m_Q^2$.}. For the $\mu$ dependence we refer to eq.~(\ref{eq:dsigmaradsup3dx}) for $m_q=0$ and to eq.~(\ref{eq:dsigmadxsup3mqinf}) for $m_q=\infty$.  The choice of $m_q$ has an effect on the overall normalization of the curves but  less on the shape. At small $\omega$, the shrinking of the phase space $\Phi$, discussed in appendix \ref{app:gbmq0sampling} -- see in particular eqs.~(\ref{eq:phifinalregime1}) and (\ref{eq:phifinalregime2})
-- tames the $\omega^{-1}$ IR divergence found in the GB spectrum and even bends it down towards zero  when $\omega \searrow m_g$, as previously discussed in \cite{Aichelin:2013mra}. By properly implementing this phase space boundary, we do not need to resort to some artificial IR regulator.
\begin{figure}[H]
    \centering
    \includegraphics[width=0.6\textwidth]{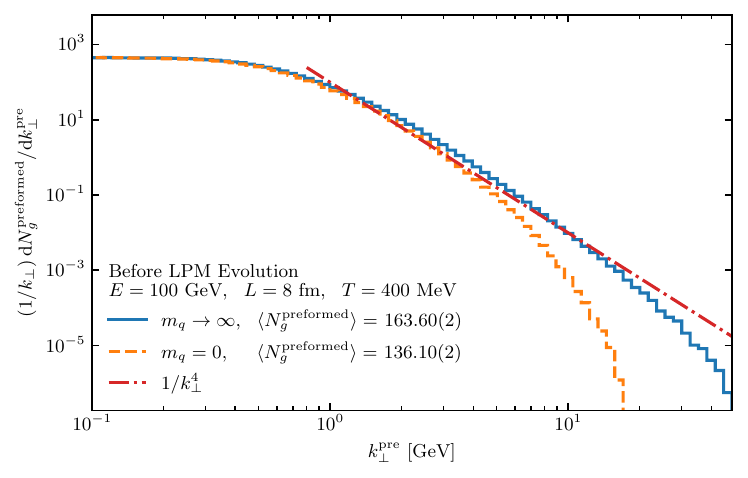}
    \caption{Gluon $k_\perp$ spectrum from the Monte Carlo. Details as in Fig.~\ref{fig:dNdomega-beforeLPM}. }
    \label{fig:dNdkT-beforeLPM}
\end{figure}
The $k_\perp$ distribution of the virtual gluons directly after their creation, is shown in Fig.~\ref{fig:dNdkT-beforeLPM}. Here we observe that for $k_{\operp} < 1$~GeV the choice of $m_{q}$ changes only the normalization of the distribution, while for the large $k_{\operp}$ energy/momentum conservation modifies the shape differently. From eqs.~(\ref{eq:ktmaxmq0}) and (\ref{eq:ktmaxmqinf}), derived in appendix~\ref{app:gbmq0sampling} and \ref{app:gbmqinfsampling}, one realizes that $k_{\operp}^{\rm max}$ can be as large as the jet energy $E_Q$ for $m_q=\infty$ case while it saturates at $\sqrt{s}\sim \sqrt{E_Q e_q}$ for $m_q=0$, with $e_q$ being the energy of the QGP parton before scattering. In the former case, one thus explores a larger range of $k_{\operp}$ values, with  $(1/k_{\operp}) \mathrm{d}N/\mathrm{d}k_{\operp}\sim k_{\operp}^{-4}$, as can clearly be seen. In the later case, the $k_{\operp}$ range is smaller,  following the thermal distribution of partons in the QGP, and no algebraic law can be observed. We display as well a reference spectrum $\propto 1/k_{\operp}^4$, the analytical result of the GB cross section for $q_{\operp}\ll k_{\operp}$,  for the later discussion of formed gluons. 

\begin{figure}\begin{minipage}[t]{.49\textwidth}
\centering
\includegraphics[width=\linewidth]{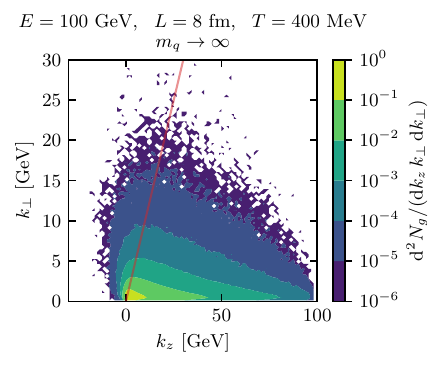}
\caption{Double differential spectrum as a function of the transverse momentum $k_{\operp}$ and longitudinal momentum $k_L$ of gluons from our Monte Carlo approach, immediately after their creation, i.e.\ before the LPM evolution, for $m_q \to \infty$. The red line stands for $k_{\operp}=k_z$. }
\label{fig:dN2_kTkz_gb_mqinft}
\end{minipage}\hfill%
\begin{minipage}[t]{.49\textwidth}
\centering
\includegraphics[width=\linewidth]{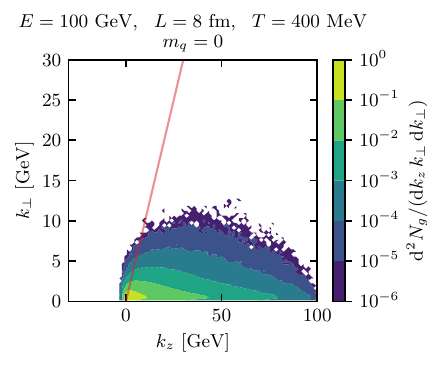}
\caption{Same as Fig. \ref{fig:dN2_kTkz_gb_mqinft}  for $m_q = 0$. }
\label{fig:dN2_kTkz_gb_mq0}
\end{minipage}
\end{figure}

In Figs. \ref{fig:dN2_kTkz_gb_mqinft} and \ref{fig:dN2_kTkz_gb_mq0},  we present the double differential spectrum $\mathrm{d}^2N/(\mathrm{d}k_z \, k_{\operp} \, \mathrm{d}k_{\operp})$ for both choices of $m_q$.  We see that for $m_q \to \infty$ a considerable fraction of gluons has even negative $k_{z}$ momenta. We see also clearly how the energy/momentum conservation limits the available phase space in the $m_q=0$ case.  In the region, which in both cases is kinematically allowed, the distributions are rather similar.

The three figures verify that our Monte Carlo algorithm to calculate the kinematic variables of the virtual gluons  after an inelastic collision  reproduces qualitative the form of the massless GB spectrum in the range where it is valid. Our  calculations include in addition to GB a finite mass as well as energy-momentum conservation. Deviation from the GB spectrum are seen when this
difference  becomes important,  at low and high $\omega$ values.  In the $m_q=0$ case the gluon phase space is considerably more limited than in the $m_q\to \infty$ case.

\subsection{Reproduction of the BDMPS-Z and GLV limits}
In order to compare our Monte Carlo results with the theoretical predictions from the BDMPS-Z and GLV approach, eqs.~(\ref{eq:bdmpsz}) and (\ref{eq:glv}), we have to adopt the conditions under which these results have been obtained. In particular we assume:
\begin{itemize}
\item The jet is represented by a energetic low virtuality quark, which radiates gluons. The quark 
suffers only inelastically collisions whereas the virtual gluons scatters only elastically.  Possible interactions of the gluons after their formation are not considered.
\item The QGP scattering centers have an infinite mass.
\item $\omega$ is conserved in the elastic scatterings of the virtual gluons.
\end{itemize}
Besides, one should keep in mind that the BDMPS-Z calculation applies under the following hierarchy: $k_\perp \ll \omega \ll E$.
\begin{figure}[H]
\centering
\includegraphics[width=0.6\textwidth]{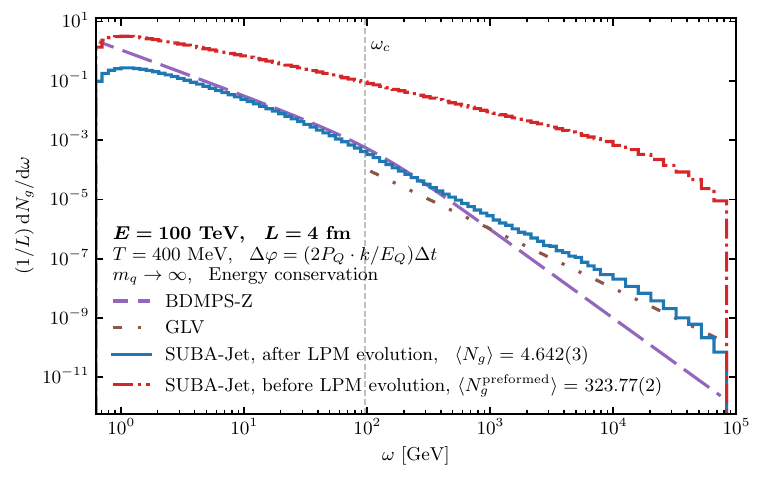}
\caption{Gluon energy spectrum of per unit length from our Monte Carlo approach. To match the BDMPS-Z conditions we assume infinitely massive scattering centers and energy conservation for elastic scattering. The gluon phase accumulation is calculated according to $\Delta \varphi = \left ( 2 P_Q \cdot k / E_Q \right ) \Delta t$. We display the spectrum for virtual gluons as well as of formed gluons, i.e before and after the BDMPS-Z evolution. The average number of finally formed  gluons $\langle N_g \rangle$ and of virtual gluons per jet are listed. The theoretical predictions for BDMPS-Z, eq.~(\ref{eq:bdmpsz}), and GLV, eq.~(\ref{eq:glv}), are also shown. }
\label{fig:dNdw_reproduction_100TeV}
\end{figure}
In Figs.~\ref{fig:dNdw_reproduction_100TeV} and  ~\ref{fig:dNdw_reproduction}
gluon energy spectra per unit length $(1/L) \,\mathrm{d}N/\mathrm{d}\omega$ are displayed.  
In Fig.~\ref{fig:dNdw_reproduction_100TeV} we consider a medium of length $L = 4$ fm and an initial projectile quark energy of $E = 100$ TeV to numerically approximate the $E \to \infty$ limit, whereas in Fig.~\ref{fig:dNdw_reproduction} the medium length is 8 fm and the initial projectile quark energy is $E = 100$ GeV. For these parameters, we have, according to eq.~(\ref{eq:limits}), $\omega_c \approx 96$ GeV and $385$ GeV, respectively. The BDMPS-Z results are calculated with eq.~(\ref{eq:bdmpsz}) and not in the limit $L\to \infty$ (eq~(\ref{eq:BDMPSLL})). 

In Fig.~\ref{fig:dNdw_reproduction_100TeV} we see that for intermediate energies $\omega< \omega_c$, the spectrum falls off roughly as $\omega^{-3/2}$, as expected by BDMPS-Z. This demonstrates that our Monte Carlo algorithm is able to reproduce the interference of the amplitudes. For $\omega \gg \omega_c$, the spectrum falls off  like $\omega^{-2}$, as predicted in  the GLV approach and not like $\omega^{-3}$ as in the BDMPS-Z approach above $\omega_c$). For this kinematic condition 
($E=100$ TeV and $L=4$ fm) around $300$ virtual gluons are produced but only a bit more than $1\%$
become real gluons.

The energy spectrum for virtual gluons right after the Gunion-Bertsch seed (i.e.\ before the LPM evolution) is also shown, in order to highlight the effect of the phase accumulation and acceptance as a function of $\omega$. The comparison between the BDMPS-Z and the GB result shows clearly how the interference --- modeled as a phase accumulation in our approach --- modifies the spectral form of the gluon emission and hence of the energy loss. We notice that this is true, in particular, for the low-$\omega$ part of the spectrum where strong deviations with respect to the BH regime are observed. This is a sign that for the chosen parameters ($T=400$ MeV and $\alpha_s=0.4$), the system is not sufficiently dilute that virtual gluons with $\omega \approx 0.5$ GeV can be formed in independent collisions. The reduction factor of the BDMPS-Z spectrum with respect to that from GB corresponds to the average number of elastic collision the virtual gluon has to suffer before it becomes an on-shell gluon.

\begin{figure}[H]
\centering
\includegraphics[width=0.6\textwidth]{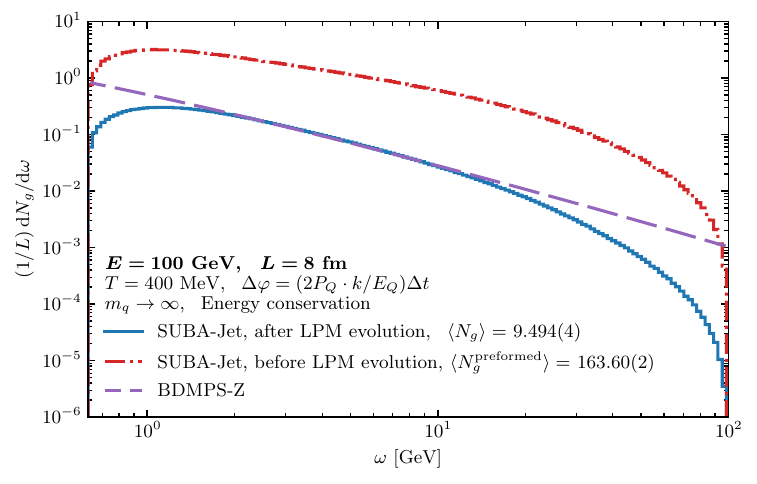}
\caption{Same as Fig.~\ref{fig:dNdw_reproduction_100TeV} for an initial jet of $E=100$~GeV and a path length $L = 8$~fm. }
\label{fig:dNdw_reproduction}
\end{figure}
\begin{figure}[H]
\centering
\includegraphics[width=0.6\textwidth]{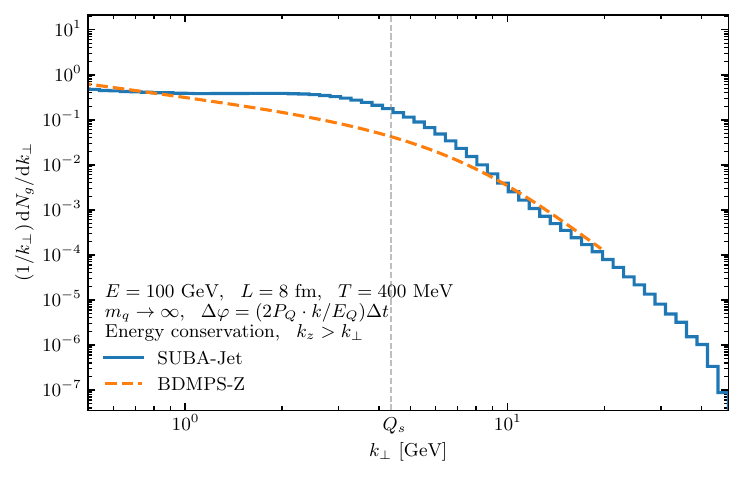}
\caption{The gluon $k_\perp$ spectrum from our Monte Carlo approach. To match the BDMPS-Z conditions we assume infinitely massive scattering centers and energy conservation for elastic scattering. The gluon phase accumulation is calculated according to $\Delta \varphi = \left ( 2 P_Q \cdot k / E_Q \right ) \Delta t$. The theoretical predictions calculated in the BDMPS-Z harmonic approximation \cite{Wiedemann:2000tf} is also shown.}
\label{fig:dNdkT_reproduction}
\end{figure}

In Fig.~\ref{fig:dNdw_reproduction}, the gluon energy spectrum per unit length
fulfills as well the BDMPS-Z expectation for intermediate energies, although the initial parton energy  is much smaller. The path length is larger than in Fig.~\ref{fig:dNdw_reproduction_100TeV}. We see a suppression of the spectrum at larger $\omega$ resembling the GLV expectation even though we have $\omega_c \gg E$. This suppression is visible  already in the distribution for the virtual gluons, before the LPM evolution, where energy and momentum conservation constraints and the $1-x$ cofactor in the GB formula suppress the radiation at large $\omega$ with respect to a  $1/\omega$ spectrum.  At low $\omega$, the agreement with the BDMPS-Z spectrum extends down to $\omega$ as small as $\omega \approx 2\,{\rm GeV} $, what should be taken with a grain of salt (see discussion of Fig.~\ref{fig:dNdkT_reproduction} ) .

The average number of radiated gluons per jet is around 6.5, while the average number of virtual gluons from the initial Gunion-Bertsch seed is around 112, which means that around $6 \%$ of all virtual gluons are realized at the end of their formation time. 

\begin{figure}[H]
\centering
\begin{minipage}[t]{.49\textwidth}
\centering
\includegraphics[width=\linewidth]{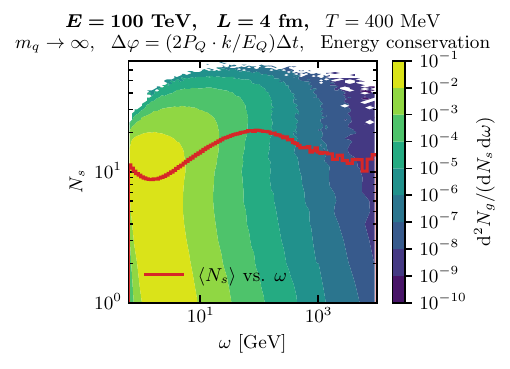}
\caption{Double differential spectrum of gluons from our Monte Carlo approach for $E = 100$ TeV and $L = 4$ fm. }
\label{fig:dN2_reproduction_100TeV}
\end{minipage}\hfill%
\begin{minipage}[t]{.49\textwidth}
\centering
\includegraphics[width=\linewidth]{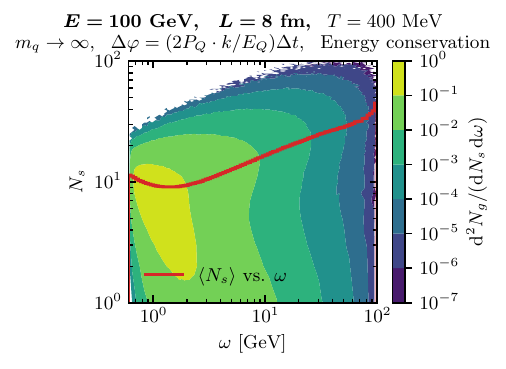}
\caption{Same as Fig.\ref{fig:dN2_reproduction_100TeV} for $E = 100$ GeV and $L = 8$ fm. }
\label{fig:dN2_reproduction}
\end{minipage}
\end{figure}

Figs.~\ref{fig:dN2_reproduction_100TeV} and \ref{fig:dN2_reproduction} show the double differential spectrum $\mathrm{d}N^2/(\mathrm{d}N_s \, \mathrm{d}\omega)$ for $L = 4$ fm, $E = 100$ TeV and $L = 8$ fm, $E = 100$ GeV, respectively. The number of elastic scatterings $N_s$ suffered by the gluon is proportional to the formation time, $N_s \sim t_f / \lambda_{\rm el }^g$. The formation time in the LPM (BDMPS-Z) is expected to behave as $t_f \sim \sqrt{\omega}$. The red lines on Figs.~\ref{fig:dN2_reproduction_100TeV} and \ref{fig:dN2_reproduction} indicate the average $N_s$ as a function of the gluon energy $\omega$, and approximates a linear dependence for intermediate energies as expected on a log-log plot. For small $\omega$, the early decrease of $\langle  N_s\rangle$ with $\omega$ reflects the opening of the accessible phase space with $\omega$ which makes these collisions more efficient in ``dephasing'' the virtual gluon. For $\omega>\omega_c$ we see that the average $N_s$ decreases. This is a clear sign that this $\omega$-range corresponds to another regime, the GLV regime. In this limit there is (at least) one hard scattering which can be accompanied by soft collisions.
Therefore, in our MC, one does not strictly reach $N_s=1$ even for $\omega \gg \omega_c$. 
One should notice the large fluctuations around the average $\langle  N_s\rangle$, which correspond to fluctuations in the formation length.

In Fig.~\ref{fig:dNdkT_reproduction}, we compare the $1/k_\perp dN_g/k_\perp$ spectrum from our MC simulation with the equivalent spectrum calculated in the BDMPS-Z harmonic approximation \cite{Wiedemann:2000tf}, restricting the $k_z$ range to $k_z > k_\perp$ in order to satisfy at best the BDMPS-Z conditions. It is known that the average $\langle k_\perp^2 \rangle $ scales like $Q_s^2= \hat{q} L$  (see the discussion in \cite{Zapp:2011ya} as well as in chapter 3 of \cite{caucal:tel-03081993}) and we indeed observe a turning point of both spectra for such a value, with $1/k_\perp dN_g/k_\perp \propto Q_s/k_\perp^4$ for $k_\perp^2 \gtrsim Q_s$ . Although our MC is in good agreement with the BDMPS-Z calculation in this region, one observes some discrepancy for $k_\perp^2 \lesssim Q_s$. One reason for this discrepancy is that the integrated $dN_g/k_\perp$ spectrum is in fact dominated by low values of $\omega$ and $k_z$ -- as can be seen from Fig. \ref{fig:dN2_kTkz_bdmpsz} --, which do not correspond to the BDMPS-Z conditions. A more realistic comparison with the BDMPS-Z calculation would require to select "large" values of $\omega$ (or $k_z$) and study both spectra for $k_\perp \ll \omega$. We plan to perform such task in a future work, comparing not only to the BDMPS-Z harmonic approximation but also to the full case \cite{Isaksen:2023nlr}, which requires a numerical calculation in its own. Finally, one may wonder why the $dN/d\omega$ spectrum displayed on  Fig.~\ref{fig:dNdw_reproduction} does not reflect, in a way or another, the discrepancy observed on Fig.~\ref{fig:dNdkT_reproduction} around $k_\perp \approx Q_s$, as we have just claimed that $dN/dk_\perp$ is mostly sensitive to regions of the double differential spectrum $d^2N/dk^+ d k_\perp$ located around $k^+ = k_\perp$. The reason is that eq.(\ref{eq:bdmpszLinf}) results from the integration of $d^2N/dk^+ d k_\perp$ over the full $k_\perp$ range, implying regions $k_\perp \gtrsim k^+$ for which the BDMPS-Z condition is not satisfied, while imposing the $k_\perp \lesssim k^+$ condition -- as done to obtain the $k_\perp$ spectrum presented in Fig.\ref{fig:dNdkT_reproduction} -- the integration leads to substantial depletion with respect to eq.~(\ref{eq:bdmpszLinf}) for $\omega \lesssim Q_s$.


\subsection{Testing the role of the prescriptions in the elastic collisions}

\begin{figure}[H]
\centering
\includegraphics[width=0.6\textwidth]{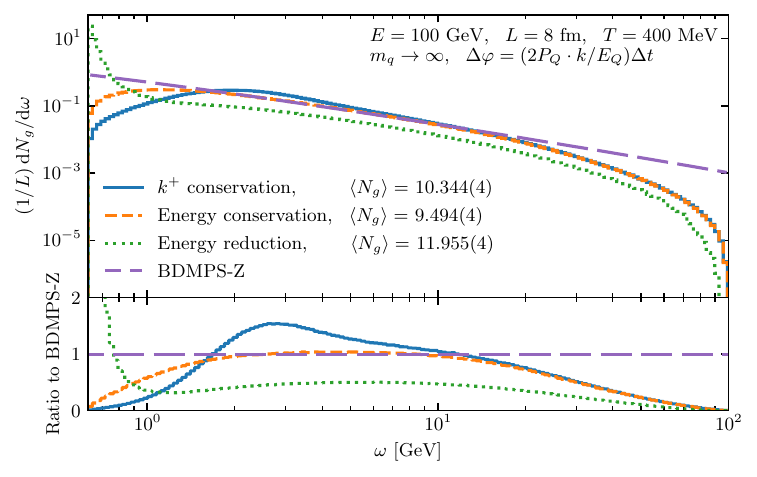}
\caption{Top: Gluon energy spectrum from our Monte Carlo approach for various prescriptions of the elastic scattering. To match the BDMPS-Z conditions, we assume infinitely massive scattering centers. The gluon phase accumulation is calculated according to $\Delta \varphi = \left ( 2 P_Q \cdot k / E_Q \right ) \Delta t$. The average number of radiated gluons per jet, $\langle N_g \rangle$, is listed. The theoretical prediction for BDMPS-Z in eq.~(\ref{eq:bdmpsz}) is shown. Bottom: Ratio of our results and the BDMPS-Z calculation.}
\label{fig:dNdw_elprescription}
\end{figure}

Fig.~\ref{fig:dNdw_elprescription} shows the energy spectrum for a medium of size $L=8$ fm and for different prescriptions of the elastic scattering of the virtual gluon as compared to the BDMPS-Z prediction. The result for the $k^+$ conservation is given as a solid blue line, that for energy conservation by a dashed orange line, and that for energy reduction by a dotted green line. The BDMPS-Z result, presented as a long-dashed purple line, falls off as $\omega^{-3/2}$. The bottom panel shows the ratio of our results with respect to the BDMPS-Z calculations.
We see that all three prescriptions follow  fairly the BDMPS-Z approach for intermediate $\omega$ energies, both in the functional form and in the overall normalization. The $k^{+}$ conservation and energy conservation prescriptions yield similar results for large $\omega$ but differ for smaller energies.\footnote{$k^+$ conservation implies that elastic collisions increase the gluon energy as $\omega=\frac{1}{2}(k^+ + m_{g,\perp}^2/k^+)$, leading to a drift of low-$\omega$ virtual gluons to larger $\omega$ values, as observed on Fig.~\ref{fig:dNdw_elprescription}.} The energy reduction prescription underpredicts the BDMPS-Z results for a large $\omega$ range. The BDMPS-Z curve is given by eq.~(\ref{eq:bdmpsz}) with a correct absolute normalization. The deviations from the BDMPS-Z predictions  at large and small $\omega$ are mainly due to energy conservation and the finite gluon mass. 

\begin{figure}[H]
\centering
\includegraphics[width=0.6\textwidth]{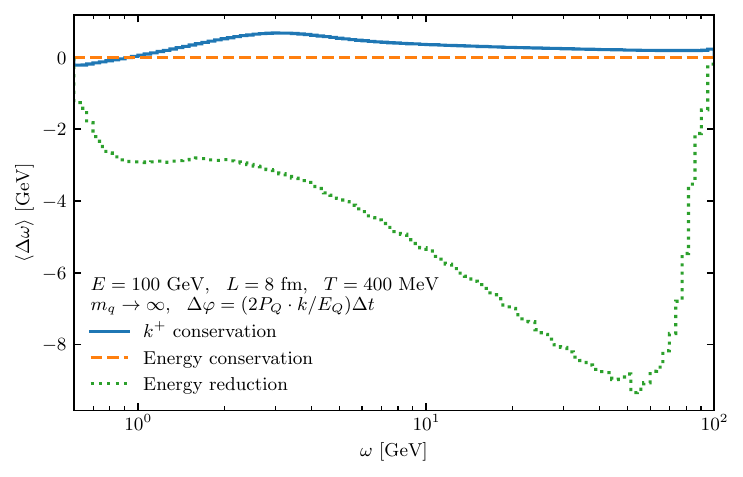}
\caption{The average energy difference of the virtual gluons, $\langle \Delta \omega \rangle$ for $\Delta \omega \equiv \omega - \omega^{\text{0}}$,  where $\omega^0$ is the gluon energy after the inelastic collisions (at beginning of LPM evolution) and $\omega$ is the energy at realization (i.e.\ end of LPM evolution), as a function of $\omega$ for the three different prescriptions of the elastic scattering. The color coding is the same as in Fig.~\ref{fig:dNdw_elprescription}. }
\label{fig:averageDeltaE}
\end{figure}

To understand better the three approaches to treat the elastic gluon scattering  we investigate them further by considering the average energy transfer to the QGP medium, $\langle \Delta \omega \rangle$ for $\Delta \omega \equiv \omega - \omega^0$ where $\omega^0$ is the energy of the virtual gluon when it is produced in the inelastic collision and $\omega$ that when the gluon is formed. The result is shown,  as a function of $\omega$, in Fig.~\ref{fig:averageDeltaE}. If the virtual gluon looses energy during its LPM evolution we find $\Delta \omega < 0$. For the energy reduction case, elastic scatterings of the virtual gluons during the LPM evolution substantially decrease the energy of produced gluons.  As observed in Fig.~\ref{fig:dNdw_elprescription}, this results in a shift of the energy spectrum to lower $\omega$ values, leading, in particular, to an excess of gluons at low $\omega$. For $k^{+}$ conservation we observe a slight increase of the gluon energy whereas in the  energy conservation cases the gluon energy remains of course unchanged. 

\begin{figure}[H]
\centering
\includegraphics[width=0.6\textwidth]{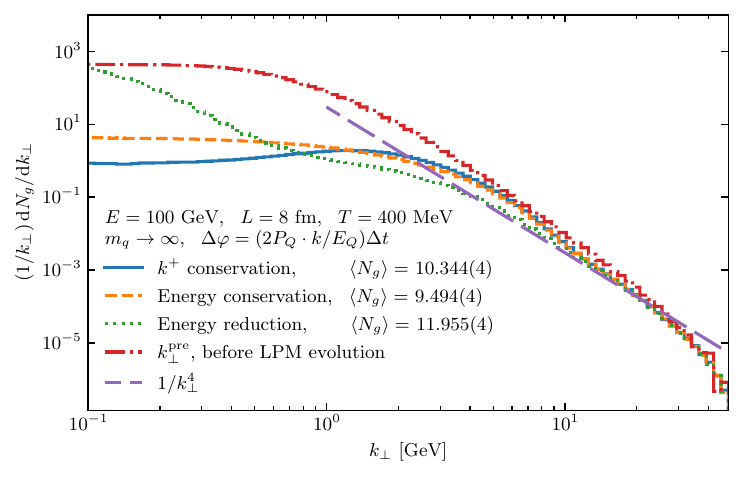}
\caption{The gluon $k_{\perp}$ spectrum in the kinematic limit of BDMPS-Z. The color coding is the same as in Fig.~\ref{fig:dNdw_elprescription}. }
\label{fig:dNdkT_prescription}
\end{figure}

Fig.~\ref{fig:dNdkT_prescription} shows the transverse momentum spectrum of the produced gluons for the different prescriptions of the elastic scattering of the virtual gluons. The color coding is the same as in Fig.~\ref{fig:dNdw_elprescription}. In addition, we add the distribution of the transverse gluon momentum from the initial GB seed. We see, as expected from eq.~(\ref{eq:tfsimple}), a large reduction with regard to the GB seed for small $k_{\perp}$, indicating that these gluons have a large formation time and are produced after  multiple scatterings --- and thus according to the BDMPS-Z scheme.

We see also a significant difference between the different prescriptions for low $k_{\perp}$. 
This is not surprising because the different prescriptions of the energy momentum conservation in the elastic scattering of virtual gluons yield different values of $k_{\perp}$ if $k_{\perp}$ is small.  In particular we observe an excess of soft gluons in the energy-reduction case whereas the $k^+$ conservation prescription shifts the gluons to larger values of $k_{\perp}$ and depopulates the low $k_{\perp}$ region. Only when $k_{\perp}$ is of the order $1$ GeV or larger, the three prescriptions yield similar results, a spectrum falling off approximately like $1/k_{\perp}^4$.  Such excess at low $k_\perp$ in the energy reduction case can be understood as follows: Most of these virtual gluons are in fact soft, implying that they do not possess enough energy to allow for the recoil of the medium parton (for the preselected $l_T$). The corresponding collisions are therefore vetoed, implying that the virtual gluon effectively propagates through a more dilute medium as for the energy conservation case, implying that they are less impacted by the LPM effect.

For $k_{\perp} > 1$ GeV, the $k_{\perp}$ spectrum of radiated gluons gradually merges with the one for the virtual gluons, which indicates that very few elastic collisions (if at all) are necessary to form these gluons, because the initial virtual gluons are already sampled with a large $k_{\operp}$, and therefore $N_s$ is small. This is the case for  $k_{\perp}^2 \gtrsim \omega \varphi_c / L$. This confirms that our model is able to reproduce the shape of the large-$k_T$ (GLV) regime.

\begin{figure}[H]
    \centering
    \includegraphics[width=0.6\textwidth]{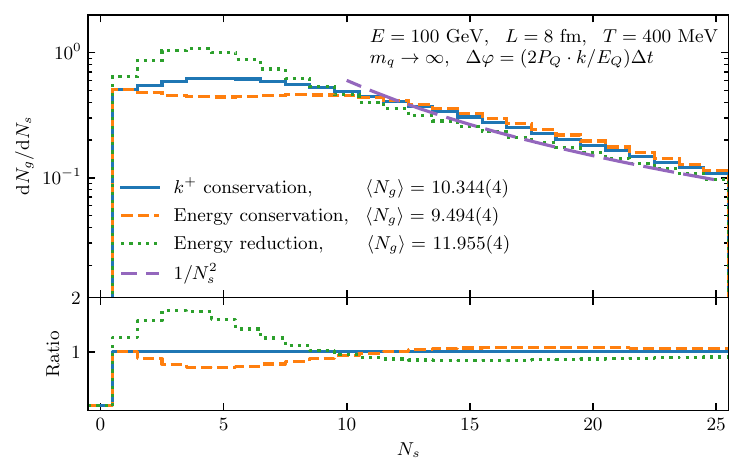}
    \caption{Distribution of the number of elastic collisions, which a gluon has to suffer in our Monte Carlo approach before it is formed. The color coding is the same as in Fig.~\ref{fig:dNdw_elprescription}. }
    \label{fig:dNdNs_prescription}
\end{figure}

The distribution of the number of elastic collisions, $N_s$, which the virtual gluon has to suffer before it gets formed, is shown in Fig.~\ref{fig:dNdNs_prescription}, as well for the three different prescriptions in the elastic scatterings. Here, all approaches give very similar results for $N_s > 10$, although the distribution for the energy-reduction option appears to be slightly depleted.
However, for values of $N_s$ around 5, typically associated with small $\omega$ values through $N_s \sim t_f \sim \sqrt{\omega}$, the energy reduction prescription gives a larger contribution. This has two origins. On the one side, the energy loss in the individual elastic collisions (see Fig.~\ref{fig:averageDeltaE}) drives some gluons to the soft sector for which the formation time (and so the associated $N_s$) is smaller.  Because for finite gluon masses the formation time is $\propto\omega/(m_g^2+k_{\operp}^2)$, the shift of the gluons to smaller $k_{\operp}$ values, seen in Fig.~\ref{fig:dNdkT_prescription}, does not lower the formation time considerably. For the ``energy reduction'' scenario we observe therefore an increase of the production for low $N_S$. This effect is reinforced by a kinematic effect: The smaller the value of $\omega$ the larger is the probability that elastic collisions of the virtual gluons are kinematically not possible. Then the collisions do not take place -- as already described above -- and $N_s$ is not increased but the phase accumulation continues. This also leads to an enhancement of the entry at $N_s=1$, as seen in Fig.~\ref{fig:dNdNs_prescription}.

\subsection{Testing the role of the phase accumulation}

\begin{figure}[H]
\centering
\begin{minipage}[t]{.49\textwidth}
\centering
\includegraphics[width=\linewidth]{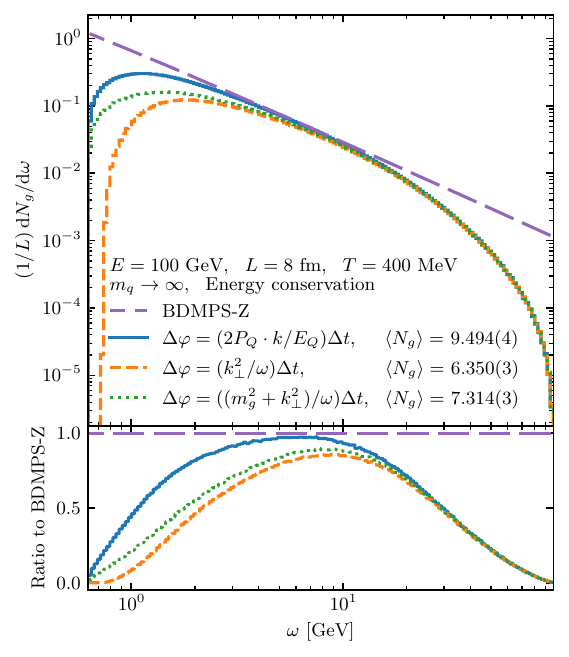}
\caption{Top: Gluon energy spectrum from the Monte Carlo as compared to the BDMPS-Z expectation, eq.~(\ref{eq:bdmpsz}), for different phase increments of the gluon radiation amplitudes. The results for the phase increment of $\Delta \varphi= \frac{2P\cdot k'}{E}\cdot \Delta t \ $ is shown as a blue solid line, that of a phase increment of $\Delta \varphi = \frac{k_{\perp}^2}{\omega}\cdot \Delta t \ $, corresponding to a vanishing gluon mass, as a dashed orange line and that of a phase increment of $\Delta \varphi = \frac{m_g^2+k_{\perp}^2}{\omega}\cdot \Delta t \ $, corresponding to the limit $E\to\infty$, as a green dotted line. The average number of radiated gluons per jet, $\langle N_g \rangle$, is listed. Bottom: Ratio of the Monte Carlo results and the BDMPS-Z result. }
\label{fig:dNdw_phaseaccum}
\end{minipage}\hfill%
\begin{minipage}[t]{.49\textwidth}
\centering
\includegraphics[width=\linewidth]{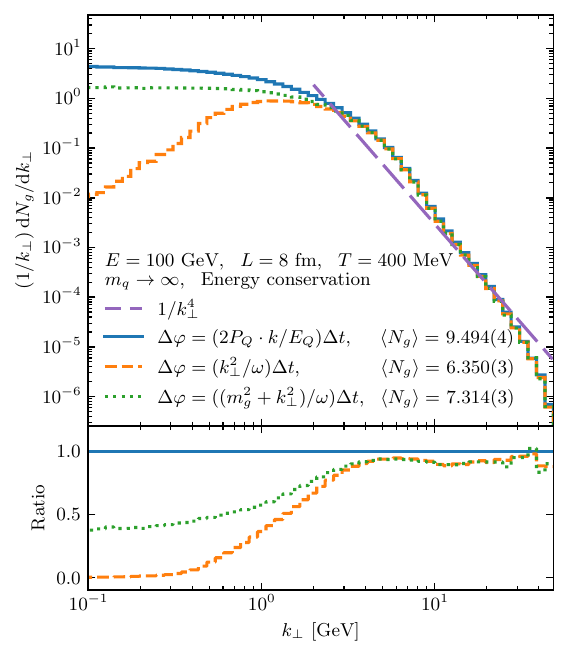}
\caption{Top: The gluon $k_{\perp}$ spectrum from the Monte Carlo calculation for different phase increments of the gluon radiation amplitudes. The color coding is the same as in Fig.~\ref{fig:dNdw_phaseaccum}. The average number of radiated gluons per jet, $\langle N_g \rangle$, is listed. Bottom: Ratio of the results for the various prescriptions and the result for the phase increment of $\Delta \varphi= \frac{2P\cdot k'}{E}\cdot \Delta t \ $. }
\label{fig:dNdkT_phaseaccum}
\end{minipage}
\end{figure}

\begin{figure}[H]
    \centering
    \includegraphics[width=0.6\textwidth]{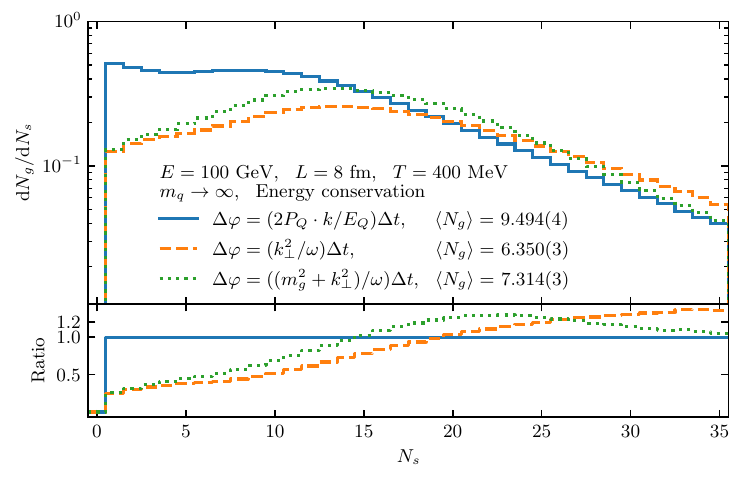}
    \caption{Distribution of the number of elastic collisions, which a virtual gluon suffers before it becomes formed. The color coding is the same as in Fig.~\ref{fig:dNdw_phaseaccum}. }
    \label{fig:dNdNs_phaseaccum}
\end{figure}

Another quantity, which is approximated in the BDMPS-Z conditions, is the phase accumulation of the virtual gluon due to elastic collisions. Figs.~\ref{fig:dNdw_phaseaccum}, \ref{fig:dNdkT_phaseaccum}, and \ref{fig:dNdNs_phaseaccum} show the influence of different forms of the phase accumulation of the gluon emission amplitudes on the gluon observables. The form of the phase accumulation cannot be inferred from more fundamental approaches.  As discussed in section~\ref{sec:lowQ} the virtual gluon will be considered as a formed particle once its phase passes the critical value of $\varphi_c = 6$. We display the results for three different descriptions of this phase accumulation. The results for the phase increment of $\Delta \varphi= \frac{2P\cdot k'}{E}\cdot \Delta t \ $ are shown as a blue solid line, that of a phase increment of $\Delta \varphi = \frac{k_\perp^2}{\omega}\cdot \Delta t \ $, corresponding to a vanishing gluon mass, as a dashed orange line and that of a phase increment of $\Delta \varphi = \frac{m_g^2+k_\perp^2}{\omega}\cdot \Delta t \ $, corresponding to the limit $E\to\infty$, as a green dotted line.

Fig.\ref{fig:dNdw_phaseaccum} shows the influence of the choice of the phase accumulation on the energy spectrum of the produced gluons. For large energies, $\omega > 10$ GeV, all curves show the same functional form because in the high $\omega$ limit all the prescriptions reduce to the same limit of $k_\perp^2/\omega$. Deviations between the different phase accumulations show up only at small $\omega$. The result for the phase accumulation prescription $\Delta \varphi = \frac{k_\perp^2}{\omega}\cdot \Delta t \ $ gives a depleted spectrum for low $\omega$ as compared to the other two prescriptions.

In Fig.~\ref{fig:dNdkT_phaseaccum}  we display the $k_\perp$ distribution of the formed gluons using the  same color coding as in Fig.~\ref{fig:dNdw_phaseaccum}. The $k_\perp$ spectra are very similar for large $k_\perp$, i.e.\ in the region where the energy loss follows the BDMPS-Z result but show substantial differences at lower $k_\perp$. The results are compared to the functional form of $1/k_\perp^4$, the analytical dependence of BDMPS-Z for large $k_\perp$. 

For both, the gluon $\omega$ and the $k_\perp$ spectra,  the high energy limit $\Delta \varphi = \frac{m_g^2+k_\perp^2}{\omega}\cdot \Delta t \ $ presents a reasonable approximation of the more general formula $\Delta \varphi= \frac{2P\cdot k'}{E}\cdot \Delta t \ $.

The distribution of the number of elastic collisions, $N_s$, which a virtual gluon suffers before being formed, is shown in Fig.~\ref{fig:dNdNs_phaseaccum}. For large $N_s > 10$ we see almost no difference between the different prescriptions of the phase accumulation, while for small $N_s$ we see a substantial enhancement for $\Delta \varphi= \frac{2P\cdot k'}{E}\cdot \Delta t \ $ with respect to the other two prescriptions. Hence, we conclude that the choice of the phase accumulation criteria has an important impact on the number of radiated gluons when considering small systems where $N_s \sim t_f$ is small. The fully boost invariant phase increase $\propto P\cdot k'$ results in faster phase accumulation for gluons, which are not colinear with the jet. This yields a  faster formation and hence a more abundant yield than expected from the asymptotic 
($E\to \infty$) approximation
$\Delta \varphi= \frac{m_g^2+k_{\operp}^2}{\omega}\cdot \Delta t \ $
.

\subsection{Testing the role of \texorpdfstring{$m_{q,g}$}{mqg}}

Another parameter of the modeling, which varies in the different approaches, is the mass of the QGP constituents with which the jet parton scatters and which influences the various rates. For the radiative process, this mass is denoted as $m_{q}$ (with the 2 options $m_q=0$ or $m_{q}=\infty$). It influences the value of the radiative rate $R_{\rm rad}$ through the thermal distribution of these QGP constituents, see appendices~\ref{app:gbmq0sampling}  and \ref{app:gbmqinfsampling},  as well as through the kinematic variables at the end of the emission of the virtual gluon. For the elastic scattering, the 2 choices we will consider here are the ``energy conservation'' and the ``energy reduction'' options, irrespective of which $m_q$ is taken for the inelastic collisions. By choosing the ``energy conservation'' option for the subsequent elastic scatterings of the gluon during its formation, we assume implicitly infinitely massive QGP constituents, while in the ``energy reduction'' option, the scattering is evaluated with $m_{q(g)}^{\rm therm}$ as the mass of the QGP constituents.

\begin{figure}[H]
\centering
\includegraphics[width=0.6\textwidth]{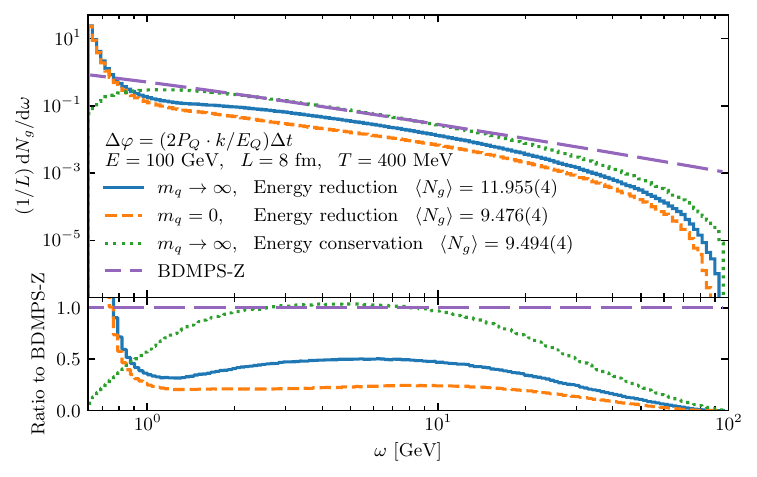}
\caption{Radiation spectrum of gluons from the Monte Carlo approach. The realistic radiation case with energy reduction in the elastic scattering and considering vanishing medium quark masses ($m_{q,g} = 0$) (orange dashed line) is compared to that for the BDMPS-Z condition ($m_{q,g} = \infty$), with both energy conservation (green dotted line) and energy reduction (blue full line) prescriptions. The average number of radiated gluons per jet, $\langle N_g \rangle$, is listed. }
\label{fig:dNdw_mq}
\end{figure}

Fig.~\ref{fig:dNdw_mq} shows the Monte Carlo results for several scenarios. The BDMPS-Z condition $m_{q/g} = \infty\;\cup$ ``energy conservation'' is compared with the more realistic $m_{q/g} = 0\;\cup$ ``energy reduction'' case and with the hybrid choice $m_{q/g} = +\infty\;\cup$ ``energy reduction''. We see that the spectrum falls off roughly as $\omega^{-3/2}$ for all combinations.  One observes that the choice of $m_{q/g}$ has little influence on the form of the energy loss spectrum above $\omega = 1$ GeV.  At low $\omega$ one sees an enhancement for the ``energy reduction'' option, observed already in Fig.~\ref{fig:dNdw_elprescription}.

\begin{figure}[H]
\centering
\begin{minipage}[t]{.49\textwidth}
        \centering
        \includegraphics[width=\linewidth]{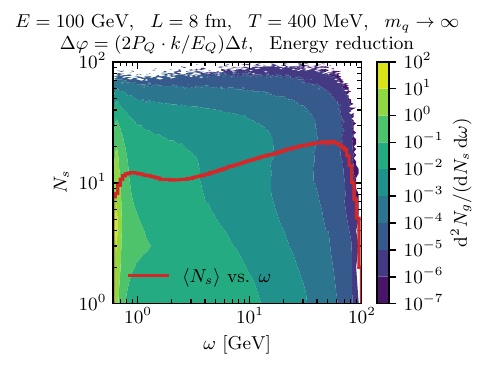}
        \caption{Double differential gluon spectrum in $\omega$ and $N_s$ from our Monte Carlo approach for $m_q\to +\infty$ in the GB radiation and for the energy reduction prescription in the elastic scattering. }
        \label{fig:dN2_Ns_w_mqinft_ered}
    \end{minipage}\hfill%
    \begin{minipage}[t]{.49\textwidth}
        \centering
        \includegraphics[width=\linewidth]{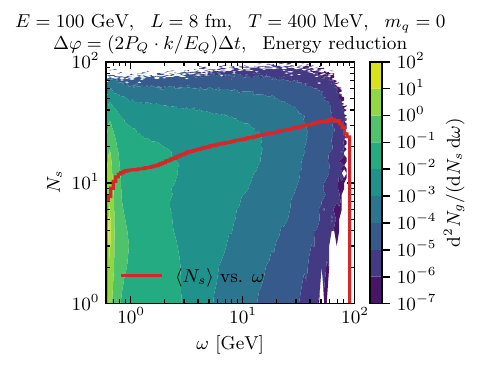}
        \caption{Same as Fig.~\ref{fig:dN2_Ns_w_mqinft_ered} for $m_q=0$. }
        \label{fig:dN2_Ns_w_mq0}
    \end{minipage}
\end{figure}

Such an independence with respect to the choice of $m_q$ is surprising because it is not observed for the GB radiation seed, shown in Fig.~\ref{fig:dNdomega-beforeLPM}, where a larger yield is observed for $m_q=0$. A finer analysis reveals that the average $\langle N_s\rangle$, needed to reach the critical $\varphi_c$, is for $\omega \gtrsim 1$~GeV larger for $m_q=0$ than for $m_q \to \infty$, as can be seen on Figs~\ref{fig:dN2_Ns_w_mqinft_ered} and \ref{fig:dN2_Ns_w_mq0}, leading to a larger LPM suppression. This originates from the 
transverse momentum given to the virtual gluon through the GB seed radiation, which is smaller for the $m_q=0$ case (see Fig.~\ref{fig:dNdkT-beforeLPM}) and hence results in a smaller phase increase at early times.

\begin{figure}[H]
\centering
\includegraphics[width=0.6\textwidth]{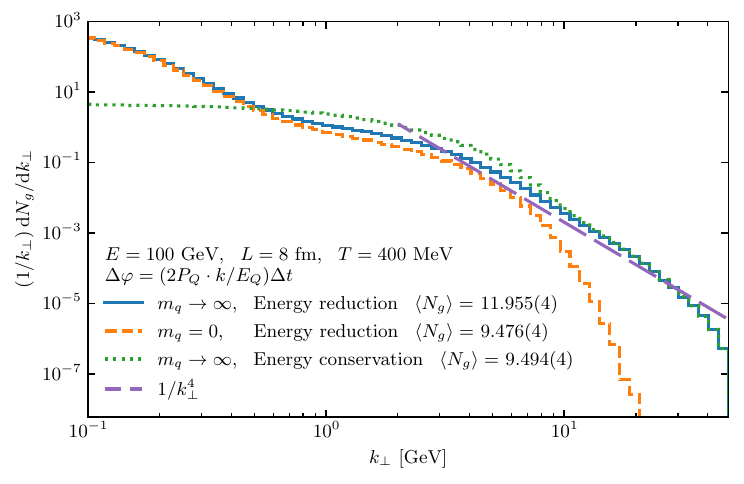}
\caption{Same as Fig.~\ref{fig:dNdkT_prescription} for 2 choices of $m_q$ in the GB radiative process and 2 prescriptions for the energy-momentum conservation in the elastic gluon collisions. }
\label{fig:dNdkT_mq}
\end{figure}

The gluon $k_\perp$-spectra, corresponding to the 3 prescriptions chosen in Fig.~\ref{fig:dNdw_mq}, are shown in Fig.~\ref{fig:dNdkT_mq}. For $m_q = 0$ we see a reduced yield at large $k_\perp$ as compared to $m_q = \infty$, which is compensated at low $k_\perp$, where $m_q = 0$ shows a larger number of radiated gluons. The difference observed in the spectra for $k_\perp\gtrsim 5$~GeV between the $m_q=0$ and the $m_q \to \infty$ scenarios is simply due to the available phase space, as has been pointed out in the discussion of Fig.~\ref{fig:dNdkT-beforeLPM}.
For $m_q\to \infty$, the natural prescription for the elastic scattering (energy conservation) leads to the extra reduction at small $k_\perp$ as compared to the ``energy reduction'' prescription, as already seen in Fig.~\ref{fig:dNdkT_prescription}.

\begin{figure}[H]
    \centering
    \begin{minipage}[t]{.49\textwidth}
        \centering
        \includegraphics[width=\linewidth]{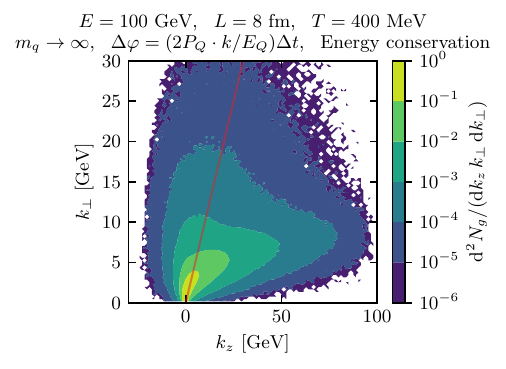}
        \caption{Double differential spectrum in transverse momentum $k_{\perp}$ and longitudinal momentum $k_z$ of gluons from the Monte Carlo approach for the kinematical BDMPS-Z conditions. The thin line corresponds to $k_\perp=k_z$, i.e.\ a jet radius $R=1$. } 
        \label{fig:dN2_kTkz_bdmpsz}
    \end{minipage}\hfill%
    \begin{minipage}[t]{.49\textwidth}
        \centering
        \includegraphics[width=\linewidth]{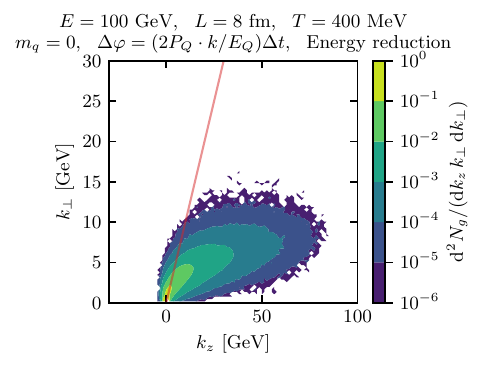}
        \caption{Double differential spectrum in transverse momentum $k_{\perp}$ and longitudinal momentum $k_z$ of gluons from the Monte Carlo approach for the most realistic conditions ($m_q=0$ and energy reduction). }
        \label{fig:dN2_kTkz_realistic}
    \end{minipage}
\end{figure}

Figs.~\ref{fig:dN2_kTkz_bdmpsz} and \ref{fig:dN2_kTkz_realistic} show the momentum distribution of the gluons produced in the $m_{q/g}=\infty$ $\cup$ ``energy conservation'' and in the  $m_{q/g}=0$ $\cup$ ``energy reduction'' scenario. They are rather different. For $m_{q/g} = 0$ one sees clearly the limits due to the available phase space.  The BDMPS-Z conditions ($m_q \to \infty$) leads even to a non-negligible backward component, which is more substantial than the one found after the GB seed (see Fig.~\ref{fig:dN2_kTkz_gb_mqinft}). This is a consequence of the elastic collisions, which the virtual gluon suffers until it gets on-shell. These figures show that the phase space for gluons is very reduced, in longitudinal as well as in transverse direction,
if we evaluate the radiation employing physical masses for the QGP partons.

\begin{figure}[H]
\centering
\begin{minipage}[t]{.49\textwidth}
\centering
\includegraphics[width=\linewidth]{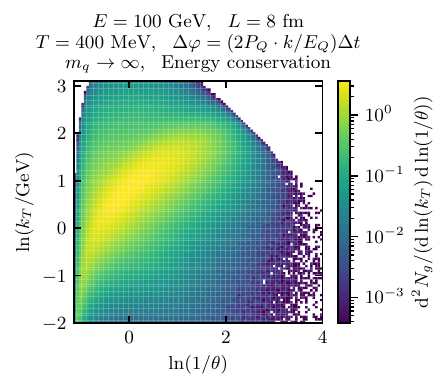}
\caption{Primary Lund jet plane for the low virtuality shower for the kinematical BDMPS-Z conditions.}

\label{fig:lowQ_bdmpsz_lund_plane}
\end{minipage}\hfill%
\begin{minipage}[t]{.49\textwidth}
\centering
\includegraphics[width=\linewidth]{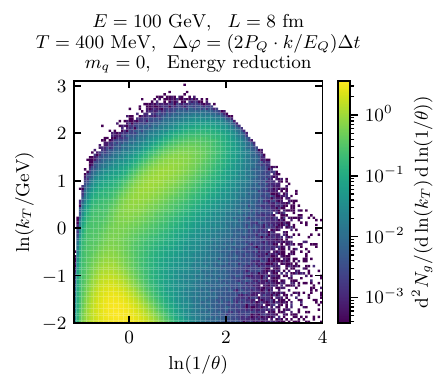}
\caption{Same as Fig~\ref{fig:lowQ_bdmpsz_lund_plane} but for the most realistic conditions. }
\label{fig:lowQ_realistic_lund_plane}
\end{minipage}
\end{figure}

In Figs. \ref{fig:lowQ_bdmpsz_lund_plane} and 
\ref{fig:lowQ_realistic_lund_plane}, we show the primary Lund plane for the low-virtuality component of our model, for the kinematical conditions of BDMPS-Z as well as for more realistic kinematical conditions. In both cases, one finds a clear structure corresponding to the growth of $k_\perp$ with the colinearity $1/\theta$. For increasing $\omega$, the BDMPS-Z radiation indeed satisfies the heuristic laws $\langle k_\perp^2  \rangle \sim \hat{q} t_f \sim \sqrt{\omega \hat{q}}$ and $1/\langle \theta^2 \rangle \sim \frac{\omega^2}{\langle k_\perp^2 \rangle }\sim \sqrt{\omega^3/\hat{q}}$, which both grow with $\omega$. Passing from BDMPS-Z to more realistic conditions, one notices a reduction of the radiation at large $k_\perp$, already seen by comparing Figs. \ref{fig:dN2_kTkz_bdmpsz} and\ref{fig:dN2_kTkz_realistic}, as well as an extra ultra soft radiation at large angle.  In an approach, which describes the expansion of the quark gluon plasma (QGP) by hydrodynamics, this ultra soft radiation would be absorbed by the  QGP.

\subsection{Testing the role of the path length \texorpdfstring{$L$}{L}}

\begin{figure}[H]
\centering
\begin{minipage}[t]{.49\textwidth}
\centering
\includegraphics[width=\linewidth]{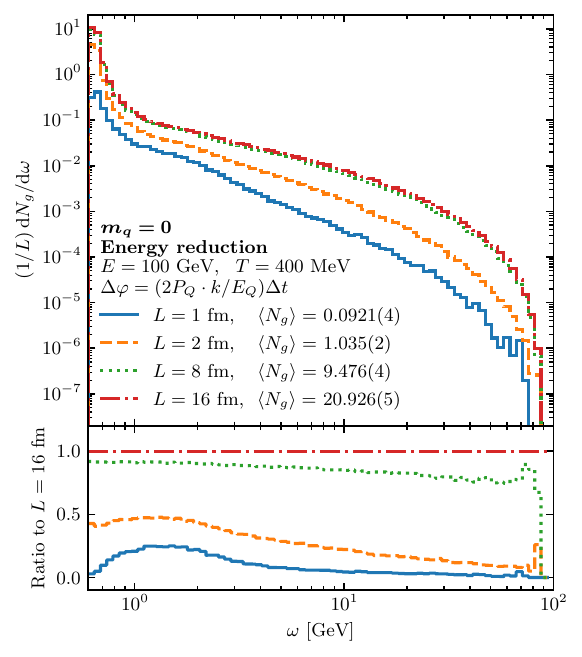}
\caption{Top: Radiation spectrum of gluons from the Monte Carlo for the realistic conditions for different path lengths $L$. The average number of radiated gluons per jet, $\langle N_g \rangle$, is listed. Bottom: Ratio of the Monte Carlo results to that for $L = 16$ fm. }
\label{fig:dNdw_pathlength_1}
\end{minipage}\hfill%
\begin{minipage}[t]{.49\textwidth}
\centering
\includegraphics[width=\linewidth]{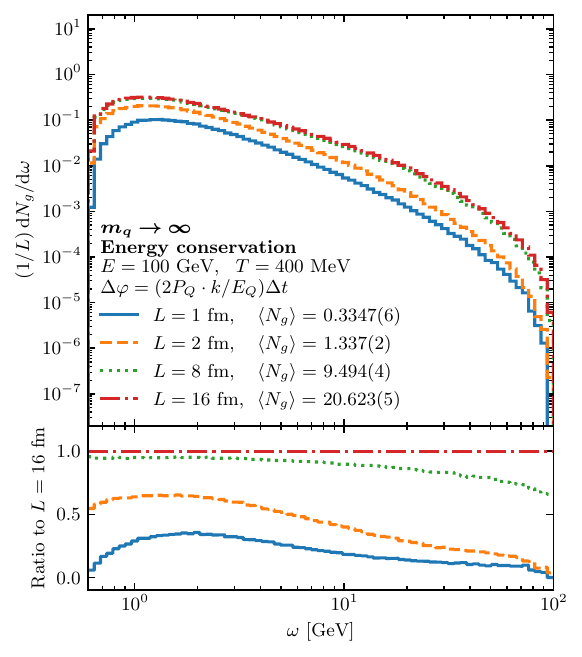}
\caption{Same as Fig.~\ref{fig:dNdw_pathlength_1} for the kinematical conditions according to the BDMPS-Z approach. }
\label{fig:dNdw_pathlength_2}
\end{minipage}
\end{figure}

Figs.~\ref{fig:dNdw_pathlength_1} and \ref{fig:dNdw_pathlength_2} show the gluon energy spectrum for
different path lengths in the QGP medium. Fig.~\ref{fig:dNdw_pathlength_1} displays this spectrum for what we consider to be the most realistic implementation of our model (energy reduction and $m_q = 0$). Fig.~\ref{fig:dNdw_pathlength_2} shows the results for the BDMPS-Z conditions (energy conservation and $m_q \to \infty$). The medium path lengths varies from very small systems ($L = 1$ and $2$ fm, with $\omega_c(L=1\text{ fm})\approx 6.25$~GeV and $\omega_c(L=2\text{ fm})\approx 25$~GeV) to very large systems ($L = 8$ and $16$ fm, for which $\omega_c(L)>E$), to highlight the scaling of the radiation spectrum with respect to $L$.

We see, first of all, a large difference between the radiation spectra for the both settings. The form of the spectra for low and high $\omega$ values is rather different and at intermediate $\omega$ under the BDMPS condition roughly $10$ times more gluons are emitted than in the realistic case. This is even overcompensated by an enhanced emission of low energy gluons ($\omega \ll 1 $ GeV) for the realistic case, as one can see from the average number of emitted gluons, which is in the realistic case about twice as large as for the BDMPS-Z condition. For a large path length the BDMPS-Z expectation, given by eq.~(\ref{eq:bdmpszLinf}) as $\mathrm{d}N/\mathrm{d}\omega\propto L$, should dominate the $\omega$-spectrum. This is indeed observed. Our MC results for large path lengths, $L = 8$ and $16$ fm, reproduce the expected linear scaling for both settings. For a smaller path length, a reduction of the spectrum per unit length is observed, with is more pronounced for larger $\omega$ and smaller $L$, in agreement with the criteria resulting from the comparison between the formation time and the path length. We note that such a reduction is also seen at low $\omega$. This is a confirmation that this region is also impacted by coherence effects for the parameter chosen to describe the QGP ($T=400$ MeV) and by its coupling to the jet ($\alpha_s=0.4$). This is true for both settings although slightly more pronounced for the realistic implementation.

\subsection{Testing the role of gluon rescatterings and reemissions}

\begin{figure}[H]
    \centering
    \begin{minipage}[t]{.49\textwidth}
        \centering
        \includegraphics[width=\linewidth]{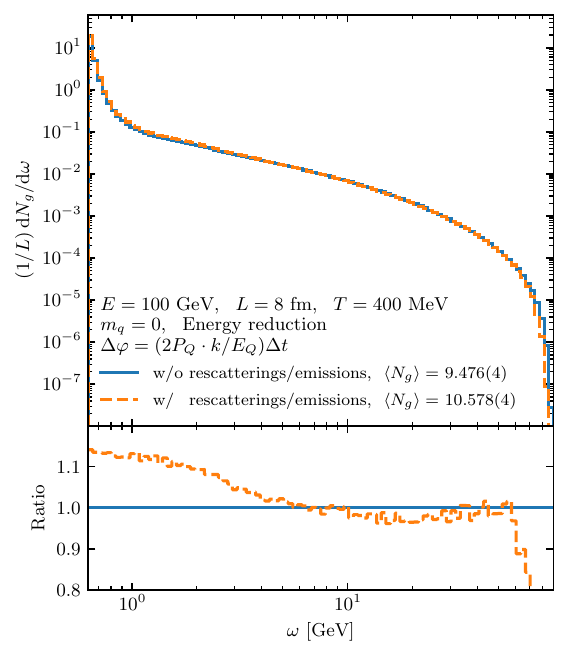}
        \caption{Top: Gluon energy spectrum from the Monte Carlo calculation for the realistic case, taking into account elastic and inelastic scattering of formed gluons. The average number of radiated gluons per jet, $\langle N_g \rangle$, is listed. Bottom: Ratio of the Monte Carlo results. }
        \label{fig:dNdw_rescattering}
    \end{minipage}\hfill%
    \begin{minipage}[t]{.49\textwidth}
        \centering
        \includegraphics[width=\linewidth]{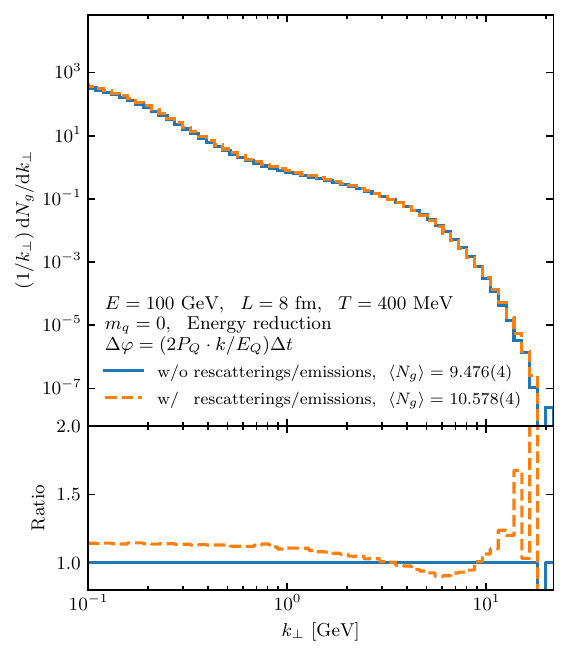}
        \caption{Top: Gluon $k_{\perp}$ spectrum from the Monte Carlo calculation for the realistic case, taking into account elastic and inelastic scattering of formed gluons. The average number of radiated gluons per jet, $\langle N_g \rangle$, is listed. Bottom: Ratio of the Monte Carlo results. }
        \label{fig:dNdkT_rescattering}
    \end{minipage}
\end{figure}

In order to come as close as possible to the BDMPS-Z condition we performed up to now calculations with the additional assumption that collisions of formed gluons as well as elastic collisions of the jet parton are turned off. These BDMPS-Z conditions are shown in Fig.~\ref{fig:rescatter1}, where only the virtual gluon (blue-line circle) undergoes elastic scatterings with the medium and this exclusively until its formation (or discarding).  No scattering of the formed gluons is allowed. The realistic case is shown in Fig.~\ref{fig:rescatter2}, where the projectile and formed gluons can have elastic collisions with the QGP and where formed gluons can emit virtual gluons themselves. 
Figs.~\ref{fig:dNdw_rescattering} and \ref{fig:dNdkT_rescattering}  show the radiation spectra if we release these BDMPS-Z assumptions. Enabling emissions from gluons naturally increases the overall number of gluons per jet, but evidently only by a small amount, and especially at small $\omega$ and $k_\perp$ values. Only small changes in the shape of the distributions are observed, on the $\sim 10\%$ level, as shown in the bottom part of the figures.

Allowing inelastic gluon scattering leads also to a some broadening of the $k_{\perp}$ distribution for $k_{\perp} > 10$~GeV, as seen on the bottom panel of Fig.~\ref{fig:dNdkT_rescattering}. As a matter of fact, gluons radiated with large $k_{\perp}$ are produced after a short formation time and are thus more likely subject to rescattering along their path length. They also benefit from a larger phase space for this process. 


\subsection{Global comparison of the BDMPS-Z and realistic conditions}

\begin{figure}[H]
    \centering
    \begin{minipage}[t]{.49\textwidth}
        \centering
        \includegraphics[width=\linewidth]{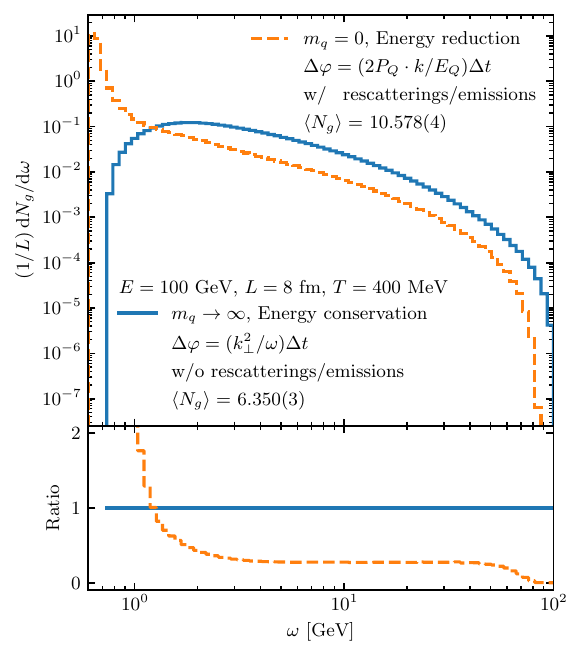}
        \caption{Top: Gluon energy spectrum from the Monte Carlo calculation for the realistic case (shown as a dashed orange line) as compared to the BDMPS-Z conditions (shown as a solid blue line). The average number of radiated gluons per jet, $\langle N_g \rangle$, is listed. Bottom: Ratio of the Monte Carlo results. }
        \label{fig:dNdw_checksum}
    \end{minipage}\hfill%
    \begin{minipage}[t]{.49\textwidth}
        \centering
        \includegraphics[width=\linewidth]{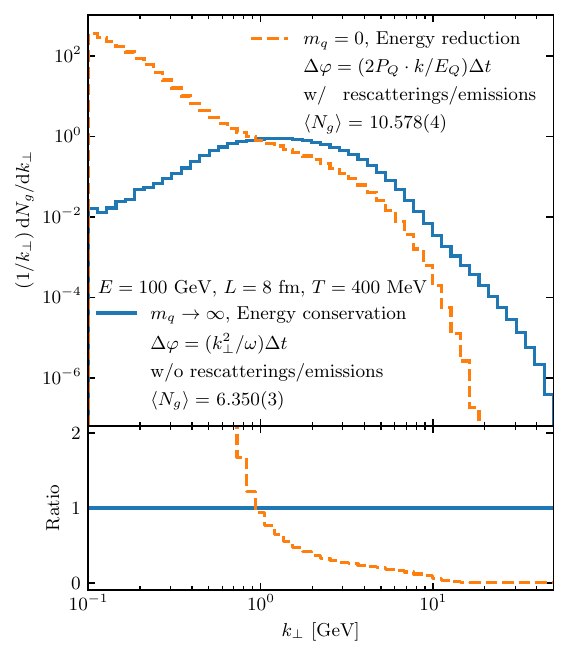}
        \caption{Top: Gluon $k_{\perp}$ spectrum from the Monte Carlo calculation for the realistic case (shown as a dashed orange line) as compared to the BDMPS-Z conditions (shown as a solid blue line). The average number of radiated gluons per jet, $\langle N_g \rangle$, is listed. Bottom: Ratio of the Monte Carlo results. }
        \label{fig:dNdkT_checksum}
    \end{minipage}
\end{figure}
In Figs.~\ref{fig:dNdw_checksum} and \ref{fig:dNdkT_checksum} we compare directly our results for the most realistic calculation [$m_q=0$, energy reduction, scattering of formed gluons and $\Delta \phi= (2P_Q\cdot k/E_Q)\Delta t$] with that which comes closest to the BDMPS-Z scenario [$m_q \to \infty$, energy conservation, no scattering of formed gluons, $\Delta \phi= (k_\perp^2/\omega)\Delta t$]. In the top figure we display the $\omega$ and $k_\perp$ distribution of both approaches, in the bottom figure we present the ratio ``realistic'' over ``BDMPS-Z like''. We see that at intermediate $\omega$ both spectra have about the same functional form, however in the realistic case the yield is reduced by a factor four as compared to the BDMPS-Z scenario. Due to the reasons discussed above the spectra differ by large factors  at low and at high gluon energies. The transverse momentum spectra differ even more, in the functional form as well as in the absolute magnitude. There the conditions for the BDMPS-Z approach are too strong to allow for a realistic description of the processes which are present when a parton traverses a block of matter.


\section{Conclusions} \label{sec:conclusions}

In this study, we have presented a new Monte Carlo model for jet evolution and quenching in a hot and dense medium formed in ultra-relativistic heavy-ion collisions. We simulate jet evolution starting from a single hard parton of energy $E$ with  a maximal virtuality scale $Q=E$ by separating two virtuality areas.
In the initial high virtuality phase, Sudakov form factors are used to sample vacuum-like jet parton splittings until a jet constituent (hard parton) arrives at a low virtuality scale  $Q_0$. The virtuality of the jet parton is modified by interactions with the medium. When the jet parton has lost its  virtuality the low virtuality domain sets in. There the jet partons interact with the surrounding medium via elastic and inelastic collisions. For the latter, the radiation of virtual gluons is first sampled according to a Gunion-Bertsch approach. We have implemented the interference between the gluon emission amplitudes, specific to the BDMPS-Z formalism, in the spirit of \cite{Zapp:2011ya}, by introducing a formation phase for the virtual radiated gluons. We took into account the exact phase space limitation what allowed us to explore the whole ($\omega,k_T$)
space. This has several consequences: a) we can go beyond the eikonal approximation
and explore the full kinematic range encountered in the study of an expanding plasma, b) we can study the recoil given to the QGP parton what allows a feed back
of the kinematical variables of the jet-QGP interaction to the QGP. c) we can explore the gluon energy distribution beyond the range of validity of the BDMPS-Z approach and recover for larger values of $\omega$ the spectral form predicted by GLV and d) we can study the influence of a finite mass of the QGP parton on the gluon spectra.  


We have presented benchmark calculations of jet energy loss in a static finite-size medium in comparison to the BDMPS-Z results. The algorithm reproduces the characteristic $\mathrm{d}N/\mathrm{d}\omega\sim\omega^{-3/2}$ behavior of the spectrum of radiated gluons when taking into account the assumptions made in  the BDMPS-Z approach. For more energetic gluons we reproduce  the GLV limit. For the parameters, which we have used in our studies ($\alpha_s=0.4$, $T=400$ MeV, $m_g^{\rm therm}\approx 600$ MeV), the BH regime is in fact not found at small values of $\omega \approx m_g^{\rm therm}$ where coherence effects are still present. In a more dilute QGP, it should however be naturally recovered in our approach. 

With these parameters, we have performed a comparison of our Monte Carlo results with
the BDMPS-Z approach. The main difference is energy/momentum conservation in the scattering processes,  as well as a thermal gluon mass. Both have not been taken into account in the BDMPS-Z calculations and  modify considerably the form of the spectra for small and large values of $k_\perp$ and $\omega$ as compared to the BDMPS-Z results. 

With these parameters, we have tested  as well the algorithms by varying input quantities, which cannot be obtained from more fundamental approaches. They include the mass of the QGP partons, which scatter inelastically with the incoming parton and elastically with the virtual gluons, as well as the functional form of the phase accumulation of the radiation amplitude. We have also investigated  different descriptions of the elastic scattering of a gluon in the QGP medium, provided in the literature.

We have found that the spectra of gluons emitted with a large transverse momentum $k_\perp$ or with a large energy are not very sensitive to a variation of these quantities, with the exception of the mass of the QGP partons, which influences the large $k_\perp$ region due to phase space constraints. At small $\omega$ and small $k_\perp$  the form and the absolute value of the gluon spectra depend on the choice of these input quantities. We have explained in detail the physical origin of these observations.

We have studied the path length dependence of our Monte Carlo results and found that the results are in agreement with the BDMPS-Z and GLV predictions. Finally we have relaxed all BDMPS-Z assumptions on the kinematic variables and on the included processes and have presented the gluon energy and transverse momentum spectra for what we consider to be the most realistic case. It turns out that the processes, which have been neglected in BDMPS-Z, modify the energy and transverse momentum spectra considerably. Whereas the functional form of the energy spectrum remains unchanged for intermediate energies (and therefore the $\omega^{-3/2}$ dependence of the spectra is also seen in the realistic case with, however, only a quarter of the gluons as compared to  the BDMPS-Z prediction), the form of energy spectra is very different at lower and higher gluon-energies $\omega$.

The form of the transverse momentum spectra 
is very different for all values of $k_\perp$ and therefore the necessary constraints to derive the   
BDMPS- Z formula are too severe to give a realistic description of the physical processes.

The present study presents a benchmark of the Monte Carlo algorithm in a static homogeneous medium. Its construction allows to embed it in realistic 3-dimensional time evolution of the medium, e.g.\ with relativistic viscous fluid dynamics. We foresee this as the next step.  The comparison with experimental data  will also help us to fix the value of some of the least-constrained parameters influencing the soft radiation sector.

In a near future, we also plan to extend the benchmarking of our approach on theoretical results going beyond the small x- approximation and including the exact $k_\perp$ dependence, such as \cite{Blaizot:2012fh,Isaksen:2023nlr}.

\section{Acknowledgements} J.A. acknowledges  valuable discussions with 
Dr. E. Bratkovskaya and  I.  Grishmanovskii.  This work is funded by the European Union’s Horizon 2020 research and innovation program under grant agreement No. 824093 (STRONG-2020).  I.K. acknowledges support by the Czech Science Foundation under project No.~22-25026S.

\appendix


\section{Conventions} \label{app:conventions}

The light cone coordinates are chosen as
\begin{equation}
    p^{+} = E + p_z \hspace{5mm} \text{ and } \hspace{5mm} p^{-} = E - p_z \,,
\end{equation}
implying that $p^2 = p^{+} p^{-} - p^2$, which leads to the light cone metric
\begin{equation}
    g_{\text{LC}} = \begin{pmatrix}
0 & \frac{1}{2} & 0 & 0 \\ 
\frac{1}{2} & 0 & 0 & 0 \\ 
0 & 0 & -1 & 0 \\ 
0 & 0 & 0 & -1
\end{pmatrix} \,,
\end{equation}
and thus to
\begin{equation}
    a \cdot b = \frac{1}{2} \left ( a^{+} b^{-} + a^{-} b^{+} \right ) - \vec{a}_\perp \cdot \vec{b}_\perp \,.
\end{equation}


\section{Sampling of the seed for the Bethe-Heitler-Gunion-Bertsch gluon emission} 
\label{app:gbmq0sampling}

In this appendix we describe the algorithm used to generate the GB pre-gluons. It is valid in the dominant region of radiation where the gluon energy is small with respect to the available invariant mass. In the opposite case of a small invariant mass, the exact measure is slightly simplified for the sake of the effectiveness of the MC sampling, which however still encompasses important physical aspects — like the shrinking of the phase space.

After integration over the energy- and momentum-conserving Dirac delta function, the inelastic cross sections depends on five variables $x$, $\vec{k}_\perp$, and $\vec{l}_\perp$ (the momentum transferred from the light quark, transversally to the jet direction). In order to model this cross section in a transport theory one uses a Monte Carlo approach. The task is to convert a set of five uncorrelated random numbers into these variables in order to reproduce the unbiased distribution $\mathrm{d}^5\sigma_{\rm rad}/(\mathrm{d}x \, \mathrm{d}^2 l_\perp \, \mathrm{d}^2 k_\perp)$.  In this appendix we describe the method presently chosen in SUBA-Jet. To draw 5 independent random numbers in the 5 dimensional $x$, $\vec{k}_\perp$, and $\vec{l}_\perp$ space and then proceed by mere rejection is very ineffective - not to say impossible - because these variables are correlated by phase space (PS) and most of the sets of random numbers would have to be discarded. To make a sampling of the cross section possible, the adopted strategy consists in finding some suitable majorant of the 5-dimensional distribution $\mathrm{d}^5\sigma_{\rm rad}/(\mathrm{d}x \, \mathrm{d}^2 l_\perp \, \mathrm{d}^2 k_\perp)$
that can be mapped best onto a flat 5D probabilistic-space and sufficiently close to the original distribution that the unavoidable rejection step preserves the efficiency of the sampling process by minimizing the number of trials, which have to be discarded. Concretely, one first preforms recursive integrations (ascending process) on this majorant (first on $\vec{k}_\perp$, then on $\vec{l}_\perp$ and finally on $x$, resorting to additional majorations when mandatory), to obtain an invariant-mass dependent radiative cross section that is further averaged over the momentum distribution of light partons in the QGP to deduce a (majorant of the) radiative rate. The complementary (descending) process consists in sampling this rate to deduce the 3-momentum of the light parton and then, recursively, the $x$, $\vec{l}_\perp$ and $\vec{k}_\perp$ variables describing the outgoing particles. In these processes, the outgoing 3-body,\footnote{The two initial partons and the radiated gluon.} phase space boundary plays a specific role. Indeed, integrating the four-momentum conservation constraint results in a $1/\sqrt{\Delta(x,\vec{l}_\perp,\vec{k}_\perp)}$ factor in the $\mathrm{d}x \, \mathrm{d}^2 l_\perp \, \mathrm{d}^2 k_\perp$ measure which leads to some superficial divergence at the PS-boundary. For finite $x$, this superficial divergence is located far beyond the ``transverse'' $(\vec{l}_\perp,\vec{k}_\perp)$ domain typically contributing to the radiation. Therefore, $\Delta(x,\vec{l}_\perp,\vec{k}_\perp)$ will be approximated to $\Delta(x,\vec{0},\vec{0})$ for practical purposes. 
As such an approximation may have consequences on the shrinking PS integral at small $x$ --- $x\approx  m_g^2/s$, with $s$ the invariant mass --- we start by scrutinizing this case. We present the majorant to the radiation cross section for two limits:\\ 
a) for the assumption that the mass of the medium quark is finite, restricting ourselves to $m_q=0$ for the concrete expressions:
$\mathrm{d}^5\sigma_{\rm rad}^{m_{q}=0}/(\mathrm{d}x \, \mathrm{d}^2 l_\perp \, \mathrm{d}^2 k_\perp)$  and \\
b) for the assumption that the mass $m_q \to \infty$: 
$\mathrm{d}^5\sigma_{\rm rad}^{m_{q}\to \infty}/(\mathrm{d}x \, \mathrm{d}^2 l_\perp \, \mathrm{d}^2 k_\perp)$\\ 
In this appendix, we will assume that the light QGP parton is a quark of mass $m_q$. The generalization to scattering on gluons is pretty direct if one assumes, as we do, that it is dominated by the $t$-channel exchange.

\subsection{Definition of the symbols}
To make it easier for the reader to follow the calculation we first present a summary of the symbols, which we will use. $\sqrt{s}$ is the center of mass energy and $m_Q$, $m_q$, $m_g$ are the mass of the fast parton, of the QGP parton and of the gluon, respectively. 

\begin{eqnarray}
\tilde{x}&=&x P^+/P_{\rm tot}^+= x P^+/(P_Q^{+} + p_q^{+})= x\;\text{for $m_q=0$}\\ 
s_-&=& s-m_Q^2 - m_q^2 \label{eq:sminus} \\
\tilde s&=& s-\frac{m_g^2}{x}\\
\bar s&=& \left ( 1-\tilde x \right ) \left ( \tilde s -\frac{k_\perp^2}{\tilde x} \right ) \\ \nonumber  \\
m_{Q,T}'&=&\sqrt{m_Q^2+(\vec{k}_\perp-\vec{l}_\perp)^2} \\
m_{q,T}&=&\sqrt{m_q^2+l_\perp^2} \\
\tilde{m}_g^2&=&(1-x) m_g^2 + x^2 m_Q^2 \label{eq:mgtilde} \\
\nonumber \\
\Delta \equiv \Delta(\tilde{x},\vec{k}_\perp,\vec{l}_\perp,m_q) &=& B^+  \times B^- \label{eq:deltadef} \\
\mathrm{where} \quad B^+ &=& (1-\tilde{x})\left[\tilde{x} s-m_{g,T}^2\right]-\tilde{x} (m'_{Q,T}+m_{q,T})^2 \label{eq:Bplusdef} \\
 B^- &=& (1-\tilde{x})\left[\tilde{x} s-m_{g,T}^2\right]-\tilde{x} (m'_{Q,T}-m_{q,T})^2 \label{eq:Bminusdef} \\ 
\Delta_a\equiv \Delta(x,\vec{k}_\perp,\vec{l}_\perp,m_q=0)&=&\bigl((1-x)(x\,s-m_g^2)-x\,m_Q^2-k_\perp^2+2x\,\vec{k}_\perp\cdot
\vec{l}_\perp\bigr)^2\nonumber \\
&&-4x(1-x)\,l_\perp^2\,(x\,s-k_\perp^2-m_g^2) \label{eq:deltamq0def} \\
\Delta_{a1}\equiv \Delta(x,\vec{k}_\perp,\vec{l}_\perp=\vec{0},m_q=0)&=&\bigl(x((1-x)\tilde{s}-m_Q^2)-k_\perp^2 \bigr)^2 \label{eq:deltaA1def} \\
\Delta_{a2}\equiv \Delta(\tilde{x},\vec{k}_\perp=\vec{0},\vec{l}_\perp=\vec{0},m_q)\label{eq:deltaa2}&=&
s^2(\tilde{x}-\tilde{x}_{\rm min})(\tilde{x}_{\rm max} -\tilde{x}) (\tilde{x}-\tilde{x}^-_{\rm min})(\tilde{x}_{\rm max}^- -\tilde{x}) \\ 
\tilde{x}_{\rm min} &=& \frac{m_g^2}{s-(m_q+m_Q)^2}  +\mathcal{O}(s^{-2})\\
\tilde{x}_{\rm max} &=& 1 - \frac{(m_q+m_Q)^2}{s} +\mathcal{O}(s^{-2})\\
\tilde{x}_{\rm min}^- &=& \frac{m_g^2}{s-(m_q-m_Q)^2}  +\mathcal{O}(s^{-2})\\
\tilde{x}_{\rm max}^- &=& 1 - \frac{(m_q-m_Q)^2}{s} +\mathcal{O}(s^{-2})\,. 
\end{eqnarray}
When $m_q=0$, we get simplified expressions  $\tilde{x}_{\rm min}^-=\tilde{x}_{\rm min}^+={x}_{\rm min}$ and same for $x_{\rm max}$ 
\begin{eqnarray}
x_{\rm min} &=& \frac{m_g^2}{s-m_Q^2}  +\mathcal{O}(s^{-2})\\
x_{\rm max} &=& 1 - \frac{m_Q^2}{s} +\mathcal{O}(s^{-2})\,. 
\end{eqnarray}
and consequently
\begin{equation}
    \Delta\equiv \Delta(\tilde{x},\vec{k}_\perp=\vec{0},\vec{l}_\perp=\vec{0},m_q=0)
= s^2 \left ( x-\frac{m_g^2}{s-m_Q^2} \right )^2 \left ( 1 - \frac{m_Q^2}{s}-x \right )^2 \\ 
\end{equation}
which agrees up to $\mathcal{O}(s^{-2})$ for $m_Q^2 << s$ with eq. \ref{eq:deltaa2}.

\subsection{Generic cross section and phase space}
For finite mass $m_Q$ of the incoming parton and finite mass $m_g$ of the radiated gluon but zero mass of the QGP partons, $m_{q}=0$, the inelastic Gunion-Bertsch scattering cross section for $Qq \to Qqg$ --- with $Q$ being the relativistic projectile quark --- can be approximated by model II of ref.~\cite{Aichelin:2013mra} which reads as 
\begin{equation}
 \frac{\mathrm{d}^5\sigma_{\rm rad}^{m_{q}=0}}{\mathrm{d}x \, \mathrm{d}^2 l_\perp \, \mathrm{d}^2 k_\perp}\approx \frac{16\pi^3 x(1-x) \left\vert\mathcal{M}_{\rm el}\right\vert^2}{16(2\pi)^5v_{qQ}} \frac{P_g}{\sqrt{\Delta_a}} \Theta(B^+)\,,
 \label{eq:mq0xsec}
\end{equation}
where $v_{qQ}=p_q\cdot p_Q = (s -m_Q^2)/2$, while the polarization-averaged elastic amplitude (in the $t$-channel) is 
\begin{equation}
\left\vert\mathcal{M}_{\rm el}\right\vert^2 = \frac{4C_F}{2N_c}
g^4\frac{s_-^2+st +\frac{t^2}{2} }{(t-\mu^2)^2}
\approx \frac{4C_F}{2N_c}
g^4\frac{s_-^2+st +\frac{t^2}{2} }{(l_\perp^2+\mu^2)^2} \,,
\label{eq:Melapproximate}
\end{equation}
with $s_-=s-m_Q^2$. The conditional probability for gluon emission $P_g$ is given by
\begin{equation}
P_g(x,\vkt,\vlt;M)=\frac{C_A\alpha_s}{\pi^2}\frac{1-x}{x}
\left(\frac{\vkt}{\vkt^2+\tilde{m}_g^2}
-\frac{\vkt-\vlt}{(\vkt-\vlt)^2 +\tilde{m}_g^2}\right)^2 \,,
\end{equation}
with
$\tilde{m}_g^2$ given by eq.~(\ref{eq:mgtilde}), 
while the measure stemming from the integration over the energy/momentum $\delta$ functions, $\Delta_a$, is given by eq.~(\ref{eq:deltamq0def}).

In the case where $m_q\neq 0$, the model given by eq.~(\ref{eq:mq0xsec}) can be extended by expressing  the full phase space in terms of $\vec{k}_\perp$, $\vec{l}_\perp$ and $\tilde{x}=xP_Q^+/P_{\rm tot}^{+}=k^+/P_{\rm tot}^{+}$, so that 
\begin{eqnarray}
\frac{\mathrm{d}^5\sigma_{\rm rad}}{\mathrm{d}\tilde{x} \, \mathrm{d}^2l_\perp \, \mathrm{d}^2k_\perp} 
&\approx &
\frac{16 \pi^3  x  (1-x)  \left\vert{\cal M}_{\rm el}\right\vert^2}{16(2\pi)^5 v_{qQ}}
\frac{\Theta(B^+)}{\sqrt{\Delta}}  P_g \,,
\label{eq:mqnot0xsec}
\end{eqnarray}
and
$\frac{\mathrm{d}^5\sigma_{\rm rad}}{\mathrm{d}x \, \mathrm{d}^2l_\perp \, \mathrm{d}^2k_\perp} =\frac{p_Q^+}{p_{\rm tot}^+}\times\frac{\mathrm{d}^5\sigma_{\rm rad}}{\mathrm{d}\tilde{x} \, \mathrm{d}^2l_\perp \, \mathrm{d}^2k_\perp}$, where the (polarization averaged) squared elastic amplitude $\left\vert{\cal M}_{\rm el}\right\vert^2$ admits the same expression as in eq.~(\ref{eq:Melapproximate}), with the general expression of $s_-=s-m_Q^2-m_q^2$, as in eq.~(\ref{eq:sminus}) and  where $v_{qQ}=p_q\cdot p_Q = s_-/2$. As for the measure, $\Delta$ can be written in an elegant factorized form: 
\begin{equation}
\Delta \equiv \Delta(\tilde{x},\vec{k}_\perp,\vec{l}_\perp,m_q) = B^+  \times B^- \,,
\label{eq:Deltafinmass}
\end{equation}
with expressions of $B^+$ and $B^-$ given by eqs.~(\ref{eq:Bplusdef}) and (\ref{eq:Bminusdef}).

This cross section (\ref{eq:mqnot0xsec}) is both IR and colinear safe thanks to the regulators $\mu$ and $\tilde{m}_g$. We work in a generic regime where $s\gg$ these scales. Then, from the $\Theta({\Delta})$ condition, one derives that $l_\perp$ should be typically $<\sqrt{s}$ while $k_\perp<\sqrt{\tilde{x} \tilde{s}}$. From the respective structures of $\left\vert\mathcal{M}_{\rm el}\right\vert^2$ and $P_g$, one understands that the typical $l_\perp\sim \mu$ while the typical $k_\perp\sim \tilde{m}_g$.

\subsection{Finite and zero $m_q$ }
\subsubsection{Phase space}
\label{appendix:phasespace}

\begin{figure}[!htb]
\centering
\includegraphics[width=0.45\textwidth]{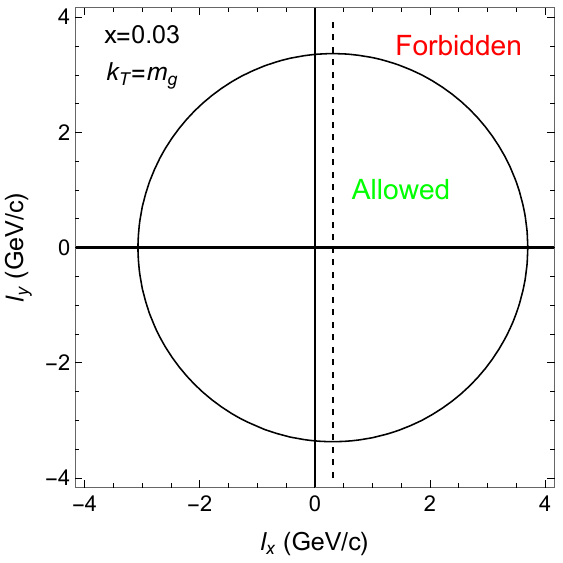}
\includegraphics[width=0.45\textwidth]{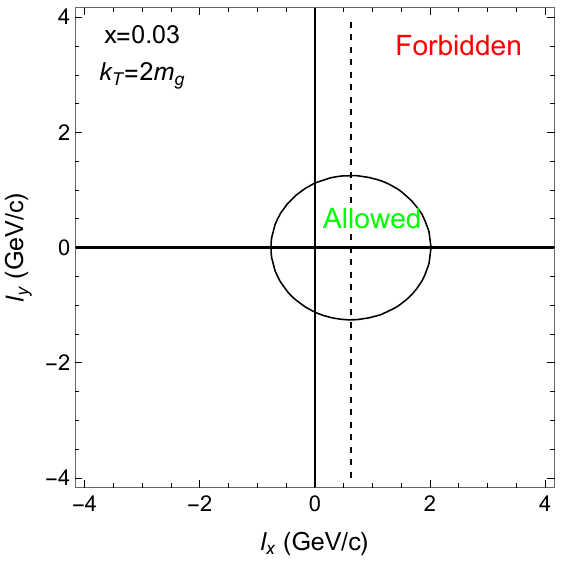}
\caption{Illustration of the $\{l_x,l_y\}$ transverse phase space for typical parameters considered in this work: $T=0.4$~GeV, $e_Q=100$~GeV, $m_g\approx 0.626$~GeV and $m_q\approx 0.370$~GeV, following eqs.~(\ref{eq:mthg}) and (\ref{eq:mthq}); $s$ is taken as $2 e_Q m_q\approx 74$~GeV$^2$ and $x=0.03 \Leftrightarrow \omega \approx 3$~GeV. The left panel corresponds to a ``small'' emission angle ($\theta \approx 0.2$) while this angle is twice as large for the right panel. The vertical dashed lines indicate the centers of both allowed disks.}
\label{fig:lTplane}
\end{figure}

We now provide a precise description of the phase space boundaries used later on. As $B^+$ is always smaller than $B^-$, the original phase space constraint $\Theta(\Delta)$ is identical to  $\Theta(B^+)$ and can be expressed as
\begin{equation}
B^+\ge 0 \hspace{5mm} \Leftrightarrow \hspace{5mm} \tilde{x} (m'_{Q,T}+m_{q,T})^2 + (1-\tilde{x}) k_\perp^2 \le (1-\tilde{x}) \tilde{x} \tilde{s} \,.
\end{equation}
For a given $x$, this condition defines a domain in the $\{\vkt,\vlt\}$ transverse phase space that shrinks for both small $x$ and $x$ close to $1$. To derive the transverse space boundary for a given $x$, the simplest method consists in fixing $k_\perp$ and then to find the $\vlt$ domain (for fixed $k_\perp$ and x) by requiring that
\begin{equation}
\left ( m'_{Q,T}+m_{q,T} \right )^2  \le \left ( 1-\tilde{x} \right ) \left ( \tilde{s} -\frac{k_\perp^2}{\tilde{x}} \right ) \equiv \bar{s} \,. 
\end{equation}
Obviously, one finds that the acceptable $k_\perp$ shrinks for small $x$. Following this procedure one finds that the $\vlt$ domain is an ellipse (slightly ex-centered along $\hat{k}_\perp$ -assumed to be in $Ox$ direction -- by an off-set\footnote{Later analysis shows that $k_\perp^2>\bar{s}$, so that the offset always keeps finite.} 
\begin{equation}
\delta=\frac{\bar{s}-{k_\perp}^2+{m_q}^2-{m_Q}^2}{
	\bar{s}-k_\perp^2} \times \frac{k_\perp}{2} \,,
\end{equation}
with small semi-axis length (along $Oy$) 
\begin{equation}
l_y^{\rm max}=\frac{1}{2}\sqrt{\frac{\left(\bar{s}-{k_\perp}^2-({m_q}-{m_Q})^2\right)
	\left(\bar{s}-{k_\perp}^2-({m_q}+{m_Q})^2\right)}
{ \bar{s}-{k_\perp}^2}} \,,
\end{equation}
and long semi-axis length (along $Ox$)
\begin{equation}
l_x^{\rm max} =\frac{l_y^{\rm max}}{\sqrt{1-\frac{k_\perp^2}{\bar{s}}}}\,,
\end{equation}
which is $\approx l_y^{\rm max}$ for small $\tilde{x}$. As $\bar{s}-k_\perp^2=(1-\tilde{x})\tilde{s}-\frac{k_\perp^2}{\tilde{x}}$, one sees that both $l_y^{\rm max}$ and $l_x^{\rm max}$ scale like 
$\sqrt{s}$ 
whenever $k_\perp\ll \sqrt{\tilde{x} \tilde{s}}$ --- a condition that is nearly always satisfied for not too small $x$ --- while the $l_\perp$ domain shrinks when $k_\perp\sim  \sqrt{\tilde{x} \tilde{s}}$. The $l_\perp$ domain shrinks to 0 when $\bar{s}-{k_\perp}^2-({m_q}+{m_Q})^2=(1-\tilde{x})\tilde{s}-\frac{k_\perp^2}{\tilde{x}}-({m_q}+{m_Q})^2=0$, as illustrated on Fig.~\ref{fig:lTplane}.

This allows to determine the upper boundary for $k_\perp^2$,
\begin{equation}
k_\perp^{\rm max} =\sqrt{\tilde{x}\left((1-\tilde{x})\tilde{s}-(m_q+m_Q)^2\right)} =\sqrt{\tilde{x}\left((1-\tilde{x}) s  -(m_q+m_Q)^2\right)-(1-\tilde{x})m_g^2}\,.
\label{eq:ktmax}
\end{equation}
In turn, this equation sets the limits for the $\tilde{x}$-range,
\begin{equation}
\tilde{x}_{\rm min} \approx \frac{m_g^2}{s-(m_q+m_Q)^2} \le \tilde{x} \le 
\tilde{x}_{\rm max} \approx 1 - \frac{(m_q+m_Q)^2}{s} \,. 
\end{equation}
For small $\tilde{x}$, one can thus write 
\begin{equation}
k_\perp^{\rm max}\approx \sqrt{s}\sqrt{\tilde{x}-\tilde{x}_{\rm min}} \,.
\label{eq:ktmaxmq0}
\end{equation}

Now we have obtained the approximate phase space boundaries $\tilde{x}_{\rm min},\tilde{x}_{\rm max}$, $k_\perp^{\rm max}(\tilde{x})$ as well as $l_x^{\rm max}(\tilde{x},k_\perp)$ and $l_y^{\rm max}(\tilde{x},k_\perp)$, we will further analyze the impact of approximations performed on the $1/\sqrt{\Delta}$ factor in sampling the cross section. Considering the $1/\sqrt{\Delta(\tilde{x},\vec{k}_\perp,\vec{l}_\perp,m_q)}$ factor in the cross section, it shows some superficial integrable divergence close to the phase space boundary in the transverse plane. The dependence on $\vec{k}_\perp$ and on $\vec{l}_\perp$ is difficult to manage when performing the integrals as well as for the Monte Carlo sampling. For these reasons we will approximate $\Delta(\tilde{x},\vec{k}_\perp,\vec{l}_\perp,m_q)$ by $\Delta(\tilde{x},0,0,m_q)$. 
Whenever $\tilde{x} s$ is large compared to the typical $l_\perp^2$ and $k_\perp^2$ values
this is a good approximation because the $k_\perp$ and $l_\perp$ dependence of $\Delta$ is small and can be neglected compared  the dependence of $\left\vert\mathcal{M}_{\rm el}\right\vert^2 \times P_g$ on these variables.  To be more precise, one can legitimately approximate $\Delta(\tilde{x},\vec{k}_\perp,\vec{l}_\perp,m_q)$ by $\Delta(\tilde{x},0,0,m_q)$ when in the $\{\vec{l}_\perp,\vec{k}_\perp\}$ space the phase space boundary is located --- for a given $\tilde{x}$ --- beyond the  $\{\vec{l}_\perp,\vec{k}_\perp\}$ values for which the matrix element is maximal,
at  $l_\perp\sim k_\perp \approx\mu\approx m_g$. Requiring $k_\perp^{\rm max}\gtrsim c_0 \times m_g$ is shown from eq.~(\ref{eq:ktmax}) to be equivalent to claim that $\tilde{x}\gtrsim (1+c_0^2)\times \tilde{x}_{\rm min}$, while $l_\perp^{\rm max}(k_\perp=0)$ increases much faster with $\tilde{x}$\footnote{For $m_q=0$ and $k_\perp=0$, one has $l_\perp^2\approx \frac{(\tilde{s}-m_Q^2)^2}{4\tilde{s}}$, which is of the order of $m_g^2$ for $\tilde{s}=2 {m_g} \sqrt{{m_g}^2+{m_Q}}+2 {m_g}^2+{m_Q}$, leading to $\tilde{x}=m_g^2/(s- m_Q^2 - 2 m_g (m_g +\sqrt{m_g^2 + m_Q^2}))=\tilde{x}_{\rm min}\times \left(1 + 
\frac{2 m_g (m_g +\sqrt{m_g^2 + m_Q2})}{s}+\mathcal{O}(s^{-2})\right)$.}. Therefore the phase space boundary in the transverse plane will be located beyond the peak region for radiation for $\tilde{x} \gtrsim \tilde{x}_{\rm min}$, legitimating the aforementioned approximation for such $\tilde{x}$-range. 

For small  $\tilde{x}\sim \tilde{x}_{\rm min}$, the opposite approach is in order.  Here the $\{\vec{l}_\perp,\vec{k}_\perp\}$ dependence of $\Delta(\tilde{x},\vec{k}_\perp,\vec{l}_\perp,m_q)$ is strong as compared to that of the matrix element $\left\vert\mathcal{M}_{\rm el}\right\vert^2 \times P_g$. We then start to analyze this case where the matrix elements can be considered as constant whereas the $\{\vec{l}_\perp,\vec{k}_\perp\}$ dependence of $\Delta(\tilde{x},\vec{k}_\perp,\vec{l}_\perp,m_q)$ has to be taken into account explicitly. One notices that $\Delta$ in eq.~(\ref{eq:deltadef}) can be written as
\begin{equation}
\Delta 
= 4 \tilde{x}^2 \bar{s} \times \left[(l_y^{\rm max})^2 - l_y^2 - (1-\frac{k_\perp^2}{\bar{s}})  (l_x-\delta)^2\right]\,.
\end{equation}
Hence, integrating to the upper limit $\sqrt{(l_y^{\rm max})^2  - (1-\frac{k_\perp^2}{\bar{s}}) (l_x-\delta)^2}$ for $l_y$, we obtain
\begin{equation}
\iint \frac{\mathrm{d}l_x \, \mathrm{d}l_y}{\sqrt{\Delta}}=\frac{\pi}{2 \tilde{x} \sqrt{\bar{s}}}\int \mathrm{d}l_x=
\frac{\pi l_x^{\rm max}}{\tilde{x}\sqrt{\bar{s}}}\,.
\end{equation}
Considering from now on the case $m_q=0$ to make the calculations easier, one has
\begin{equation}
\frac{l_x^{\rm max}}{\sqrt{\bar{s}}}= \frac{(k_\perp^{\rm max})^2-k_\perp^2}{2 x (\bar{s}-k_\perp^2)}
=
\frac{(k_\perp^{\rm max})^2-k_\perp^2}{2 \left(x m_Q^2+(k_\perp^{\rm max})^2-k_\perp^2\right)} \,,
\end{equation}
so that the integral over the transverse momenta gets
\begin{equation}
\Phi=\iint \frac{\mathrm{d}l^2_\perp \, \mathrm{d}^2k_\perp}{\sqrt{\Delta}} = 
\frac{\pi^2}{2 x}\int_0^{(k_\perp^{\rm max})^2} \mathrm{d}k_\perp^2 
\frac{k_\perp^2}{x m_Q^2+k_\perp^2} \,.
\label{eq:phiexact}
\end{equation}
One can then distinguish two regimes. If $(k_\perp^{\rm max})^2\lesssim x m_Q^2$, one approximates $x m_Q^2+k_\perp^2$ by $x m_Q^2$. This is equivalent to require
\begin{equation}
x s - m_g^2 - x m_Q^2 \lesssim x m_Q^2 \Leftrightarrow
x  \lesssim \frac{m_g^2}{s- 2 m_Q^2} \,,
\end{equation}
and represents a very small $x$-range. For this tiny range, one thus gets
\begin{equation}
\Phi\approx  \frac{\pi^2 (k_\perp^{\rm max})^4}{4x^2 m_Q^2}
\approx  \frac{\pi^2 s^2(x-x_{\rm min})^2}{4 x^2 m_Q^2}
\approx  \frac{\pi^2 s^4}{4 m_Q^2 m_g^4}\times (x-x_{\rm min})^2 \,.
\label{eq:phifinalregime1}
\end{equation}
For slightly larger $x$ value (still of the order of $x_{\rm min}$), one has $(k_\perp^{\rm max})^2 \approx x s$ so that $m_Q^2+k_\perp^2\approx k_\perp^2$ and
\begin{equation}
\Phi\approx \frac{\pi^2}{2 x }\int_0^{(k_\perp^{\rm max})^2} \mathrm{d}k_\perp^2 \approx \frac{\pi^2 (k_\perp^{\rm max})^2}{2x }\approx \frac{\pi^2 s^2}{2 m_g^2} \times (x-x_{x_{\rm min}}) \,. 
\label{eq:phifinalregime2}
\end{equation}

We now inspect the consequence of various approximations for $\Delta$. If one neglects the $\vec{l}_\perp$ dependence in $\Delta$ and one approximates $\Delta\approx\Delta_{a1}=\Delta(x,\vec{k}_\perp,0)$ (eq.~(\ref{eq:deltaA1def}))
one obtains, still for $m_q=0$
\begin{equation}
\int \frac{\mathrm{d}l_x \, \mathrm{d}l_y}{\sqrt{\Delta_{a1}}} \approx \frac{\pi l_x^{\rm max} l_y^{\rm max}  }{x((1-x)\tilde{s}-m_Q^2)-k_\perp^2} =
\pi\frac{\sqrt{\bar{s}}(\bar{s}-k_\perp^2-m_Q^2)^2}{4(\bar{s}-k_\perp^2)^{3/2}((k_\perp^{\rm max})^2-k_\perp^2)} \,.
\end{equation}

Using eq.~(\ref{eq:ktmax}) we find
\begin{equation}
\bar{s}=\frac{(k_\perp^{\rm max})^2+x m_Q^2-(1-x)k_\perp^2}{x} \,,
\end{equation}
and thus the integral on $\vec{l}_\perp$ yields
\begin{equation}
\int \frac{\mathrm{d}l_x \, \mathrm{d}l_y}{\sqrt{\Delta_{a1}}} \approx
\frac{\pi}{4x}\frac{\sqrt{(k_\perp^{\rm max})^2+x m_Q^2-(1-x)k_\perp^2}}{\left((k_\perp^{\rm max})^2+x m_Q^2-k_\perp^2\right)^{3/2}}\times ((k_\perp^{\rm max})^2-k_\perp^2) \,.
\end{equation}
One sees that approximating $\Delta(x,\vec{k}_\perp,\vec{l}_\perp,0)$ by  $\Delta(x,\vec{k}_\perp,\vec{l}_\perp=0,0)$ does not lead to any divergence. For $x\ll 1$, one obtains 
\begin{equation}
\int \frac{\mathrm{d}l_x \, \mathrm{d}l_y}{\sqrt{\Delta_{a1}}} \approx
\frac{\pi}{4x}\frac{(k_\perp^{\rm max})^2-k_\perp^2}{(k_\perp^{\rm max})^2+x m_Q^2-k_\perp^2} \,.
\end{equation}
and 
\begin{equation}
\Phi_{a1}=\iint \frac{\mathrm{d}l^2_\perp \, \mathrm{d}^2k_\perp}{\sqrt{\Delta_{a1}}} \approx 
\frac{\pi^2}{4 x}\int_0^{(k_\perp^{\rm max})^2} \mathrm{d}k_\perp^2 
\frac{k_\perp^2}{x m_Q^2+k_\perp^2} \,.
\end{equation}
as compared to the expression eq.~(\ref{eq:phiexact}), one obtains a mere reduction of a factor $2$ --- as the measure is now flat at the boundaries of the $\vec{l}_\perp$ transverse space, but the $x$-scaling is preserved. 

If one proceeds with an even more drastic approximation by neglecting both $\vec{l}_\perp$ and $\vec{k}_\perp$ dependencies in $\Delta$, $\Delta\approx \Delta_{a2}= \bigl(x((1-x)\tilde{s}-m_Q^2)\bigr)^2$, one finds, for $m_q=0$,
\begin{equation}
\int \frac{\mathrm{d}l_x \, \mathrm{d}l_y}{\sqrt{\Delta_{a2}}} = 
\pi\frac{\sqrt{\bar{s}}(\bar{s}-k_\perp^2-m_Q^2)^2}{4(\bar{s}-k_\perp^2)^{3/2}(k_\perp^{\rm max})^2}
\approx
\frac{\pi}{4x}\frac{\left((k_\perp^{\rm max})^2-k_\perp^2\right)^2}{(k_\perp^{\rm max})^2\,\left((k_\perp^{\rm max})^2+x m_Q^2-k_\perp^2\right)} \,,
\end{equation}
and 
\begin{equation}
\Phi_{a1}\approx 
\frac{\pi^2}{4 x (k_\perp^{\rm max})^2}\int_0^{(k_\perp^{\rm max})^2} \mathrm{d}k_\perp^2 
\frac{k_\perp^4}{x m_Q^2+k_\perp^2}
\underset{(k_{\rm max})^2\gg x m_Q^2}{\approx}
\frac{\pi^2 (k_{\rm max})^2}{8x} \,,
\end{equation}
which is a factor $4$ smaller then the exact phase space eq.~(\ref{eq:phifinalregime2}). The conclusion is that the approximate phase space $\Phi_{a1}$ or  $\Phi_{a2}$ scales like $x-x_{\rm min}$ for all approximations of $\Delta$, but with a smaller slope. We have checked that this conclusion stays valid for finite $m_q$.

From now on, we will then approximate 
$\Delta(\tilde{x},\vec{k}_\perp,\vec{l}_\perp,m_q)$ by $\Delta_{a2}(\tilde{x},\vec{0},\vec{0},m_q)$ -- given by eq.~(\ref{eq:deltaA1def}) -- in the expressions (\ref{eq:mq0xsec}) and (\ref{eq:mqnot0xsec}) of the radiative cross section.\footnote{An intermediate way would be to consider $\Delta(\tilde{x},\vec{k}_\perp,\vec{0},m_q)$ in the measure.}. This approximation has no impact when $\tilde{x} \gtrsim \tilde{x}_{\rm min}$ -- since the transverse phase space is open enough for $\left\vert\mathcal{M}_{\rm el}\right\vert^2 \times P_g$ to vary strongly as compared to  $\Delta(\tilde{x},\vec{k}_\perp,\vec{l}_\perp,m_q)$ -- and limited impact on the shrinking PS when $\tilde{x}\simeq \tilde{x}_{\rm min}$, as we have just demonstrated. Notice however that the PS condition $\Theta(B^+)$ is imposed exactly. 

\subsubsection{Explicit radiative cross section for $m_q=0$}

We present now the explicit expression for the radiative cross section for $m_q=0$, implying $\tilde{x}=x$\footnote{The general $m_q \neq 0$ case can be treated similarly but implies more tedious expressions.}. Writing 
\begin{equation}
\sqrt{\Delta_{a2}}=s(x-x_{\rm min})(x_{\rm max}-x) \,,
\end{equation}
one obtains from eq.~(\ref{eq:mq0xsec}): 
\begin{equation}
\frac{\mathrm{d}^5\sigma_{\rm rad}^{a2}}{\mathrm{d}x \mathrm{d}^2 l_\perp \mathrm{d}^2 k_\perp}\approx \frac{2\pi^3  \left\vert\mathcal{M}_{\rm el}\right\vert^2}{(2\pi)^5 s\,s_-} \underbrace{\frac{x(1-x)}{(x-x_{\rm min})(x_{\rm max}-x) }}_{c_1(x)} P_g \Theta(B^+)\,.
\label{dsigmarad1}
\end{equation}
This can also be written in terms of the differential elastic cross section as
\begin{equation}
\vert{\cal M}_{\rm el}\vert^2 =64 \pi s p_{cm}^2	\frac{\mathrm{d}\sigma_{\rm el}}{\mathrm{d}t}= 
16 \pi s_-^2	\frac{\mathrm{d}\sigma_{\rm el}}{\mathrm{d}t}
	\quad\text{ with }\quad p_{\rm cm}=\frac{s-m_Q^2}{2\sqrt{s}}\,,
\end{equation}
leading to
\begin{equation}
\frac{\mathrm{d}^5\sigma_{\rm rad}^{a2}}{\mathrm{d}x \, \mathrm{d}^2l_\perp \, \mathrm{d}^2k_\perp} 
\approx 
\frac{s_-}{s} \times
\frac{\mathrm{d}\sigma_{\rm el}}{\mathrm{d}^2l_\perp} \times P_g \times
\Theta(B^+)\times {c_1(x)}\,.
\end{equation} 
For later purposes we define the cross section without phase space limitations
\begin{equation}
\frac{\mathrm{d}^5\sigma_{\rm rad}^{{\rm sup}_1}}{\mathrm{d}x \, \mathrm{d}^2l_\perp \, \mathrm{d}^2k_\perp} 
\approx 
\frac{s_-}{s} \times
\frac{\mathrm{d}\sigma_{\rm el}}{\mathrm{d}^2l_\perp} \times P_g 
\times {c_1(x)}\,.
\label{eq:sup1}
\end{equation} 
The factor $c_1$ in eq.~(\ref{dsigmarad1}) is close to unity for $\sqrt{s}\gg$ parton masses. 

In order to generate a sampling method, the generic strategy consists in integrating successively over more and more of the five independent variables on which the cross section depends until we obtain the integrated cross section and the rate. In this process
we generate data tables or recursive formulas. Then we start from the rate and determine in inverse order the different variables. 
For the given case of the cross section, eq.~(\ref{dsigmarad1}), we integrate successively over the azimuthal angle in the transverse plane between $\vec{l}_\perp$ and $\vec{k}_\perp$, $k_\perp$, $\vec{l}_\perp$, and $x$. We also favour a method which does not require the storage of numerical tables as they are tedious to generate in an exploration stage during which the parameters are often modified. Because the integrals do not systematically admit analytical expression, we will proceed to a series of functions whose values are larger than the function we want to sample and apply corresponding rejections to obtain a sampling corresponding to the desired function.     

We start out with the integration over $\vec{k}_\perp$ and get
\begin{equation}
\frac{\mathrm{d}^3\sigma_{\rm rad}}{\mathrm{d}x \, \mathrm{d}^2l_\perp} 
\approx   \frac{c_1(x)\left\vert{\cal M}_{\rm el}\right\vert^2}{16\pi^2 s s_-}\int \mathrm{d}^2k_\perp  P_g \Theta(B^+)\,, 
\end{equation}
which is of course smaller then 
\begin{equation}
\frac{\mathrm{d}^3\sigma_{\rm rad}^{{\rm sup}_1}}{\mathrm{d}x \, \mathrm{d}^2l_\perp} 
\approx   \frac{c_1(x)\left\vert{\cal M}_{\rm el}\right\vert^2}{16\pi^2 s s_-}\underbrace{\int \mathrm{d}^2k_\perp  P_g}_{I_1}\,.
\end{equation}
Removing the phase space boundary obviously facilitates the evaluation of this integral. The exact expression of $I_1$, after integration up to $k_\perp \to \infty$  is given in ref.~\cite{Aichelin:2013mra}, 
\begin{equation}
I_1=\frac{C_A\alpha_s}{\pi}\frac{1-x}{x}
\left(\frac{l_\perp^2+2 \tilde{m}_g^2}{\sqrt{l_\perp^2(l_\perp^2+4\tilde{m}_g^2)}}
\ln\left(\frac{l_\perp^2\left(\sqrt{1+4 \tilde{m}_g^2/l_\perp^2}+1\right)
+4 \tilde{m}_g^2}{\tilde{m}_g^2\left(
\sqrt{1+4 \tilde{m}_g^2/l_\perp^2}-1\right)}+1\right)-2\right) \,.
\label{klimno}
\end{equation}
The next step is the $\vec{l}_\perp$ integration. One can still get an exact relation defining 
\begin{equation}
I_1^{\rm sup} \coloneqq 2\frac{C_A\alpha_s}{\pi}\frac{1-x}{x}\ln\left(1+\frac{l_\perp^2}{e\tilde{m}_g^2}\right) \,,
\label{eq:I1}
\end{equation}
where $e$ is the Euler constant not to be confused with an energy, and realizing that $I_1^{\rm sup}>I_1$. This implies
\begin{equation}
\frac{\mathrm{d}^3\sigma_{\rm rad}^{{\rm sup}_1}}{\mathrm{d}x \, \mathrm{d}^2 l_\perp}<
\frac{\mathrm{d}^3\sigma_{\rm rad}^{{\rm sup}_2}}{\mathrm{d}x \, \mathrm{d}^2 l_\perp}=	
 \frac{4 C_F\alpha_s^3}{\pi} \times\underbrace{\frac{s_-}{s}\frac{c_1(x)(1-x)}{x}}_{c_2(x)} 
\frac{1+\frac{t}{s_-} +\frac{t^2}{2 s_-^2} }{(l_\perp^2+\mu^2)^2} \ln\left(1+\frac{l_\perp^2}{e\tilde{m}_g^2}\right) \,,
\end{equation}
with $c_2(x)=\frac{(1-x)^2 x_{\rm max}}{(x-x_{\rm min})(x_{\rm max}-x)}$. In other terms, 
$\frac{\mathrm{d}^3\sigma_{\rm rad}^{{\rm sup}_2}}{\mathrm{d}x \, \mathrm{d}^2 l_\perp}$ is $\frac{\mathrm{d}^3\sigma_{\rm rad}^{{\rm sup}_1}}{\mathrm{d}x \, \mathrm{d}^2 l_\perp}$  where $I_1$ is replaced by $I_1^{\rm sup}$.
Although simpler, this expression still cannot be integrated over $l_\perp$ or $x$. Therefore  one defines 
$\frac{\sigma_{\rm rad}^{{\rm sup}_3}}{\mathrm{d}x \, \mathrm{d}^2 l_\perp}$ with
\begin{equation}
\frac{\mathrm{d}^3\sigma_{\rm rad}^{{\rm sup}_2}}{\mathrm{d}x \, \mathrm{d}^2 l_\perp} <
\frac{\mathrm{d}^3\sigma_{\rm rad}^{{\rm sup}_3}}{\mathrm{d}x \, \mathrm{d}^2 l_\perp}=
 \frac{4 C_F\alpha_s^3}{\pi} \times c_2(x) 
\frac{\ln\left(1+\frac{l_\perp^2}{e\tilde{m}_g^2}\right)}{(l_\perp^2+\mu^2)^2}  \,,
\label{eq:d3sigmaradsup3}
\end{equation}
because  $1+\frac{t}{s_-} +\frac{t^2}{2 s_-^2} \le 1$ in the allowed  $l_\perp$-range,  $\frac{\mathrm{d}^3\sigma_{\rm rad}^{{\rm sup}_3}}{\mathrm{d}x \, \mathrm{d}^2 l_\perp}$ can be integrated over $l_\perp$ in $\mathbb{R}^2$.
\begin{eqnarray}
\frac{\mathrm{d}\sigma_{\rm rad}^{{\rm sup}_3}}{\mathrm{d}x} &=& 
 4 C_F\alpha_s^3 \times c_2(x)
\int \mathrm{d}l_\perp^2  
\frac{\ln\left(1+\frac{l_\perp^2}{e\tilde{m}_g^2}\right)}{(l_\perp^2+\mu^2)^2} 
\nonumber\\
&=&
4 C_F\alpha_s^3c_2(x) \times \frac{\ln e \tilde{m}_g^2(x) - \ln \mu^2  }{e \tilde{m}_g^2(x) -\mu^2} \,.
\label{eq:dsigmaradsup3dx}
\end{eqnarray}
The last step (the $x$-integration) is less trivial. On the one side, the $x$-dependence in the last factor is not tractable. Defining $\tilde{m}_g^{\rm min}$ as the smallest value\footnote{We here assume that $m_g\ne 0$, implying the existence of such finite value.} of $\tilde{m}_g(x)$ on the $[x_{\rm min},x_{\rm max}]$, one obtains
\begin{equation}
\frac{\mathrm{d}\sigma_{\rm rad}^{{\rm sup}_3}}{\mathrm{d}x} < \frac{\mathrm{d}\sigma_{\rm rad}^{{\rm sup}_4}}{\mathrm{d}x} \coloneqq 
4 C_F\alpha_s^3 \times \frac{\ln e (\tilde{m}_g^{\rm min})^2/\mu^2}{e (\tilde{m}_g^{\rm min})^2-\mu^2}\times c_2(x) \,.
\end{equation}
The next difficulty is the fact that $\int_{x_{\rm min}}^{x_{\rm max}}$ does not converge. This is an artificial divergence because the various phase space rejections (that have been removed) lead to finite results. To regulate the cross section, one can approximate $c_2(x)$ by $1/x$ (which is the traditional growth considered for small $x$ radiation). In this case one gets
\begin{equation}
\frac{\mathrm{d}\sigma_{\rm rad}^{{\rm sup}_4}}{\mathrm{d}x} \approx  
4 C_F\alpha_s^3 \times \frac{\ln e (\tilde{m}_g^{\rm min})^2/\mu^2}{e (\tilde{m}_g^{\rm min})^2-\mu^2}\times \frac{1}{x} \,.
\label{eq:dsigmaradsup4dxapprox}
\end{equation}
and
\begin{equation}
\sigma_{\rm rad}^{{\rm sup}_4}\approx  
4 C_F\alpha_s^3 \times \frac{\ln e (\tilde{m}_g^{\rm min})^2/\mu^2}{e (\tilde{m}_g^{\rm min})^2-\mu^2}\times \ln \frac{s}{m_g^2} \,.
\label{eq:sigmaradsup4}
\end{equation} 
For more rigor, one can proceed to the exact integral of $c_2$ on $[x_{\rm min}+\frac{\Lambda_r^2}{s},x_{\rm max}-\frac{\Lambda_r^2}{s}]$  and consider small numerical values of $\Lambda_r$ to preserve most of the allowed region of phase space. Then one finds
\begin{eqnarray}
\int_{x_{\rm min}+\frac{\Lambda_r^2}{s}}^{x_{\rm max}-\frac{\Lambda_r^2}{s}} c_2(x) \mathrm{d}x
&=&
x_{\rm max}\times\left[\left(\frac{(1-x_{\rm max})^2+(1-x_{\rm min})^2}{x_{\rm max}-x_{\rm min}} \right) \ln \frac{s(x_{\rm max}-x_{\rm min})}{\Lambda_r^2}- \right.
\nonumber\\
&& \left. (x_{\rm max}-x_{\rm min}) +
\mathcal{O}(\Lambda_r^{-2})\right] \,.
\end{eqnarray}
This is identical to the previous results when $s\gg$ all masses, with $s(x_{\rm max}-x_{\rm min})\approx s-m_Q^2-m_g^2$.

\subsubsection{The rate for $m_q=0$}
We now proceed to the evaluation of the radiation rate. We take as an example the scattering on quarks (and antiquarks)\footnote{Gluon contribution is obviously added when compared to experiment.}. We assume that the QGP quarks are distributed according to the Boltzmann-Jüttner distribution and start from the usual invariant radiation rate (volumic density) 
\begin{equation}
\frac{\mathrm{d}R_{\rm rad}}{\mathrm{d}V} = \int \frac{\mathrm{d}^3p_Q}{e_Q} \frac{\mathrm{d}^3p_q}{p_q} v_{qQ} f_Q f_q \times \sigma_{\rm rad}\,,
\end{equation}
where $e_Q$ is the energy of the incoming projectile, $f_Q$ its distribution normalized $f_Q=\delta^{(3)}(\vec{x}_Q-t \vec{v}_Q)\delta^{(3)}(\vec{p}_Q-m_Q \vec{u}_Q)$ such that $\int \frac{\mathrm{d}^3p_Q}{e_Q} f_Q p_Q^\mu=m_Q u_Q^\mu$, with $u_Q$, the projectile four-velocity and $f_q$ will be taken as the local Boltzmann-Jüttner distribution $f_q=\frac{g_q}{(2\pi)^3} e^{-p_q\cdot u_f}$, with $g_q=4\times N_c \times N_f$ the quark-antiquark degeneracy and $u_f$ the local QGP four-velocity. 

After integration over the volume as well as over $\vec{p}_Q$, one derives
\begin{equation}
R_{\rm rad}^{{\rm sup}_4}=\frac{g_q}{(2\pi)^3}\times \frac{{2} C_F \alpha_s^3}{e_Q} \times  
\frac{\ln\frac{e (\tilde{m}_g^{\rm min})^2}{\mu^2}}{e (\tilde{m}_g^{\rm min})^2 -\mu^2}
\int_{s>s_{\rm min}} \frac{\mathrm{d}^3 p_q}{e_q} \,e^{-\frac{p_q\cdot u_{f}}{T}}\,s_-\,\ln \frac{s}{m_g^2} \,,
\end{equation}   
where the factor $\ln \frac{s}{m_g^2}$ directly comes from eq.~(\ref{eq:sigmaradsup4}), while $s_{\rm min}=(m_Q+m_g)^2$. $s$ is the center of mass energy, $s=(p_q+p_Q)^2$ and $v_{qQ}=p_q\cdot p_Q = (s -m_Q^2)/2$. The condition $s>s_{\rm min}$ avoids the sampling of too small values of $s$ which would be rejected afterwards. The calculation of the invariant integral is done in the rest frame of the incoming projectile.  Because no simple expression exists for this integral, one introduces a last inequality, based on $\ln \frac{s}{m_g^2}<
\frac{s}{m_g^2}-1$,
\begin{equation}
R_{\rm rad}^{{\rm sup}_4}< R_{\rm rad}^{{\rm sup}_5}=
\frac{g_q}{(2\pi)^3}\times 
\frac{{2} C_F \alpha_s^3}{e_Q m_g^2} \times  
\frac{\ln\frac{e (\tilde{m}_g^{\rm min})^2}{\mu^2}}{e (\tilde{m}_g^{\rm min})^2 -\mu^2} \times \tilde{\Sigma} \,,
\label{eq:rsup5}
\end{equation}  
where
\begin{equation}\tilde{\Sigma}=
\int_{s>s_{\rm min}} \frac{\mathrm{d}^3 p_q}{e_q} \,e^{-\frac{p_q\cdot u_{f}}{T}}\,(s-m_Q^2)\,(s-m_g^2)\,.
\label{Sigma}
\end{equation}
To evaluate the integral, one introduces $s_= =s-s_{\rm min}$.  Hence,
\begin{equation}
(s-m_g^2)\,(s-m_Q^2) =(s_= + m_Q^2 + 2 m_g m_Q) \times (s_= + m_g^2 + 2 m_g m_Q)\,.
\end{equation}
In the rest frame of the incoming parton $u_f=(u_Q^0,\vec{u}_Q)$, where $u_Q^0=e_Q/mQ$ and $\vec{u}_Q=\vec{p}_Q/m_Q$. Moreover $s=m_Q^2 + 2 e'_q m_Q+ m_q^2 \approx m_Q^2 + 2 e'_q m_Q $ and consequently
$s_= \coloneqq s-s_{\rm min} = 2 e'_q m_Q -m_g^2 -2 m_g m_Q$ and
$ \Theta(s-s_{\rm min})
\Leftrightarrow e_q' \ge \underbrace{m_g + \frac{m_g^2}{2m_Q}}_{e'_{\rm min}}.$
Introducing the difference
$\tilde{e}_q=e_q' -\left(m_g + \frac{m_g^2}{2m_Q}\right)$ which ranges from 0 to $+\infty$ $\Rightarrow s_= = 2 m_Q  \tilde{e}_q $ and 
\begin{equation}
	(s-m_Q^2) \times (s- m_g^2)=
	4 m_Q^4\left[  \frac{\tilde{e}_q^2}{m_Q^2} +  (1+\bar{m}_g)^2 \frac{\tilde{e}_q}{m_Q} +
	\bar{m}_g   (1+\frac{\bar{m}_g}{2})(\bar{m}_g+\frac{1}{2})\right] \,, 
\end{equation}
where $\bar{m}_g=\frac{m_g}{m_Q}$. The product is written as a sum of three positive terms, which is convenient for the MC sampling. One thus defines 
\begin{eqnarray}
	\tilde{\Lambda}_n &\coloneqq& 
	\int_{\tilde e_q >0} \frac{\mathrm{d}^3 p_q}{e_q}\,e^{-\frac{p_q\cdot u_{f}}{T}}\times \frac{\tilde{e}_q^n}{m_Q^{n+2}}
\\
& =&
\frac{2\pi\,n! }{u_Q} \times 
\left(\frac{e^{-\frac{u_Q^0-u_Q}{T} e_{\rm min}'}}
{(u_Q^0-u_Q)^{n+1}}-\frac{e^{-\frac{u_Q^0+u_Q}{T}e_{\rm min}'}}
{(u_Q^0+u_Q)^{n+1}}\right)\times \left( \frac{T}{m_Q}\right)^{n+2} \,,
\label{eq:Lambdan}
\end{eqnarray}
where $u_Q=\|\vec{u}_Q\|$, and finds
\begin{equation}
	\tilde{\Sigma}= 4 m_Q^6\times \left[\tilde{\Lambda}_2 +  (1+\bar{m}_g)^2 \tilde{\Lambda}_1 +
	\bar{m}_g  (1+  \frac{ \bar{m}_g}{2})(\frac{1}{2} + \bar{m}_g) \tilde{\Lambda}_0 \right] \,,
\end{equation}
which completes the evaluation of the radiative rate $R_{\rm rad}^{{\rm sup}_5}$. 

\subsubsection{The sampling process for $m_q=0$}

The sampling is organized as follows: The probability for a leading heavy quark of mass $m_Q$ and momentum $\vec{p}_Q$, to interact with a parton of momentum $\vec{p}_q$ and mass $m_q=0$ from the QGP of temperature $T$ during some time step $\Delta t$ and to emit a gluon of mass $m_g$, is obtained as follows:

\begin{enumerate}
    
    \item Sample the potential scattering centers $n_{\rm rad}$ during $\Delta t$ using a Poissonian distribution with an average $R_{\rm rad}^{{\rm sup}_5} \Delta t$, eq.~(\ref{eq:rsup5}), (which should not be\footnote{The fact of having several candidates at this level is not a big issue as only a small fraction will be eventually accepted.} $\gg 1$). Then, for each radiation candidate:

    \item Sample $\vec{p}_q$ according to the integrand of eq.~(\ref{Sigma}) in the expression of $R_{\rm rad}^{{\rm sup}_5}$, eq.~(\ref{eq:rsup5}) more specifically chose the $n$ of $\Lambda_n$ and sample $\vec{p}_q$ with the help of eq.~(\ref{eq:Lambdan}). As $\tilde{e}_q$ is required, the condition $s>(m_Q+m_g)^2)$ will be systematically satisfied.

    \item Accept the $s$ value according to the probability $\frac{\ln \frac{s}{m_g^2}}{\frac{s}{m_g^2}-1}$; If not accepted, reject the radiation candidate. Candidates are thus accepted at this stage according to $\sigma_{\rm rad}^{{\rm sup}_4}$ at eq.~(\ref{eq:sigmaradsup4}).  

    \item Sample $x$ according to the distribution $\frac{\mathrm{d}\sigma_{\rm rad}^{{\rm sup}_4}}{\mathrm{d}x}$ (approximated by eq.~(\ref{eq:dsigmaradsup4dxapprox})), with $x_{\rm min}< x <1$, and accept this value according to the ratio $\frac{\mathrm{d}\sigma_{\rm rad}^{{\rm sup}_3}}{\mathrm{d}x}$ --- eq.~(\ref{eq:dsigmaradsup3dx}) --- over $\frac{\mathrm{d}\sigma_{\rm rad}^{{\rm sup}_4}}{\mathrm{d}x}$ --- eq.~(\ref{eq:sigmaradsup4}). $x$ is then distributed according to $\frac{\mathrm{d}\sigma_{\rm rad}^{{\rm sup}_3}}{\mathrm{d}x}$.

    \item Sample $l_\perp$ according to $\frac{\mathrm{d}^3\sigma_{\rm rad}^{{\rm sup}_3}}{\mathrm{d}x \, \mathrm{d}^2 l_\perp}$ and reject it according to the ratio $\frac{\mathrm{d}^2\sigma_{\rm rad}^{{\rm sup}_1}}{\mathrm{d}x \, \mathrm{d}^2l_\perp}/\frac{\mathrm{d}^3\sigma_{\rm rad}^{{\rm sup}_3}}{\mathrm{d}x \, \mathrm{d}^2l_\perp}<1$.

    \item Sample $\vec{k}_\perp$ according to $P_g$ for fixed $x$ and $l_\perp$. This corresponds to a sampling of $\frac{\mathrm{d}\sigma_{\rm rad}^{{\rm sup}_1}}{\mathrm{d}x \, \mathrm{d}^2l_\perp \, \mathrm{d}^2k_\perp}$ in eq.~(\ref{eq:sup1}).

\item Check whether the sampled variables satisfy the boundary conditions $B^+>0$. If not, reject the radiation candidate.

\item At this stage, we are dealing with accepted (pre)gluons and one just needs to generate the final kinematics assuming that the particles are on-shell. One recalls that $\vec{l}_\perp$ and $\vec{k}_\perp$ are interpreted as transverse momenta (only) in the frame where $\vec{p}_Q$ and $\vec{p}_q$ are collinear. One can either generate the final kinematics in this frame and then boost back to the local QGP rest frame or evaluate invariant quantities that will be used to directly generate the final four-momenta in the local QGP frame. In the eikonal limit where $p_Q^+$ is the largest scale and $k^+$ the second largest scale, the transferred four-momentum $l$ is dominated by its transverse component while $l^+$ and $l^-$ are both $\propto 1/p^+_Q$. Assuming $l^+=l^-=0$ however implies small violations of the energy-momentum conservation. These can be cured by an explicit evaluation of $l^+$ and $l^-$. 

    \item If one deals properly with the kinematics, one can demonstrate that $l^+$ and $l^-$ become of the same order as $l_\perp$ whenever $x \lesssim \frac{m_g^2+k_\perp^2}{{p_Q^+}^2}$, what corresponds to gluons radiated \emph{backwards}. For this kinematic range, approximating the exchanged gluon propagator by $(l_\perp^2+\mu^2)^{-1}$ as done in eq.~(\ref{eq:Melapproximate}) leads to a large excess of gluon radiated backwards. A pragmatic way to cure this problem --- without building a new sampling strategy --- consists in oversampling the initial rate by a factor $\times 2$ and permuting the role of $\vec{p}_Q$ and $\vec{p}_q$ once placed in the CM frame with $50\%$ probability (thus considering that the gluon is radiated ``forward'' by the target parton in this frame) while rejecting gluons that are radiated in the opposite direction of their ``radiator'' ($y<0$ for gluons radiated by the projectile parton and $y>0$ for gluons radiated by the target parton).    

\end{enumerate}


\subsection{Sampling the Bethe-Heitler-Gunion-Bertsch seed for \texorpdfstring{$m_{q}=+\infty$}{mq=+inf}} 
\label{app:gbmqinfsampling}

We now adapt the strategy to the case of infinite mass, $m_q\to \infty$. Since this case is solely developed in order to perform the comparison with the BDMPS-Z benchmark results, we will assume an ultra-relativistic projectile and $x$ not too close to $1$. 


One starts from the definition~(\ref{eq:mqnot0xsec}), with 
$\Delta(\tilde{x}, \vec{k}_\perp,  \vec{l}_\perp,m_q) \to \Delta(\tilde{x}, \vec{0}_\perp,  \vec{0}_\perp,m_q)$ as a good approximation of the measure, as argued previously in this Appendix: \ref{appendix:phasespace}. 

\subsubsection{Phase space for $m_q\to \infty$}

To derive the phase space and the radiation cross section in the $m_q\to \infty$ limit, we naturally retain the leading terms of each factor, placing ourselves in the rest frame of the static scattering centers. $B^+$ (eq.~(\ref{eq:Bplusdef})) can be reformulated as
\begin{equation}
B^\pm \underset{m_q \rightarrow +\infty}{=}
x p_Q^+ \left((1-x)p_Q^+ + p_Q^-   \mp 2  m'_{Q,T}\right)  - (m_g^2 + k_\perp^2) \,.
\end{equation}
One can convince oneself that the $B^+\ge 0$ condition simply expresses that $p'_{Q,z}$, the projectile momentum along $Oz$ after the collision is real, using light-cone variables as defined in appendix~\ref{app:conventions}.\footnote{Neglecting $p_Q^-$,
$B^+>0\Leftrightarrow (1-x)\frac{p_Q^+}{2}  - \frac{k^-}{2} >m_{Q,T}'
\Leftrightarrow e_Q - x\frac{p_Q^+}{2} - \frac{k^-}{2} >m_{Q,T}'\Leftrightarrow e_Q - \omega >m_{Q,T}'$, which is precisely the onshellness condition for the outgoing projectile.} This translates, for a given $x$, into the condition for  ${\vec{l}_\perp,\vec{k}_\perp}$ in the transverse space 
\begin{equation}
\frac{m_g^2+ k_\perp^2}{x p_Q^+} +2 \sqrt{m_Q^2+(\vec{k}_\perp-\vec{l}_\perp)^2}\le (1-x) p_Q^+ + p_Q^- \,.
\label{eq:C3}
\end{equation} 
For a given $\vec{k}_\perp$, the $\vec{l}_\perp$ domain is a disk centered at $\vec{k}_\perp$ and of radius 
\begin{equation}
l_\perp^{\rm max}(k_\perp)=\sqrt{\left(\frac{(1-x)p_Q^++p_Q^-}{2} - \frac{m_g^2 + k_\perp^2}{2 x p_Q^+}  \right)^2- m_Q^2} \,. 
\end{equation}
Hence $l_\perp^{\rm max}(k_\perp)\propto \frac{p_Q^+}{2}$ for moderate $k_\perp$. This corresponds to a hard scattering shifting the direction of the incoming projectile by $\pi/2$. Conversely, the largest $k_\perp$ is obtained when $\vec{l}_\perp=\vec{k}_\perp$. Neglecting all masses eq.~(\ref{eq:C3}) reduces to 
\begin{equation}
k_\perp^{\rm max} \propto \sqrt{x}\sqrt{1-x} p_Q^+ \,,
\label{eq:ktmaxmqinf}
\end{equation}
for an ultra-relativistic projectile.
As already announced, $\Delta_{\rm a2}=\Delta(x,0,0,m_q)$  -- where the finite-$m_q$ expression of $\Delta$ is given in eq.~(\ref{eq:deltadef}) -- is used 
in the expression of $\sigma_{\rm rad}$, with the $m_q\to\infty $ limit of $\Delta_{\rm a2}$
\begin{equation}
\Delta_{\rm a2} \underset{m_q\to\infty}{=} (p_Q^+)^{4} \left( 
x  \left((1-x)  + \frac{2  m_Q}{p_Q^+}\right)  - \frac{m_g^2}{{p_Q^+}^2} \right)\times
\left( 
x  \left((1-x)  -  \frac{2  m_Q}{p_Q^+}\right)  - \frac{m_g^2}{{p_Q^+}^2} \right) \,.
\end{equation}

At leading order in $\mathcal{O}(m_g/p_Q^+)$ and $\mathcal{O}(m_Q/p_Q^+)$, the $x$ boundaries $x'_{\rm min}$ and $x'_{\rm max}$ obtained by setting $\Delta_{\rm a2}=0$ are 
\begin{equation}
x'_{\rm min}=\left(\frac{m_g}{p_Q^+}\right)^2  \quad\text{and} \quad
x'_{\rm max}=1-\frac{2 m_Q}{p_Q^+}.
\end{equation}
It should be noted that these definitions differ from the ones obtained previously for finite $m_q$. This is the consequence of the non-commutativity between the $e_Q\to \infty$ and $m_q\to \infty$ limits. With these boundaries, one can write 
\begin{equation}
\sqrt{\Delta_{\rm a2}} \approx  (p_Q^+)^{2} (x-x'_{\rm min}) \sqrt{(x'_{\rm max}-x)(2-x'_{\rm max}-x)} \approx (p_Q^+)^{2}  (x-x'_{\rm min})(1-x) \,,
\end{equation}
where the $p_Q^{+}\to \infty $ limit was taken for $x'_{\rm max}$, as will be assumed in the rest of this appendix.

\subsubsection{Explicit radiative cross section for $m_q\to \infty$}

As $p_{\rm tot}^+  \underset{m_q\to\infty}{ \approx} m_q$ while $p_Q^+\approx 2 e_Q$, one has 
\begin{equation}
\frac{p_Q^+}{p_{\rm tot}^+\sqrt{\Delta}} \approx \frac{1}{2 m_q e_Q (x-x'_{\rm min})\times(1-x)} \,.
\label{eq:fact1mqinf}
\end{equation}
Injecting this result in eq.~(\ref{eq:mqnot0xsec}) yields
\begin{equation}
\frac{\mathrm{d}^5\sigma_{\rm rad}^{m_q\to \infty}}{\mathrm{d}x \mathrm{d}^2 l_\perp \mathrm{d}^2} k_\perp\approx \frac{\pi^3  \left\vert\mathcal{M}_{\rm el}\right\vert^2}{(2\pi)^5 v_{qQ}} \frac{1}{2 m_q e_Q }\frac{x}{(x-x'_{\rm min})} P_g \Theta(B^+)\,,
\label{dsigmarad1m0}
\end{equation}
an expression which is formally similar to eq. \ref{dsigmarad1} for the $m_q=0$ case.
Coming to the elastic amplitude (eq.~(\ref{eq:Melapproximate})) one has $s_-=2 m_q e_Q$ for a static quark $q$, as well as $v_{qQ}=m_q e_Q$ and thus
\begin{equation}
\frac{\left\vert\mathcal{M}_{\rm el}\right\vert^2}{v_{qQ}} \approx (m_q e_Q) \times \frac{8C_F g^4}{N_c}
\times \frac{1-\frac{l_\perp^2}{4e_Q^2}}{(l_\perp^2+\mu^2)^2} \,.
\label{eq:fact2mqinf}
\end{equation}
Combining both eqs.~(\ref{dsigmarad1m0}) and (\ref{eq:fact2mqinf}),
we obtain 
\begin{eqnarray}
\frac{\mathrm{d}^5\sigma_{\rm rad}^{m_q\to \infty}}{\mathrm{d}x \, \mathrm{d}^2l_\perp \, \mathrm{d}^2k_\perp} 
&\approx& 
\frac{1}{64 \pi^2}
 \frac{8C_F g^4}{N_c}
\times \frac{1-\frac{l_\perp^2}{4e_Q^2}}{(l_\perp^2+\mu^2)^2}	
\frac{x}{(x-x'_{\rm min})}  P_g \times \Theta(B^+)
\nonumber\\
&\approx&
\frac{2 C_F \alpha_s^2}{N_c}
\times \frac{1-\frac{l_\perp^2}{4e_Q^2}}{(l_\perp^2+\mu^2)^2}	
\frac{x}{(x-x'_{\rm min})}  P_g \times \Theta(B^+) \,. \label{eq:mqinfxsec}
\end{eqnarray}

We will proceed now to the integration of this cross section similarly to the $m_q=0$ case. Ignoring the $\Theta(B^+)$ phase space condition, one arrives at 
(the $m_q\rightarrow +\infty$ is now implicit in the notation):
\begin{equation}
	\frac{\mathrm{d}^3\sigma_{\rm rad}^{{\rm sup}_1}}{\mathrm{d}x \, \mathrm{d}^2l_\perp } 
	\approx \frac{2 C_F \alpha_s^2\left(1-(\frac{l_\perp}{2 e_Q})^2 \right)x}{N_c (l_\perp^2+\mu^2)^2(x'_{\rm min}-x)}
	\int \mathrm{d}^2k_\perp P_g = \frac{2 C_F \alpha_s^2\left(1-(\frac{l_\perp}{2 e_Q})^2 \right)x}{N_c (l_\perp^2+\mu^2)^2(x'_{\rm min}-x)} \times
	I_1 \,,
\end{equation}
where $I_1$ is defined by eq.~(\ref{klimno}). We thus proceed with the inequalities (see eq.~(\ref{eq:I1}))
\begin{equation}
	I_1 < I_1^{\rm sup} \coloneqq \frac{2C_A\alpha_s}{\pi}\frac{1-x}{x}
	\ln\left(1+\frac{l_\perp^2}{e\tilde{m}_g^2}\right) \quad\text{and}\quad
	1-\left (\frac{l_\perp}{2 e_Q} \right )^2 \le 1 \,,
\end{equation}
 to define 
\begin{eqnarray}
\frac{\mathrm{d}^3\sigma_{\rm rad}^{{\rm sup}_1}}{\mathrm{d}x \, \mathrm{d}^2l_\perp} <
\frac{\mathrm{d}^3\sigma_{\rm rad}^{{\rm sup}_3}}{\mathrm{d}x \, \mathrm{d}^2l_\perp}&=&
\frac{2 C_F \alpha_s^2}{N_c(l_\perp^2+\mu^2)^2} \times
\frac{2C_A\alpha_s}{\pi}\frac{1-x}{x-x'_{\rm min}}
\ln\left(1+\frac{l_\perp^2}{e\tilde{m}_g^2}\right)
\nonumber\\
&=&
\frac{4 C_F \alpha_s^3}{\pi} \times \frac{1-x}{x-x'_{\rm min}} \times \frac{\ln\left(1+\frac{l_\perp^2}{e\tilde{m}_g^2}\right)}{(l_\perp^2+\mu^2)^2} \,,
\end{eqnarray} 
as $C_A=N_c$. 
One should notice that this expression is quite similar to the one obtained in the $m_q\rightarrow 0$ limit --- see eq.~(\ref{eq:d3sigmaradsup3}) --- with $x'_{\rm max}\to 1$ and $x_{\rm min} \to x'_{\rm min}$ in the expression of $c_2(x)$. One can thus proceed as in the $m_q=0$ case, which ends with the definition of 
\begin{equation}
\frac{\mathrm{d}\sigma_{\rm rad}^{{\rm sup}_3}}{\mathrm{d}x}\approx  
4 C_F\alpha_s^3 \times \frac{\ln e (\tilde{m}_g)^2/\mu^2}{e (\tilde{m}_g)^2-\mu^2}\times \frac{(1-x)}{x-x'_{\rm min}} \,,
\label{eq:dsigmadxsup3mqinf}
\end{equation} 
and next
\begin{equation}
\sigma_{\rm rad}^{{\rm sup}_4}\approx  4 C_F\alpha_s^3 \times \frac{\ln e (\tilde{m}_g^{\rm min})^2/\mu^2}{e (\tilde{m}_g^{\rm min})^2-\mu^2}\times \underbrace{
\int_{x'_{\rm min}}^1 \frac{(1-x)\mathrm{d}x}{x-x'_{\rm min}}}_{I_2} \,,
\label{eq:sup4}
\end{equation} 
where $\tilde{m}_g^{\rm min}$ is here the smallest value of $\tilde{m}_g$ on $[x'_{\rm min},1]$. We find 
\begin{equation}
I_2\approx \ln \frac{1}{x'_{\rm min}} =\ln \frac{(p_Q^+)^2}{m_g^2} \,,
\end{equation} 
if one approximates $1/(x-x'_{\rm min})$ by $1/x$, leading to
\begin{equation}
\sigma_{\rm rad}^{{\rm sup}_4}\approx  4 C_F\alpha_s^3 \times \frac{\ln e (\tilde{m}_g^{\rm min})^2/\mu^2}{e (\tilde{m}_g^{\rm min})^2-\mu^2}\ln \frac{(p_Q^+)^2}{m_g^2} \,.
\label{eq:sigmaradsup4mqinfty}
\end{equation} 
Comparing this expression with the equivalent one for the $m_q=0$ case (eq.~\ref{eq:sigmaradsup4}) we see that for the $m\to \infty$ limit s is replaced by $(P_Q^+)^2$, a relation which we found already when comparing the $k_\perp^{\rm max}$ values for the $m_q=0$ (eq. \ref{eq:ktmaxmq0}) and the $m_q\to \infty$ (eq. \ref{eq:ktmaxmqinf}) case.

Keeping the exact expression and adding a small offset $x'_{\rm min}\to \frac{m_g^2+\Lambda_r^2}{(p_Q^+)^2}$, one would get, at leading order in $\Lambda_r$,
\begin{equation}
I_2\approx (1-x'_{\rm min}) \left( \ln \frac{(p_Q^+)^2(1-x'_{\rm min})}{\Lambda_r^2} -1 \right) 
\approx (1-x'_{\rm min}) \left( \ln \frac{(p_Q^+)^2-m_g^2}{\Lambda_r^2} -1 \right) \,,
\end{equation}
which is however not used in the present implementation relying on eq.~(\ref{eq:sigmaradsup4mqinfty}).

\subsubsection{The total radiative rate for $m_q \to \infty$}

The rate expression simplifies drastically as all scatterers are considered to be infinitely massive, leading to 
\begin{equation}
R_{\rm rad}^{{\rm sup}_4}(p_Q^+)  = n_q v_Q \sigma_{\rm rad}^{{\rm sup}_4}(p_Q^+) 
\approx  n_q \times 8 C_F \alpha_s^3  \times 
\frac{\ln\frac{e (\tilde{m}_g^{\rm min})^2}{\mu^2}}{e  (\tilde{m}_g^{\rm min})^2 -\mu^2}\,\ln \frac{p_Q^+}{m_g} \,,
\end{equation}
where $n_q$ should be considered as the density of partons (calculated in the $m_q=0$ approximation with a MB distribution\footnote{We are indeed just probing the consequence of the $m_q=0$ vs  $m_q=+\infty$ choice for the radiation, assuming the same density of scattering centers.}):
\begin{equation}
n_q = \frac{g_q}{(2\pi)^3} \int \mathrm{d}^3 p_q e^{-\frac{p_q}{T}}=
\frac{4 \pi g_q}{(2\pi)^3} \int p_q^2  dp_q e^{-\frac{p_q}{T}}= \frac{g_q T^3}{\pi^2} \,.
\end{equation}

Combining eq.~(\ref{eq:dsigmadxsup3mqinf}) -- considered as a {\it bona fide} approximation of the differential radiative cross section --  and the definition of the -- spin averaged -- elastic $Qq$ cross section within the same hypothesis,
$\sigma_{\rm el}=\frac{2 C_F \alpha_s^2}{N_c \mu^2}$, one gets
\begin{equation}
\frac{\omega \frac{\mathrm{d}\sigma_{\rm rad}}{\mathrm{d}\omega}}{ \sigma_{\rm el}}
\approx   \frac{2C_A\alpha_s}{\pi}  \times \frac{\ln e (\tilde{m}_g)^2/\mu^2}{e (\tilde{m}_g)^2/\mu^2-1}. 
\end{equation}
The power spectrum of gluons radiated over a path length $L$ can therefore be written as 
\begin{equation}
\omega\frac{\mathrm{d}N^{\rm GB}}{\mathrm{d}\omega}=
\omega \frac{\mathrm{d}\sigma_{\rm rad}}{\mathrm{d}\omega} \,L n_q \approx  
\frac{2C_A\alpha_s}{\pi}  \times \frac{\ln e (\tilde{m}_g)^2/\mu^2}{e (\tilde{m}_g)^2/\mu^2-1}\times \frac{L}{\lambda_{\rm el}} \,,
\label{eq:approxradrate} 
\end{equation} 
as $\frac{L}{\lambda_{\rm el}}=n_q \sigma_{\rm el}$.
We find here the elastic mean free path because in every elastic collisions there is a probability that a gluon is emitted (see eq.~(\ref{eq:mq0xsec}))

\subsubsection{The sampling algorithm for $m_q \to \infty$}

The sampling algorithm consists in a simplified version of the one for $m_q=0$. The sampling strategy is as following: For a leading quark of mass $m_Q$ and momentum $\vec{p}_Q\gg m_Q$ interacting with some QGP of temperature $T$ during some time step $\Delta t$:
\begin{enumerate}
\item Sample the number of radiation candidates $n_{\rm rad}$ during $\Delta t$ using a Poissonian distribution of average $R_{\rm rad}^{{\rm sup}_4} \Delta t$ (which should not be $\gg 1$). Then, for each radiation candidate:
\item Sample $x$ according to $1/x$ -- that is $\approx \mathrm{d}\sigma_{\rm rad}^{{\rm sup}_4}/\mathrm{d}x$ eq.~\eqref{eq:sup4} with $x'_{\rm min}< x <1$.
\item Accept $x$ according to the ratio $\frac{\mathrm{d}\sigma_{\rm rad}^{{\rm sup}_3}}{\mathrm{d}x}/\frac{\mathrm{d}\sigma_{\rm rad}^{{\rm sup}_4}}{\mathrm{d}x}$.
\item Sample $l_\perp$ 	according to $\frac{\mathrm{d}^3\sigma_{\rm rad}^{{\rm sup}_3}}{\mathrm{d}x \, \mathrm{d}^2l_\perp}$.
\item Accept the $l_\perp$ value according to the ratio $\frac{\mathrm{d}^2\sigma_{\rm rad}^{{\rm sup}_1}}{\mathrm{d}x \, \mathrm{d}l_\perp^2}/\frac{\mathrm{d}^3\sigma_{\rm rad}^{{\rm sup}_3}}{\mathrm{d}x \, \mathrm{d}l_\perp^2}<1$.
\item Sample $\vec{k}_\perp$ according to the corresponding $\frac{\mathrm{d}\sigma_{\rm rad}^{{\rm sup}_1}}{\mathrm{d} x \,  \mathrm{d}^2l_\perp \, \mathrm{d}^2k_\perp}$, means according to $P_g$.
\item Check whether the sampled variables satisfy the boundary condition $B^+>0$. If not, reject the radiation candidate.
\item Generate the  four-momentum of the outgoing gluon as $k^+=\omega + k_z = x p_Q^+ = x (p_Q+e_Q)\approx 2 e_Q$ and $k^-=\frac{m_g^2+k_\perp^2}{k^+} \Rightarrow \omega = \frac{k^+ + k^-}{2}=x e_Q + \frac{m_{g,T}^2}{4 x e_Q}$.
\item Generate the outgoing four-momentum of the parton applying energy conservation ($e'_Q=e_Q-\omega$), while $\vec{p}\,'_{Q\perp}=\vec{l}_\perp-\vec{k}_\perp$ and $p'_{Qz}$ follows from requiring that the parton is on-shell. Let us recall that in our MC we have two options to shift from $p_Q\to p'_Q=p_Q+l-k$: This shift can be performed at the time when the virtual gluon is emitted or when it is effectively formed as a real gluon, see discussion in section \ref{subsubsect:phaseaccumalgo}.

\end{enumerate}

\section{Calibration} \label{app:calibration}

As explained in subsection \ref{subsubsect:phaseaccumalgo}, our Monte Carlo (MC) model for jet energy loss has two parameters which needs to be adjusted:

\begin{itemize}

    \item The gluon phase accumulation critera $\varphi_c$. The $\omega$ power spectrum of BDMPS-Z is created by the interference between the forward going and the backward going amplitudes for gluon radiation. Interference terms cannot be easily formulated in usual MC based transport theories which propagate particles and where the collisions among these particles are independent, but they can be introduced in transport approaches with the help of phase accumulation. The gluon radiation amplitude involves phases $\varphi_j$~\cite{Baier:1996kr} which are accumulated from the vertex where the virtual gluon is created until the last rescattering of the gluon in the medium. The total phase which has to be accumulated to consider the gluon as formed, $\varphi_c$, is a free parameter of the model, and will affect the transition from the LPM regime to the GLV regime.

    \item The regulating gluon mass $m_g^{\mathrm{reg}}$ for energetic gluons appearing in the expression for $\tilde{m}_{g}$ in eq.~(\ref{eq:mgtilde}) in the initial Gunion-Bertsch seed. We note that the Gunion-Bertsch power spectrum in the LPM region, given in eq.~(\ref{eq:gb}) or (\ref{eq:approxradrate}), is reduced by the $N_s$ factor, stemming from the Monte Carlo sampling, $N_s \simeq \frac{1}{\lambda} \sqrt{\frac{\omega}{\hat{q}}\,\varphi_c}$, and shows the same scaling w.r.t.\ $\omega$, $\alpha_s$, and $T$ as the seeked BDMPS-Z power-spectrum itself (up to some subleading logs). However, it can differ by some numerical factors, as the adopted MC procedure, described in subsection \ref{subsubsect:phaseaccumalgo}, does not have the ambition to reproduce exactly eq.~(\ref{eq:bdmpsz}). To achieve such a goal, we have chosen to consider the gluon mass $m_g$ in the GB ``seed'' as a tunable parameter for large-$\omega$ gluons. As a matter of fact, the genuine gluon thermal mass is a subdominant scale for such gluons\footnote{Since $k_\perp^2 \propto  \sqrt{\omega \hat{q}} \gg m_g^2$.}, but it still appears in our two steps MC procedure through the gluon seed, as can be seen from the factor  $\frac{\ln e (\tilde{m}_g)^2/\mu^2}{e (\tilde{m}_g)^2/\mu^2-1}$ in eq.~(\ref{eq:approxradrate}). We thus take advantage of this factor, which ranges from 0 to $+\infty$, to fix a regulating mass $m_g^{\rm reg}$ for energetic gluons --- the ones that are emitted coherently --- in the expression of $\tilde{m}_g$ in eq.~(\ref{eq:mgtilde}). Due to the aforementioned scalings, one expects that the ``corrective'' factor $\frac{\ln e (\tilde{m}_g)^2/\mu^2}{e (\tilde{m}_g)^2/\mu^2-1}$ should be fairly independent of both $T$ and $\alpha_s$ in order to warrant a good matching between our Monte Carlo approach and the BDMPS-Z reference, we therefore assume that the scaling of $m_g^{\rm reg}$ and $\mu$ in $T$ and $\alpha_s$ should be similar, hence 
\begin{equation}
m_g^{\rm reg} \propto \sqrt{\alpha_s} \times T,
\label{eq:mgregofT}
\end{equation}
which assures a good robustness for all values of $T$ and $\alpha_s$.
\end{itemize}

Both parameters, $\varphi_c$ and $m_g^{\mathrm{reg}}$, needs to be calibrated in order to match theoretical predictions. We consider, as reference for the calibration, the improved opacity expansion (IOE)~\cite{Mehtar-Tani:2019tvy, Mehtar-Tani:2019ygg}, which includes the next-to-leading-order expansion in opacity around the harmonic oscillator solution. The IOE spectrum is able to reproduce both the BDMPS-Z behavior in the LPM regime, as well as the GLV behavior in the fully coherent regime. The IOE spectrum is given for a medium of finite length $L$ as the sum of a leading-order (LO) and next-to-leading-order (NLO) contribution,
\begin{equation}
    \frac{\mathrm{d}N^{\text{IOE}}}{\mathrm{d}\omega} = \frac{\mathrm{d}N^{\text{IOE}}_{\text{LO}}}{\mathrm{d}\omega} + \frac{\mathrm{d}N^{\text{IOE}}_{\text{NLO}}}{\mathrm{d}\omega} \,,
\end{equation}
where the LO contribution is given by the BDMPS-Z result in eq.~(\ref{eq:bdmpsz}), while the NLO correction takes the form~\cite{Mehtar-Tani:2019tvy, Mehtar-Tani:2019ygg}
\begin{equation}
    \omega \frac{\mathrm{d}N^{\text{IOE}}_{\text{NLO}}}{\mathrm{d}\omega} \simeq \frac{1}{2} \frac{\alpha_s C_R}{\pi} \hat{q}_0 \, \text{Re} \int_{0}^{L} \mathrm{d}s \, \frac{1}{k^2(s)} \left [ \ln \frac{k^2(s)}{Q^2} + \gamma_E \right ] \,,
\end{equation}
with the Euler-Mascheroni constant $\gamma_E \approx 0.55786$, and where we have, for a medium with constant density,
\begin{equation*}
    k^2(s) = i \frac{\omega \Omega}{2} \big [ \cot(\Omega s) - \tan(\Omega (L - s)) \big ] \,,
\end{equation*}
with $\Omega$ given in eq.~(\ref{eq:complexfreq}). The gluon transport coefficient stripped of its Coulomb logarithm, $\hat{q}_0$, is in our case given by
\begin{equation}
    \hat{q}_0 = \frac{8 C_A}{\pi} (N_f + N_C) \alpha_s^2 T^3 \,.
\end{equation}
The scale $Q^2$ is energy-dependent in the LPM regime, $Q^2 \simeq \sqrt{\omega \hat{q}}$.

The calibration is done for a medium path length $L = 4$ fm and a very large projectile energy, $E = 100$ TeV, to ensure that $\omega_c < E$, and that the IOE spectrum converges nicely to the GLV result in the fully coherent regime. Furthermore, the underlying assumption in the IOE result is to assume a large $q_{\perp, \mathrm{max}}$ in the elastic scatterings, hence the same is done in our Monte Carlo algorithm, though this will only have a small effect on the $\omega$ distribution at small $\omega$. For the calibration, the strong coupling constant has been fixed to $\alpha_s = 0.3$, such that we can consider $m_g^{\text{reg}}(T)/T$ a constant.

\begin{figure}[H]
    \centering
    \includegraphics[width=0.7\textwidth]{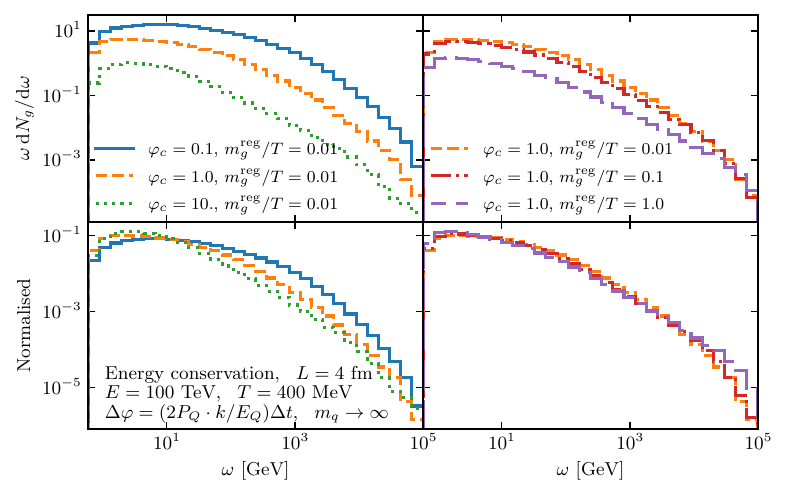}
    \caption{Top: The energy-weighted radiation spectrum for various values of $\varphi_c$ (left) and $m_g^{\text{reg}}/T$ (right). Bottom: The distributions normalized to $1$, in order to highlight the shape differences. }
    \label{fig:dNdw_varphic_mgreg}
\end{figure}

Fig.~\ref{fig:dNdw_varphic_mgreg} shows the behavior of the $\omega$ spectrum for varying values of $\varphi_c$ and $m_g^{\text{reg}}/T$ respectively. Both parameters changes the overall normalization of the spectrum, where larger values of $\varphi_c$ and larger values of $m_g^{\text{reg}}$ both decreases the normalization. It is mainly only $\varphi_c$ which changes the shape of the spectrum, especially the transition from the LPM regime to the fully coherent regime. Hence, for practical purposes, the shape can first be fixed with a suitable $\varphi_c$, and then the normalization can be fixed with a corresponding $m_g^{\text{reg}}$.

\begin{figure}[H]
\centering
\includegraphics[width=0.7\textwidth]{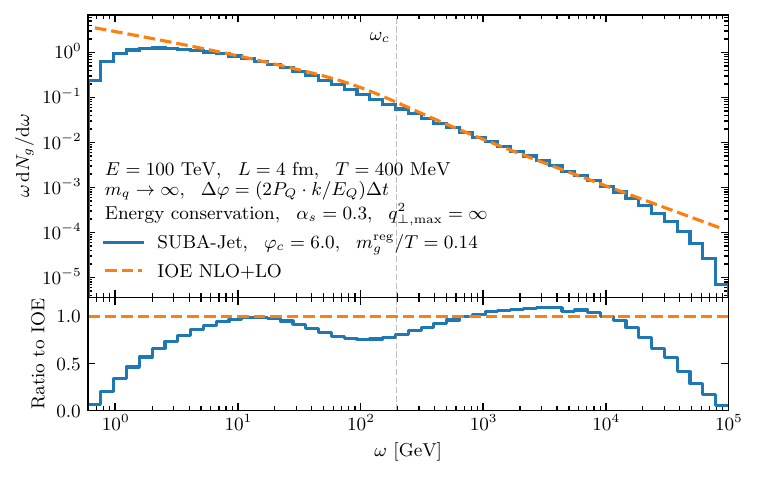}
\caption{The energy-weighted radiation spectrum comparing the Monte Carlo result with the IOE for the optimal choice of the parameters $\varphi_c$ and $m_g^{\text{reg}}$. }
\label{fig:dNdw_IOE}
\end{figure}

\noindent The calibration to the IOE spectrum (see Fig.~\ref{fig:dNdw_IOE}) reveals that the optimal choice for the gluon phase accumulation criteria is $\varphi_c \approx 6.0$. For $T=400$~MeV and $\alpha_s = 0.3$, the optimal prescription for $m_g^{\text{reg}}$ is found to be $56$~MeV. Following eq.~(\ref{eq:mgregofT}), one thus takes
\begin{equation}
    m_g^{\rm reg}(k^+\gg T) \approx 0.14\,\sqrt{\frac{\alpha_s}{0.3}}\times T \,.
\end{equation}
In the GB seed algorithm, $m_g$ is thus taken as an interpolation between the thermal gluon mass for very small $x$ (such that $k^+\approx T$) and $m_g^{\rm reg}$ for finite $x$.

Fig.~\ref{fig:dNdw_IOE} shows the comparison between our MC and the IOE spectrum for our specified parameters and with the optimal prescription for $\varphi_c$ and $m_g^{\text{reg}}$. The energy-weighted spectrum $\omega \, \mathrm{d}N/\mathrm{d}\omega$ is shown, to better illustrate the transition from the LPM regime to the fully coherent regime, happening around $\omega \sim \omega_c \equiv \hat{q}_0 L^2 / 2 \approx 196$~GeV, where $\hat{q}_0$ is the transport coefficient stripped of its Coulomb logarithm. Both our MC and the IOE reproduce the BDMPS-Z expectation in the LPM regime and the GLV result in the fully coherent limit. The agreement between our MC and the IOE is fulfilled nicely in the regions where the IOE is expected to be valid. For both very low (Bethe-Heitler) and very large ($z \to 1$) energies, the IOE is not expected to hold. Around the transition region, $\omega \sim \omega_c $, the IOE spectrum exhibits a sharp transition, which seems unphysical. On the other hand, our MC shows a smoother transition.




\bibliographystyle{utphys}
\bibliography{refs-jets}

\end{document}